\documentclass[twocolumn, twocolappendix]{aastex631} 

\usepackage{savesym}
\savesymbol{tablenum}
\usepackage{siunitx}
\usepackage{amsmath}
\usepackage{graphicx}
\usepackage{tabularx,colortbl}
\usepackage{makecell}
\usepackage{booktabs}
\usepackage{hyperref}
\usepackage{boldline}
\usepackage{mathtools}
\usepackage{lipsum}
\usepackage{float}
\usepackage{footmisc}
\usepackage{pifont}  
\usepackage{xurl}
\usepackage[T1]{fontenc}
\usepackage[ruled,vlined,linesnumbered]{algorithm2e}
\usepackage{comment}
\usepackage{subfigure}
\usepackage{makecell}
\SetArgSty{textnormal}
\DontPrintSemicolon
\allowdisplaybreaks
\usepackage{_custom_defs}  
\usepackage{_settings}  

\newcommand{\citea}[1]{\citeauthor{#1} (\citeyear{#1})}

\begin{document}

\title{
Deep Attention-Based Supernovae Classification of Multi-Band Light-Curves
}
\author{Óscar Pimentel}
\affiliation{Department of Electrical Engineering, Universidad de Chile, Av. Tupper 2007, Santiago 8320000, Chile.}
\affiliation{Millennium Institute of Astrophysics (MAS), Nuncio Monseñor Sótero Sanz 100, Providencia, Santiago, Chile.}

\author[0000-0001-9164-4722]{Pablo A. Estévez}
\affiliation{Department of Electrical Engineering, Universidad de Chile, Av. Tupper 2007, Santiago 8320000, Chile.}
\affiliation{Millennium Institute of Astrophysics (MAS), Nuncio Monseñor Sótero Sanz 100, Providencia, Santiago, Chile.}

\author[0000-0003-3459-2270]{Francisco Förster}
\affiliation{Data and Artificial Intelligence Initiative (D$\&$IA), University of Chile.}
\affiliation{Millennium Institute of Astrophysics (MAS), Nuncio Monseñor Sótero Sanz 100, Providencia, Santiago, Chile.}
\affiliation{Center for Mathematical Modeling, Universidad de Chile, Beauchef 851, North building, 7th floor, Santiago 8320000, Chile.}
\affiliation{Departamento de Astronomía, Universidad de Chile, Casilla 36D, Santiago, Chile.}

\received{}
\revised{}
\accepted{}
\submitjournal{AJ}

\begin{abstract}
In astronomical surveys, such as the Zwicky Transient Facility, supernovae (SNe) are relatively uncommon objects compared to other classes of variable events. Along with this scarcity, the processing of multi-band light-curves is a challenging task due to the highly irregular cadence, long time gaps, missing-values, few observations, etc. These issues are particularly detrimental to the analysis of transient events: SN-like light-curves. We offer three main contributions: 1) Based on temporal modulation and attention mechanisms, we propose a Deep attention model (TimeModAttn) to classify multi-band light-curves of different SN types, avoiding photometric or hand-crafted feature computations, missing-value assumptions, and explicit imputation/interpolation methods. 2) We propose a model for the synthetic generation of SN multi-band light-curves based on the Supernova Parametric Model, allowing us to increase the number of samples and the diversity of cadence. Thus, the TimeModAttn model is first pre-trained using synthetic light-curves. Then, a fine-tuning process is performed. The TimeModAttn model outperformed other Deep Learning models, based on Recurrent Neural Networks, in two scenarios: late-classification and early-classification. Also, the TimeModAttn model outperformed a Balanced Random Forest (BRF) classifier (trained with real data), increasing the balanced-$F_1$score from $\approx.525$ to $\approx.596$. When training the BRF with synthetic data, this model achieved similar performance to the TimeModAttn model proposed while still maintaining extra advantages. \highlighttext{3) We conducted interpretability experiments. High attention scores were obtained for observations earlier than and close to the SN brightness peaks. This also correlated with an early highly variability of the learned temporal modulation.}
\end{abstract}


\section{Introduction}\label{sec:intro} 
The study of transient astronomical events, specifically supernovae (SNe), has played a critical role in astronomy. Type Ia SNe (thermonuclear SNe) are standardizable candles and have become important tools for cosmological distance determinations \citep[][]{Wright2017}, leading to the discovery of the accelerated expansion of the universe \citep[][]{Schmidt1998, Riess1998} and its precise characterization with projects such as the Dark Energy Survey \citep[DES;][]{Sanchez2006, DarkEnergySurveyCollaboration2016}. At the same time, the study of Type Ib/c and Type II supernovae (core-collapse SNe) has helped astronomers to understand the evolution and explosion mechanisms of stars, including insights into the formation of stellar mass black holes \citep[][]{Sukhbold2020}.

These expanded opportunities to study the cosmos are a consequence of the constant efforts to develop new telescopes that collect massive amounts of data every night, creating a new Big Data paradigm for astronomy. High-volume data collection is managed by astronomical surveys such as the Zwicky Transient Facility survey \citep[ZTF;][]{Bellm2019} and experiments such as the High Cadence Transient Survey \citep[HiTS;][]{Forster2016}. These surveys are preparing us for the Vera C. Rubin Observatory and its Legacy Survey of Space and Time \citep[LSST;][]{Ivezic2019}. The LSST survey is expected to gather approximately 15 terabytes of raw data per night by observing up to 37 billion astronomical objects in 10 years, including several millions of SNe \citep[][]{Ivezic2019}.

\subsection{Previous Works} 
Historically, SNe have been studied and classified into different types through optical spectroscopy. However, the use of this technique requires an immense investment of time and human effort. Given this limitation, only a marginal proportion of the SN candidates, reported from high-volume data streams, are being effectively studied and followed-up.

The Big Data paradigm challenge has motivated the scientific community to search for alternative methods for classification other than spectroscopic observations. In particular, several methods have been proposed to classify different types of SNe using the discovery images and light-curves. Most existing methods are based on features extracted from the light-curves by using parametric models \citep[][]{Karpenka2013, Arnett2016, Lochner2016, Villar2019}, PCA and Kernel PCA reductions \citep[][]{Ishida2013, Lochner2016}, Wavelet-based features \citep[][]{Varughese2015, Lochner2016}, Gaussian processes light-curve augmentation \citep[][]{Boone2019}, and different hand-crafted features \citep[][]{Villar2019}.

A successful example for the processing of discovery images and light-curves is the Automatic Learning for the Rapid Classification of Events broker \citep[ALeRCE;][]{\alercerefs}. The ALeRCE broker considered a vast collection of features\footnote{\url{http://alerce.science/features/}.} based on prior astrophysical expert knowledge \citep{Sanchez-Saez2021}, including an SN parametric model. These extracted features are used to classify SNe (or other astronomical events) along with classical Machine Learning models, such as the Balanced Random Forest (BRF), Multi-Layer Perceptron (MLP), Support Vector Machine (SVM), and Gradient Boosting.

A notable difficulty is an intrinsic scarcity in the number of empirical SN light-curves, especially for certain SN types such as the Superluminous SNe (SLSN), which also leads to a high class imbalance within the SN classes. These difficulties have motivated the release of several simulated SN light-curve datasets as part of data classification challenges, such as the Supernova Photometric Classification Challenge \citep[SPCC;][]{Kessler2010} and the Photometric LSST Astronomical Time-Series Classification Challenge \citep[PLAsTiCC;][]{PLAsTiCC2018}.

Several Deep Learning models have been motivated by these challenges. In \citea{Charnock2017}, SN light-curve classifiers based on Recurrent Neural Networks (RNNs) were proposed using models such as the Gated Recurrent Unit (GRU) and Long Short-Term Memory (LSTM). To deal with the multi-band missing-values, a light-curve imputation was performed using random values between the last and next valid light-curve observations. In \citea{Moss2018}, a Phased-LSTM model was used to include the time information as a new memory gate, computing averages between the last and next observations to deal with missing-values.

\highlighttext{The PELICAN project \citep[][]{Pasquet2019} proposed an autoencoder (encoder-decoder) architecture based on the use of Convolutional Neural Networks (CNNs) that are adapted to process time series, such as SN multi-band light-curves. This work dealt with the irregular cadence by using a missing-value assumption, where a set of additional loss functions were proposed to attenuate the overfitting risk associated with zero mask values. In \citea{Brunel2019}, an adapted CNN based model was also proposed for the processing of SN light-curves along with a promising Siamese network architecture.}

The RAPID project \citep[][]{Muthukrishna2019} used a GRU model to classify different transient and SN types, including a new pseudo-class to characterize the SN pre-explosion region. The irregular cadence and multi-band misalignments were treated using a grid linear interpolation. In \citea{Moller2020} a Bayesian RNN model was developed, where the time difference information between the current and last observation was included to describe the irregular cadence information as model input.

In addition, notable efforts have been made for the classification of other astronomical light-curves, such as variable stars and stochastic events. Deep Learning encoder and autoencoder models (encoder-decoder), based on RNN models \citep[][]{Naul2018, Jamal2020, Tachibana2020, Donoso-Oliva2021} and Temporal CNN (TCNN) models \citep[][]{Jamal2020, Zhang2021}, have been proposed for the automatic feature extraction from light-curves. Moreover, the direct processing of image-stamp sequences has been also proposed using Recurrent CNNs (RCNNs) \citep[][]{Carrasco-Davis2019, Gomez2020}.

As a competitive alternative to RNNs, CNNs, and TCNNs, light-curve classification models based on attention mechanisms have started to emerge. In \citea{Ibsen2020}, a GRU model was jointly used with a self-attention mechanism to improve the early-classification performance. Recently, in \citea{Allam2021}, a model to classify light-curves using an adapted Transformer model was developed, where a Gaussian process interpolation method was used to deal with the irregular cadence. However, both works used simulated light-curves from the PLAsTiCC dataset in a completely supervised learning scheme and heavily relying on light-curve interpolation methods.

\subsection{Main Contributions} 
In this work, we propose an attention-based model for the classification of different types of SN by using empirical multi-band light-curves from the ZTF survey. Our main contributions are the following: 1) We propose a Deep Attention model (TimeModAttn), based on temporal modulation (TimeFiLM) and attention mechanisms (MHSelfAttn), to process and classify SN multi-band light-curves. The proposed model avoids the computation of any time-consuming photometric or hand-crafted features, as well as the use of missing-value assumptions and explicit light-curve imputation/interpolation methods. From our experiments, we found that the TimeModAttn model achieved higher performance than other classical baselines: a feature-based BRF model trained with real data, and RNN-based models (GRU, LSTM). 2) To support the optimization of the tested Deep Learning models, we propose a new method to generate synthetic SN multi-band light-curves as an effort to increase both, the total number of samples and the diversity of the irregular cadence population from the original dataset. 3) We conduct several interpretability experiments for SN multi-band light-curves to explore, evaluate, and validate the proposed model.

\subsection{Organization of This Work} 
This \docname is structured as follows: First, in section \refsec{sec:notation}, we introduce the mathematical notation used in this work. In section \refsec{sec:background}, a brief background about relevant theoretical concepts is presented. Next, in section \refsec{sec:dataset}, we introduce the light-curve dataset used in this work and describe the dataset pre-processing procedures. In section \refsec{sec:synth}, we describe our methodology for generating synthetic SN multi-band light-curves. In section \refsec{sec:brf_baseline}, the classifier baseline used for comparison purposes is described, which is based on photometric features and the Balanced Random Forest (BRF) baseline. In section \refsec{sec:deep}, we describe the complete methodology associated with our proposed model (TimeModAttn) and the optimization process for the classification of SN light-curves. We also describe other baselines based on RNN models. In section \refsec{sec:results}, the results from our experiments are reported using several metrics to compare the performance of the TimeModAttn model w.r.t. the tested baselines. In addition, we conduct several interpretability experiments based on the proposed model. Finally, in section \refsec{sec:conclusions}, we conclude and propose guidelines for future work in this research line.
\section{Notation}\label{sec:notation} 
\subsection{Multi-Band Light-Curve} 
An arbitrary $i$-th multi-band light-curve $\Phi_i$, from a light-curve dataset, is defined as follows:
\begin{align}
\Phi_i\newdef\set{\tuple{\obs_{i,j},\obse_{i,j}, t_{i,j}, b_{i,j}}}{j\seq 1}{L_i} & | t_{i,j'}>t_{i,j},\forall j'>j,
\end{align}
where the light-curve $\Phi_i$ is defined as a sequence set\footnote{$\set{\phi_{i,j}}{j\seq 1}{L_i}=\keys{\phi_{i,1},\dots,\phi_{i,L_i}}$, where $\phi_{i,j}$ is an arbitrary object.} with an arbitrary (variable-length) number of $L_i$ photometric multi-band observations. Each observation contains photometric information, such as the observation-flux $\obs_{i,j}$ (flux) and the observation-error $\obse_{i,j}$ (flux error). Also, each observation is associated with an observation-time $t_{i,j}$ (days) and an observation-band indicator $b_{i,j}\in \keys{1,\dots,B}$, where $B$ is the total number of photometric bands available on the survey dataset. \highlighttext{Note that the subscript notation $i,j$ represents the $j$-th observation of the $i$-th light curve $\Phi_i$}. The light-curve sequence object is defined to be causally sorted over time, i.e., the observation-time $t_{i,j}$ increases monotonically if the sequence step $j$ also increases: $t_{i,j'}>t_{i,j}, \forall j'>j$.

In addition, given a target band $b$, a single-band operator $\tuple{\cdot}^\ob$, for an arbitrary $i$-th multi-band light-curve $\Phi_i$, is defined as follows\footnote{Note that the band indicator $b_{i,\job}$ is redundant after applying the band operator.}:
\begin{align}
\Phi_i^\ob&\newdef\tuple{\Phi_i}^\ob\newdef\tuple{\tupleset{\obs_{i,j},\obse_{i,j}, t_{i,j},b_{i,j}}{j=1}{L_i}}^\ob,\nonumber\\
&\newdef\tupleset{\obs_{i,\job},\obse_{i,\job}, t_{i,\job},b_{i,\job}}{\job=1}{L_i^\ob},
\end{align}
where the resulting single-band light-curve $\Phi_i^\ob$ is defined as a sequence set collection of all the photometric observations from the multi-band light-curve $\Phi_i$ that are associated with the selected band $b$. In this case, the light-curve $\Phi_i^\ob$ has an arbitrary (variable-length) number of $L_i^\ob\leq L_i$ photometric observations\footnote{The total variable-length of the multi-band light-curve is the sum from all band observations as $L_i = \sum_{b\seq 1}^{B} L_i^\ob$.}. For simplicity, if a sequence step has the form $j^\ob$, it is then related with the single-band light-curve $\Phi_i^\ob$. This single-band operator is used in the following sections to define operations exclusively over the target band $b$.

\subsection{First and Last Sequence Steps} 
The sequence step $j=1$ is associated with the very first observation from a multi-band light-curve $\Phi_i$ (which can occur at any band). The sequence step $j=L_i$ is associated with the very last observation from a multi-band light-curve $\Phi_i$ (at any band). As a simplified notation, we use $1$ and $-1$ for the first and last sequence steps, respectively (e.g., $\obs_{i,1}$, $\obs_{i,-1}$).

The sequence step $j^\ob=1$ is associated with the very first observation from a single-band light-curve $\Phi_i^\ob$ (first observation, occurring in band $b$, from the multi-band light-curve $\Phi_i$). The simplified sequence step $j^\ob=L_i^\ob$ is used to denote the sequence step that is associated with the very last observation from a single-band light-curve $\Phi_i^\ob$ (last observation, occurring in band $b$, from the multi-band light-curve $\Phi_i$). As a simplified notation, we use $1^\ob$ and $-1^\ob$ for the first and last sequence steps, respectively (e.g., $\obs_{i,1^\ob}$, $\obs_{i,-1^\ob}$).

\subsection{Time Difference}\label{sec:time_diff} 
Given a multi-band light-curve $\Phi_i$, an arbitrary time difference is defined as follows:
\begin{align}
\Delta t_{i,j} &\newdef
\begin{cases}
0, & \text{if }j=1,\\
t_{i,j} - t_{i,j-1}, & \text{otherwise },\\
\end{cases}
\end{align}
where the time difference associated with the first observation is $\Delta t_{i,1}=0$. Moreover, the time difference between the current observation (at the sequence step $j$) and the previous observation (at the sequence step $j-1$) is denoted as $t_{i,j} - t_{i,j-1}$.

In addition, given a single-band light-curve $\Phi_i^\ob$, an arbitrary time difference is defined as follows:
\begin{align}
\Delta t_{i,\job}^\ob &\newdef
\begin{cases}
t_{i,\job}-t_{i,1}, & \text{if }\job=1,\\
t_{i,\job} - t_{i,\job-1}, & \text{otherwise },\\
\end{cases}
\end{align}
where the time difference associated with the first observation is $t_{i,\job}-t_{i,1}$. Thus, the first time difference is $\Delta t_{i,1^\ob}^\ob=0$ only if the first observation of the multi-band light-curve $\Phi_i$ occurs in the target band $b$. Moreover, the time difference between the current observation (at the sequence step $j^\ob$) and the previous observation (at the sequence step $j^\ob-1$) is denoted as $t_{i,j^\ob} - t_{i,j^\ob-1}$.
\section{Theoretical Background}\label{sec:background} 
In this section, a theoretical background is presented. First, the multi-head self-attention mechanism (MHSelfAttn) is described, explaining key concepts for the understanding of the model proposed in section \refsec{sec:deep}. Second, the Supernova Parametric Model (SPM) is introduced, supporting the synthetic generation method presented in section \refsec{sec:synth}.

\subsection{Multi-Head Self-Attention Mechanisms}\label{sec:background/attn} 
The attention mechanisms were initially conceived as strategies to support the processing of word sequences (tokens) in the Natural Language Processing (NLP) research field. One of the first architectures to include an attention mechanism was proposed as an alternative to improve the performance of an LSTM-RNN model \citep[][]{Hochreiter1997} for the language translation task \citep{Bahdanau2014}. In general, this composite architecture design, based on RNNs and supported by attention mechanisms, has been vastly used in the NLP research field for several years.

Later on, the idea of implementing Deep Learning models based solely on attention mechanisms was introduced by the Transformer model \citep[][]{Vaswani2017}. An increasing interest in the use of attention mechanisms has been developed over the last few years along with notable examples, such as the BERT model \citep{Devlin2018} or the GTP model \citep{Radford2019} for different NLP tasks. Recently, this interest has also spread among other research fields outside NLP, with attention-based models used in general multi-variate time-series classification \citep{Lin2020}, healthcare and clinical time-series processing \citep{Horn2019, Lee2021, Shukla2021}, financial time-series \citep{Kim2019}, or simulated photometric transient light-curves classification \citep{Ibsen2020, Allam2021}.

\subsubsection{Multi-Head Dot-Attention} 
\highlighttext{Here, the multi-head dot-attention mechanism is described, which is the principal mechanism of the Transformer model \citep[][]{Vaswani2017}. The attention mechanisms allow us to process and extract information from a collection of an arbitrary number of vectors. Moreover, this capability can be extended to process a causal temporal sequence of vectors. In this work, a sequence of vectors, each one composed of an observation-flux, an observation-time and/or a band indicator, is used to represent a variable-length light-curve.}

\highlighttext{The definition of the multi-head dot-attention mechanism is based on concepts such as the query, key, value, and context vectors. In particular, in self-attention, we would like to process different sequence steps of a sequence of vectors. The query is a vector related with the current sequence step from the input sequence. The keys are all vectors from the previous sequence steps. An attention mechanism computes the alignment weights (attention scores) of the relative importance of all the keys for the given query. Typically, the alignment weights are computed using the dot-product operation between the query vector and all the key vectors. Then, the alignment weights are multiplied with the input sequence (the value vectors) to get a new weighted sequence. Finally, a single context vector is obtained as the sum of this weighted sequence. Note that the attention mechanism could also be used to work with two different sequences. In such a case, the keys are obtained from the first sequence, while the values are from the second sequence.}

\highlighttext{Thus, the attention mechanism can \quo{pay high attention to} and retrieve all the best-matched values according to the best-matched associated keys given the current query. Additionally, a summarized description (context) can be constructed by summarizing all the common and relevant information among all the best-matched retrieved values. Note that the above operation (single-head attention mechanism) could be performed multiple times independently, leading to a multi-head attention mechanism, where, depending on the query, different contexts can be constructed. This means, for example, that one head may pay attention to the previous sequence step, while another could pay attention to the early sequence steps in the sequence. As a final step, the context vectors obtained from each attention head are concatenated and linearly projected, gathering the information from all the attention heads.}

\begin{figure}[!t]
\centering
\includegraphics[width=\onecolumnscale]{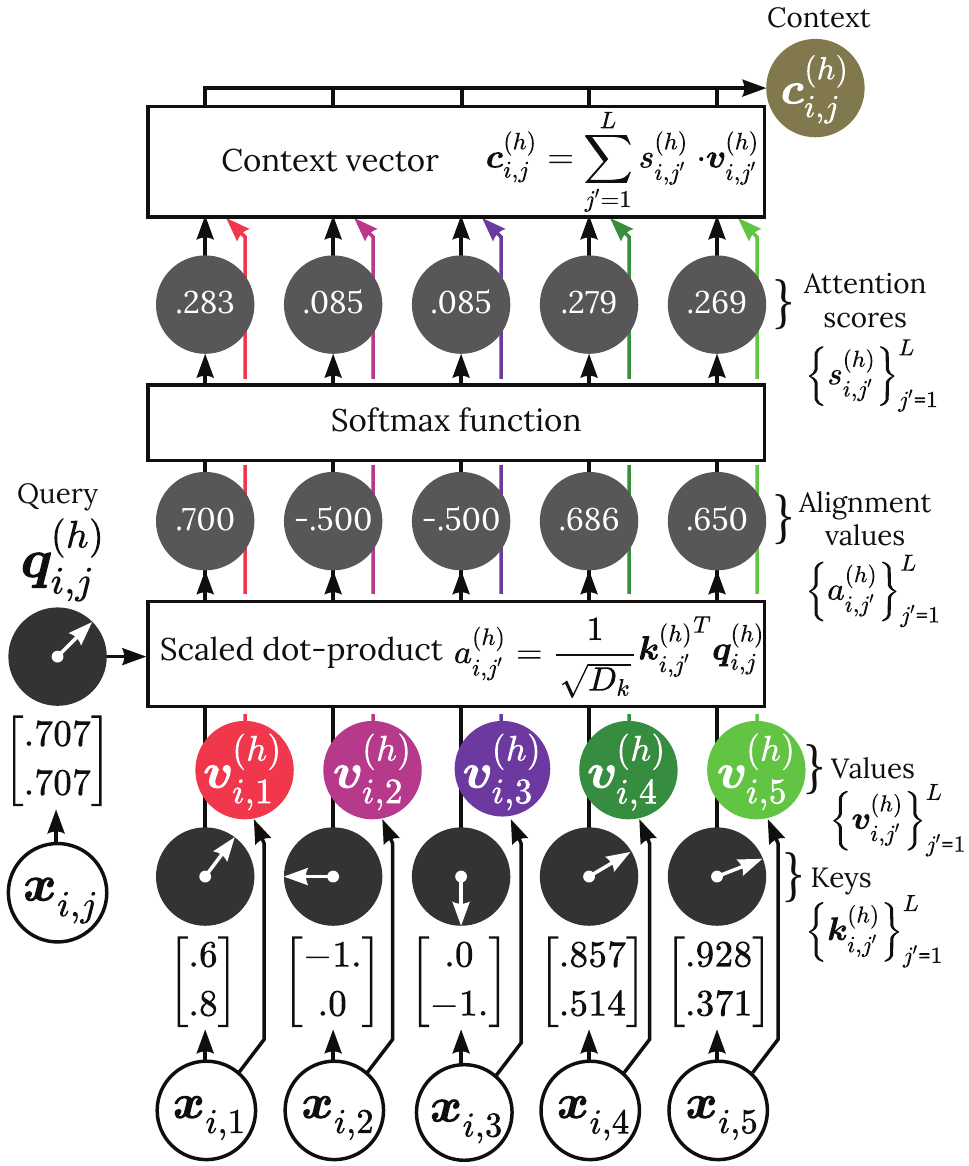}
\caption{
\highlighttext{A simplified example of a dot-attention mechanism given an arbitrary attention head $h$. To better illustrate the alignments, the query and key vectors are represented as pointing arrow 2D vectors (e.g., $\q_{i,j}^{(h)}=\vect{.707, .707}$). Those 2D vectors (and the value vectors) are the projected vectors, from the input vectors, obtained by using the associated learnable linear projections and bias vectors for the attention head $h$. The attention scores are based on the scaled dot-product between the query and key vectors. The value vectors are represented as color codes, explaining the final color used for the context vector $\vtc_{i,j}^{(h)}$. This operation can be extended and parallelized for an arbitrary number $H$ of attention heads.}
}
\label{fig:dot_attn}
\end{figure}

\highlighttext{Let's assume a sequence set of arbitrary input vectors $\keys{\x_{i,1},\dots,\x_{i,L}}=\set{\x_{i,j'}}{j'\seq 1}{L}$ (associated with the key and value vectors) and an arbitrary input vector $\x_{i,j}$ (associated with the query vector), where the sequence of vectors is analogous to an arbitrary light-curve $\Phi_i$. As it is fully detailed in section \refsec{sec:deep}, each input vector $\x_{i,j}$ is a high-dimensional representation that is expected to automatically contain information about the current observation, such as the observation-flux $\obs_{i,j}$, the band indicator $b_{i,j}$, and the observation-time $t_{i,j}$ (days). The equations that describe the multi-head dot-attention mechanism are the following\footnote{According to the implementation in \url{https://pytorch.org/docs/stable/generated/torch.nn.MultiheadAttention.html}. For extra details, see the original Transformer model in \citea{Vaswani2017}.}:}
\begin{align}
&a_{i,j'}^{(h)} = \frac{1}{\sqrt{D_k}} \underbrace{\tuple{\mtw_k^{(h)^T}\x_{i,j'}+\b^{(h)}_k}^T}_{\text{key: }\k^{(h)}_{i,j'}}\underbrace{\tuple{\mtw_q^{(h)^T}\x_{i,j}+\b^{(h)}_q}}_{\text{query: }\q^{(h)}_{i,j}} ,&&\customtext{Aligment values}\label{eq:attn_a}\\
&s_{i,j'}^{(h)} = \frac{\expfun{a^{(h)}_{i,j'}}}{\sum_{j'\seq1}^L \expfun{a^{(h)}_{i,j'}}},&&\customtext{Attention score}\label{eq:attn_s}\\
&\vtc_{i,j}^{(h)} = \sum_{j'\seq1}^L s_{i,j'}^{(h)} \cdot \underbrace{\tuple{\mtw_v^{(h)^T}\x_{i,j'}+\b^{(h)}_v}}_{\text{value: }\vtv^{(h)}_{i,j'}},&&\customtext{Context vector}\label{eq:attn_ch}\\
&\vtc_{i,j} = \mtw_c^{T}\tuple{\cat{\underbrace{\vtc_{i,j}^{(1)},\dots,\vtc_{i,j}^{(H)}}_{\text{head contexts}}}}+\b_c,&&\customtext{Multi-head context vector}\label{eq:attn_c}
\end{align}
\highlighttext{where eqs. \refeq{eq:attn_a}-\refeq{eq:attn_ch} follow the example scheme shown in Fig. \ref{fig:dot_attn}: given a query vector $\q_{i,j}^{(h)}$, the goal of the attention mechanism is to compute a context vector $\vtc_{i,j}^{(h)}$ based on the set of key vectors $\set{\k_{i,j'}^{(h)}}{j'\seq 1}{L}$ and the set of value vectors $\set{\vtv_{i,j'}^{(h)}}{j'\seq 1}{L}$. Note that the computation of the context vector can be performed independently and in parallel for each attention head $h\in\keys{1,\dots,H}$, where $H$ is an arbitrary number of context vector computations: the number of attention heads (denoting $h$ as an arbitrary attention head). Thus, in eq. \refeq{eq:attn_c}, the final context vector $\vtc_{i,j}$ is computed by using the information from all the $H$ parallel attention heads (the operator $\cat{\dots}$ stands for the concatenation operator for vectors). A detailed explanation of the multi-head dot-attention mechanism is as follows:}

\begin{enumerate}
\item \textbf{Alignment values}: given an arbitrary attention head $h$, in eq. \refeq{eq:attn_a}, the alignment values $\set{a_{i,j'}^{(h)}}{j'\seq 1}{L}$ are computed as the dot product between a query vector $\q_{i,j}^{(h)}$ and the key vectors in the set $\set{\k_{i,j'}^{(h)}}{j'\seq 1}{L}$. Both, the query and key vectors, are projected from an input vector $\x_{i,j}$ and a set of input vectors $\set{\x_{i,j'}}{j'\seq 1}{L}$, respectively. A learnable linear projection $\mtw^{(h)}_q\in\real{D\times D_q}$, and a bias vector $\b_q$, are used for the query vector; and a learnable linear projection $\mtw^{(h)}_k\in\real{D\times D_k}$, and a bias vector $\b_k$, are used for the key vectors. \highlighttext{The dimensions $D$, $D_q$, and $D_k$ are associated with the input, query, and key vectors, respectively}\footnote{\highlighttext{The dimensions $D_q$ and $D_k$ are selected as $D_q=D/H$ and $D_k=D/H$, respectively \citep[][]{Vaswani2017}. Note that the dimensions for the query and key values must be the same in order to properly compute the dot-product between both vectors.}}. The alignment values are computed using a scaled dot-product operation between the query and key vectors as $\frac{1}{\sqrt{D_k}} \k_{i,j'}^T\q_{i,j}$ (scaled dot-attention mechanism). This operation represents an estimation of the linear correlation between the query and key vectors, where a high correlation implies a high alignment value.

\item \textbf{Attention scores}: given an arbitrary attention head $h$, in eq. \refeq{eq:attn_s}, the attention scores $\set{s_{i,j'}^{(h)}}{j'\seq 1}{L}$ are computed using the alignment values $\set{a_{i,j'}^{(h)}}{j'\seq 1}{L}$. This operation is performed using the softmax function over the set of alignment values, constructing a discrete distribution, where the following condition holds: $\sum_{j'\seq 1}^{L}s_{i,j'}^{(h)}=1,s_{i,j'}^{(h)}\in (0,1),\forall h\in\keys{1,\dots,H}$. Note that the best-matched key vectors, w.r.t. the query vector, achieve higher attention scores.

\item \textbf{Context vector}: given an arbitrary attention head $h$, in eq. \refeq{eq:attn_ch} the context vector $\vtc_{i,j}^{(h)}$ is computed as the vector aggregation from the value vectors $\set{\vtv_{i,j'}^{(h)}}{j'\seq 1}{L}$ weighted by the attention scores $\set{s_{i,j'}^{(h)}}{j'\seq 1}{L}$. These value vectors are projected from the same input vectors used for the key values, through the learnable linear projection $\mtw^{(h)}_v\in\real{D\times D_v}$ and a bias vector $\b_v$, where $D_v=D/H$ is the value vector dimension. Recalling that the attention scores represent a discrete distribution, this operation is like the estimation of the vector expectation over the set of value vectors. Finally, if a given query $\q_{i,j}^{(h)}$ gets a high alignment with the key $\k_{i,j'}^{(h)}$; then, the associated vector $\vtv_{i,j'}^{(h)}$ will be weighted higher in the resulting context vector $\vtc_{i,j}^{(h)}$.

\item \textbf{Multi-head context vector}: one of the novel architecture ideas introduced by the Transformer model is the multi-head attention capability. This configuration allows the model to distribute the attention computation, as described above, among several heads running independently and in parallel. This usually helps to increase the model performance, as each head can specialize in different tasks, paying attention to different patterns along the input sequence. Thus, in eq. \refeq{eq:attn_c}, a final context vector $\vtc_{i,j}$ is projected from the concatenation of all the $H$ parallel attention heads context vectors $\keys{\vtc_{i,j}^{(1)},\dots,\vtc_{i,j}^{(H)}}$. This final context is computed through the learnable linear projection $\mtw_c\in\real{(H \cdot D_v)\times D}$ and a bias vector $\b_c$. Finally, this operation allows the model to capture relevant information from all the $H$ parallel context vectors $\vtc_{i,j}^{(h)}$.
\end{enumerate}

\subsubsection{Multi-Head Self-Attention} 
\begin{figure}[!t]
\centering
\includegraphics[width=\onecolumnscale]{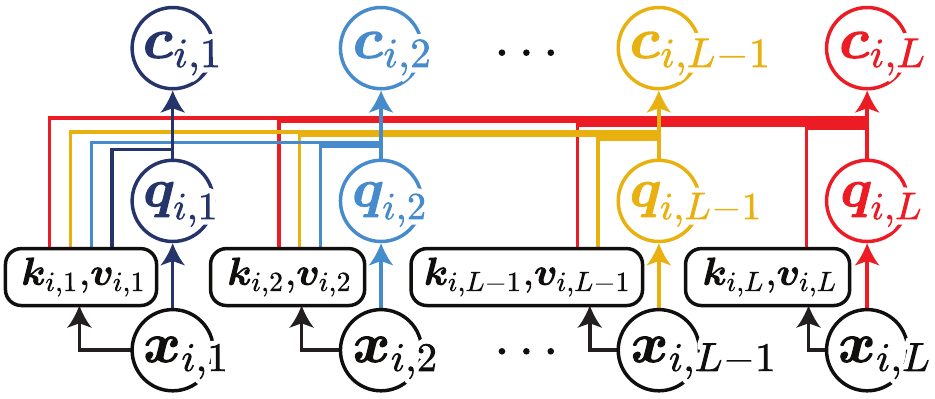}
\caption{
Self-attention example diagram. The key $\k_{i,j}$, value $\vtv_{i,j}$, and query $\q_{i,j}$ vector relationships are shown for the 1st, 2nd, $(L-1)$-th, and $L$-th sequence steps. For simplicity, we assume a single attention head mechanism and omit the attention head superscript $\tuple{\cdot}^{(h)}$. This operation is causal as each context vector $\vtc_{i,j}$ computation only depends on the current and previous sequence steps. All query, key, and value vectors come from the same sequence.
}
\label{fig:self_attn}
\end{figure}

A self-attention scenario arises when the query vectors are projected from the same sequence domain as the key and value vectors. Given a sequence of input vectors $\set{\x_{i,j'}}{j'\seq 1}{L}$, the context vector $\vtc_{i,j}$ (at an arbitrary sequence step $j$) is computed using query vectors also projected from the current sequence step $j$, with key and value vectors projected from the current and previous sequence steps: $\keys{1,\dots,j}$. An example diagram for this operation is shown in Fig. \ref{fig:self_attn}, where the causality of the self-attention mechanisms can be observed.

This configuration directly produces a causal context vector computation for each intermediate sequence step $j$, obtaining a new context vector sequence from the original input vector sequence, where both share the same variable-length $L$. This process reminds the causal structure in the output of the RNN sequential processing: an output sequence where each memory-vector computation strictly depends on the current and previous memory-vectors.

\subsubsection{Additional Operations} 
Additional Residual Connections (in the form of an arbitrary function $f(\x_{i,j})$ are added to the current function input $\x_{i,j}$), plus a nonlinear operation. The Transformer model formulation is as follows\footnote{For simplicity of the explanation, we omit the Layer Normalization \citep[][]{Ba2016} used in the original Transformer model.}:
\begin{align}
&\vtc_{i,j}' = \underbrace{\vtc_{i,j}}_{\mathrlap{\text{MHSelfAttn: }f(\x_{i,j})}}+\x_{i,j},&&\text{}\label{eq:attn_res1}\\
&\vtc_{i,j}'' = \underbrace{\mtw_2^{T}\tuple{\nonlinearfun{ReLU}{\mtw_1^{T}\vtc_{i,j}'+\b_1}}+\b_2}_{\text{MLP: }f(\vtc'_{i,j})}+\vtc_{i,j}',&&\text{}\label{eq:attn_res2}
\end{align}
where a 1-hidden-layer Multi-Layer Perceptron model \citep[MLP;][]{Rumelhart1986} is used to induce nonlinear interactions among the context vectors from all heads. The terms $\mtw_1\in\real{D\times (k_\text{mlp} \cdot D)}$ and $\mtw_2\in\real{(k_\text{mlp} \cdot D)\times D}$ are the MLP linear projections and bias vectors $\b_1$ and $\b_2$, where $k_\text{mlp}=2$ controls the number of the MLP hidden units used: $k_\text{mlp} \cdot D$. The expression $\nonlinear{ReLU}$ stands for the Rectified Linear Unit function (ReLU).

Recalling that self-attention can compute a new causal sequence of context vectors, this operation can be stacked into a multi-layer sequence processing architecture using an arbitrary number of $N_L$ layers. For simplicity, in the following sections of this \docname, we use the term Multi-Head Self-Attention (MHSelfAttn) for the aforementioned and extended formulation, calling the processed vector $\vtc_{i,j}''$ for the context vector $\vtc_{i,j}$.

\subsubsection{Extra Properties and Limitations} 
Some extra properties of the attention mechanisms are the following:
\begin{enumerate}
\item \textbf{Long sequences}: empirically, attention-based models have shown higher performance than the RNN models for long sequence processing in NLP tasks \citep{Vaswani2017}. For RNN models, the maximum length between long-term dependencies (maximum path length) results in $\complexity{L}$ as the maximum path connection requires passing over the complete sequence length. In contrast, for attention mechanisms, this maximum path length is $\complexity{1}$ given the existence of a direct shortcut path between each context vector and each value vector, connecting the context with the entire sequence at a constant cost. These direct paths facilitate the learning of long-term sequence dependencies \citep{Hochreiter2001}.

\item \textbf{Parallelizable computation}: attention mechanisms have a complexity, per layer, of $\complexity{L^2 \cdot D}$ operations, while RNN models have a complexity of $\complexity{L \cdot D^2}$ operations. This implies that attention mechanisms have lower computational complexity than the RNN models when $L$ is lower than the embedding dimension $D$ \citep{Vaswani2017}. However, the RNN models require a number of $\complexity{L}$ strictly non-parallelizable sequential operations as each memory-vector computation requires the previously computed memory-vector. In contrast, for attention mechanisms, all context vectors for both, the entire sequence and $H$ parallel heads, can be computed simultaneously in a highly parallelizable operation which is optimal for the GPU usage (constant number of $\complexity{1}$ non-sequential operations).

\item \textbf{Attention masks}: the variable-length and explicit causality of the self-attention formulation are directly implemented with attention score masks over the alignment values. A negative infinite value $-\infty$ can be used to replace the alignments values over invalid sequence steps, where null attention is required (zero attention score). Optionally, null attention scores can be randomly imputed in the attention mask, during the model training process, as a sequence dropout regularization technique \citep{Vaswani2017}.

\item \textbf{Positional encoding}: one of the main limitations of the attention mechanisms is the loss of the sequential information, i.e., the explicit information of which sequence step comes before or after another sequence step. This is a direct consequence of the value vector aggregation using the attention scores: an operation that is invariant to the order of the vectors. \highlighttext{As a solution, the Transformer model proposed the use of a positional encoding vector, which is a collection of fixed sinusoidal waves that aims to preserve the sequential information for models based solely on attention mechanisms}.
\end{enumerate}

\highlighttext{Table \ref{tab:sequential_models_comparison} shows a comparison between the self-attention mechanism w.r.t. RNNs and other sequential models.}

\def\tabsrule{\rule{0pt}{0pt}\rule[0pt]{0pt}{0pt}}
\def\tabbline{\Xcline{1-5}{1.5pt}\tabsrule}
\begin{table*}[!t]
\centering
\caption{
Comparison between different approaches for processing sequences (e.g., time series, light-curves): Recurrent Neural Networks (RNNs), Convolutional Neural Networks (CNNs), Temporal CNNs (tCNNs), and the multi-head dot-attention mechanism. We denote the dimensionality of the model embedding by $D$, the variable-length of a processed sequence by $L$, and the kernel size of convolutions by $k$. For all models, a causal configuration is assumed.
}
\label{tab:sequential_models_comparison}\vspace{.1cm}
\scriptsize\footnotesize\small\normalsize
\scriptsize
\begin{tabularx}{\textwidth}{p{0.1\textwidth}XXXX}
\tabbline
Feature & RNNs (e.g., LSTM, GRU)  & CNNs & TCNNs & Multi-head self-attention \tabsrule\\ 
\cmidrule{2-5}
Maximum path length & Complete: a cost of $\complexity{L}$ is required as the model is required to pass over the complete sequence & Medium: a stack of $\complexity{L/k}$ convolutions are required to reach the furthest sequence step & Medium: a stack of $\complexity{\log_k(L)}$ convolutions are required to reach the furthest sequence step (less than those required for CNNs) & Direct: a cost of $\complexity{1}$ is required as a direct path exists to each sequence step  \tabsrule\\
Complexity per layer & $\complexity{L\cdot D^2}$  & $\complexity{k\cdot L \cdot D^2}$  & $\complexity{k\cdot L \cdot D^2}$ & $\complexity{L^2\cdot D}$ \tabsrule\\
Number of sequential non-parallelizable operations & $\complexity{L}$: RNNs have a strictly sequential optimization process. This may result in a computational cost bottleneck  & $\complexity{1}$: it is highly parallelizable and optimal for GPU usage & same as CNNs & $\complexity{1}$: it is highly parallelizable and optimal for GPU usage  \tabsrule\\
Handling of variable length & Variable length can be directly handled by performing a recurrent graph unrolling for RNNs & The standard implementation is not designed to directly handle variable length input: other strategies, after the processing of CNNs, must be used, e.g., pooling operations & Same as CNNs & Variable length can be directly handled by inducing null attention in the attention score masks \tabsrule\\
Interpretability & The flow of the information in the memory vector of RNN could be hard to interpret, especially in experiments that use real data & Even if interpretability experiments can be performed based on the learned convolution kernels, this is not direct for convolution over time series data & Same as CNNs & Different and explicit interpretability experiments can be designed by exploring the attention scores that can help us to understand the model \tabsrule\\
\tabbline
\end{tabularx}
\end{table*}
\subsection{Supernova Parametric Model (SPM)} 
The Supernova Parametric Model \citep[SPM;][]{Villar2019, Sanchez-Saez2021} is an analytical function that attempts to describe the typical behavior of a SN light-curve. The SPM function is defined as follows:
\begin{align}
\fsne(t;\thetabf) &= \fearly(t) \cdot (1-\fearlylate(t))+\flate(t) \cdot \fearlylate(t),\label{eq:spm}\\
\fearlylate(t) &= \sigma\tuple{s\cdot\tuple{t-(\gamma + t_0)}},\label{eq:spm_g}\\
\fearly(t) &=\frac{A\cdot\tuple{1-\spmbeta\frac{\tuple{t-t_0}}{\gamma}}}{1+\expfun{\frac{-(t-t_0)}{\trise}}},\label{eq:spm_early}\\
\flate(t) &= \frac{A\cdot(1-\spmbeta)\cdot\expfun{\frac{-(t-(\gamma+t_0))}{\tfall}}}{1+\expfun{\frac{-(t-t_0)}{\trise}}},\label{eq:spm_late}
\end{align}
where the SPM function $\fsne(t):\funmap{\real{}}{\real{}}$ allows us to construct a light-curve (flux) for a SN by evaluating the SPM model, given a vector of SPM parameters $\thetabf=\vect{A,t_0,\gamma,\spmbeta,\trise,\tfall}$, along with a collection of arbitrary and continuous-time values $\set{t_{j}}{j\seq 1}{L}$ (days). This analytical function is defined as a smooth transition between an early function $\fearly(t):\funmap{\real{}}{\real{}}$, in eq. \refeq{eq:spm_early}, and a late function $\flate(t):\funmap{\real{}}{\real{}}$, in eq. \refeq{eq:spm_late}. The use of these functions aims to characterize a typical SN behavior: the brightness abruptly increases (SN-rise) up to a maximum (SN-peak), followed by a decrease (SN-fall) where a plateau or radioactive tail (SN-plateau) could be observed, and ending with the final dimming of the transient event (SN-dimming). The transition between the early and late functions is controlled by the function $\fearlylate(t):\funmap{\real{}}{(0,1)}$, where $\sigma$ is the logistic sigmoid function and $s=.2$ is a transition smoothness control factor.

Intuitions behind the 6 SPM parameters are given by:
\begin{enumerate}
\item  $A\in\real{+}$: affects the brightness scale for the SN light-curve.
\item  $t_0\in\real{}$: acts as a temporal shift for the light-curve. Even though this value is close to the light-curve maximum brightness, it does not exactly correspond to the SN-peak time.
\item  $\gamma\in\real{+}$: controls the time duration of the SN-plateau region.
\item  $\spmbeta\in\boxx{0,1}$: controls the slope of the SN-plateau region.
\item  $\trise\in\real{+}$: controls the required time to reach the maximum brightness along the light-curve.
\item  $\tfall\in\real{+}$: controls the brightness decay time along and after the SN-plateau region.
\end{enumerate}
\section{Dataset and Pre-Processing}\label{sec:dataset} 
\subsection{Dataset} 
\begin{figure}[!t]
\centering
\includegraphics[width=\onecolumnscale]{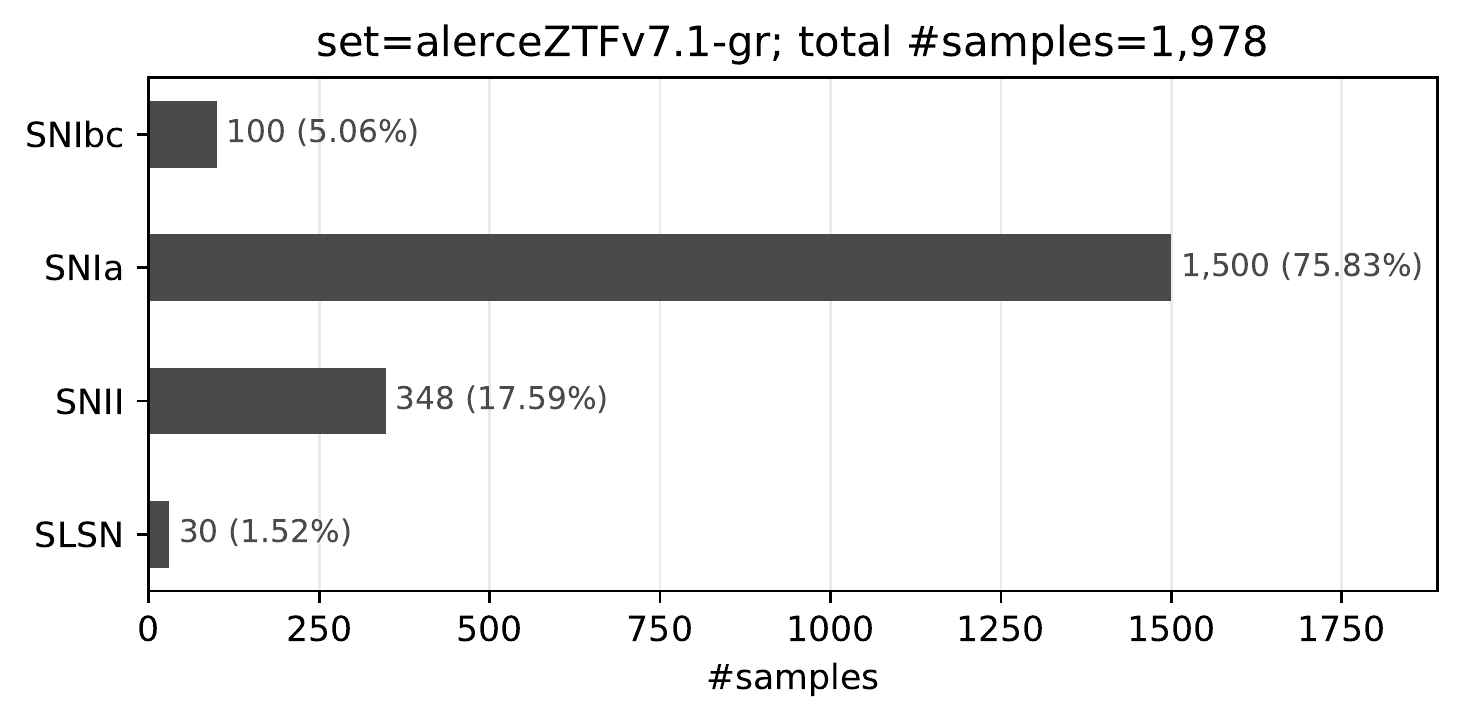}
\caption{
Class population distribution of SN types (from the original dataset $\dataset{})$.
}
\label{fig:classes}
\end{figure}

In this work, we use a dataset $\dataset{}$ that consists of a collection of flux SN multi-band light-curves from the Zwicky Transient Facility survey \citep[ZTF;][]{Bellm2019}, composed of two bands: g and r. These SN events have been confirmed spectroscopically and reported in the Transient Name Server (TNS) catalog\footnote{\url{https://wis-tns.weizmann.ac.il}.}. As we aim to classify different types of SNe, the following SN types are used from the dataset $\dataset{}$: SNIa, SNIbc, SNII, and SLSN, as researched in \citea{Sanchez-Saez2021}. We remove some short-length SN multi-band light-curves: only multi-band light-curves having at least $L_i^\ob\geq 5$ observations, in any of the $B$ bands, are preserved in the dataset $\dataset{}$. Fig. \ref{fig:classes} shows the class distribution, where a high class imbalance can be observed with majority classes (SNIa, SNII) and minority classes (SLSN, SNIbc).

\subsection{Pre-Processing}\label{sec:preprocessing} 
To prepare the dataset and light-curves, the following pre-processing procedures are implemented.

\subsubsection{Stratified 5-Fold Cross-Validation} 
A non-stochastic 5-fold cross-validation procedure is performed. The dataset $\dataset{}$ is split into 5 different variations of training/validation/test sets, following the proportion $\cardi{\dataset{train}}\allowbreak/\cardi{\dataset{val}}\allowbreak/\cardi{\dataset{test}}=60/20/20$. The imbalance of classes is similar for all sets as this split is stratified. When performing the splits, we aim to ensure that each fold configuration is unique, ensuring that each SN light-curve appears at least once in some of the test-set variations. This methodology aims to correctly preserve the representativeness of each split, which is important for the minority classes.

\subsubsection{Simultaneous Multiple Observations}\label{sec:filter_obs} 
Given a single-band light-curve $\Phi_i^\ob$, all photometric observations reported within a short-range time-window $\Delta t=12\tunits{hours}$ are merged into a single observation. Close observations are merged because same night observations can be redundant and can harm the performance of some algorithms. Given an arbitrary group of close observations, the merging process is performed by using weight factors $w_{i,\job}\in[0,1]$ defined as follows:
\begin{align}
w_{i,\job} &= \frac{e^{-\logfun{\obse_{i,\job}+\varepsilon}}}{\sum_{\job\in \Delta J_{i}^\ob} e^{-\logfun{\obse_{i,\job}+\varepsilon}}},\forall \job\in \Delta J_{i}^\ob\label{eq:w},
\end{align}
where $\obse_{i,\job}$ is the observation-error and $\Delta J_{i}^\ob$ is a subset of the sequence steps associated with all the observations, in the single-band light-curve $\Phi_i^\ob$, that are sharing the same time-window $\Delta t$. A new observation-flux can be generated as a weighted sum of all close observations within $\Delta t$, as $\obs_{i,\job} \gets\sum_{\job\in \Delta J_{i}^\ob} w_{i,\job}\cdot \obs_{i,\job}$. \highlighttext{The proposed weighting method allows observations with lower observation-errors (lower uncertainty) to be more represented in the final weighted sum. These weight factors are also used to generate the new observation-times and observation-errors. By merging close observations, the total number of observations in our ZTF dataset is reduced by approximately $13\%$\customfootnote{$13.151\%$.}.}

\subsubsection{Sigma Clipping Error Filter}\label{sec:sigma_clipping} 
Sigma clipping is used to remove highly uncertain photometric observations from the datasets. As this is related with the observation-error, the sigma clipping is applied to remove observations, from a single-band light-curve $\Phi_i^\ob$, with observation-errors $\obse_{i,\job}$ above a threshold of $5\sigma^\ob$. The standard deviation $\sigma^\ob$ is computed using all the observation-errors from the band $b$ in the training-set $\dataset{train}$.
\section{Generation of Supernova Synthetic Multi-Band Light-Curves}\label{sec:synth} 
In this section, a method to generate synthetic multi-band light-curves for SNe, based on the SPM model, is described. Given the scarcity of spectroscopically confirmed SN light-curves in the ZTF survey, we perform this artificial generation to support the training of several Deep Learning models with a stable and well-behaved optimization scenario.

\subsection{Estimation of the Posterior Distribution of SPM Parameters}\label{sec:synth.mcmc} 
To generate a new SN light-curve, a method to sample an optimal and well-behaved set of SPM parameters is required. We use a Bayesian framework to estimate the posterior distribution of the SPM parameters $\thetabf_{i}^\ob$, given an empirical single-band light-curve $\Phi_{i}^\ob$, as $p\tuple{\thetabf_{i}^\ob|\Phi_{i}^\ob}\propto p\tuple{\Phi_{i}^\ob|\thetabf_{i}^\ob}  p\tuple{\thetabf_{i}^\ob}$.

In this framework, a correct estimation of the posterior distribution allows us to sample SPM parameters given a set of empirical observations from a single-band light-curve as $\thetabf_{i}^{\ob*}\sim p\tuple{\thetabf_{i}^\ob|\Phi_{i}^\ob}$. For estimating this distribution, the Markov Chain Monte Carlo (MCMC) Ensemble Samplers algorithm \citep{Goodman2010} is used. To estimate the distribution with the MCMC algorithm, given an arbitrary single-band light-curve $\Phi_{i}^\ob$, the likelihood and prior distributions are defined as follows:
\begin{align}
&p\tuple{\Phi_{i}^\ob|\thetabf_{i}^\ob}= \nonumber\\
&\prod_{\job=1}^{L_{i}^\ob} \frac{1}{{\sigma'}_{i,\job}\sqrt{2\pi}} \expfun{\frac{-1}{2{\sigma'}_{i,\job}^2}\tuple{\obs_{i,\job}-\fsne\tuple{t;\thetabf_{i}^\ob}}^2},\label{eq:mcmc_likelihood}\\
&p\tuple{\thetabf_{i}^\ob} = \normdis{\thetabf_{i}^\ob;\thetabf_{i}^{(b'\neq b)},\Sigmabf}.\label{eq:mcmc_prior}
\end{align}

The explanation for the choices made above are the following:
\begin{enumerate}
\item \textbf{Likelihood}: in eq. \refeq{eq:mcmc_likelihood}, the likelihood distribution is defined. This formulation is based on the assumption of a Gaussian distribution for the empirical observation-fluxes $\obs_{i,\job}$, where the standard deviation is proportional to the empirical observation-errors $\obse_{i,\job}$. \highlighttext{The standard deviation is defined as ${\sigma'}_{i,\job} = \gamma\cdot\obse_{i,\job}^2+\beta$, where $\gamma=10$ and $\beta=1$ were empirically selected to adjust the influence of the observation-error in the standard deviation}. This setting seeks to prevent that observations with near zero observation-errors completely control the likelihood of the light-curve.

\item \textbf{Prior}: in eq. \refeq{eq:mcmc_prior}, the prior distribution is defined. This formulation implies that the SPM prior selection is based on the SPM optimal parameters from the companion band $b'\neq b$ of the single-band light-curve $\Phi_{i}^\ob$ (a companion band within the multi-band light-curve $\Phi_{i}$). The prior is defined as an isotropic multivariate Gaussian distribution centered in the companion band $b'$ optimal SPM parameters, where $\Sigmabf\in\real{6 \times 6}$ is a diagonal matrix for the standard deviation. This prior selection attempts to induce information from the companion band $b'$ in the optimization of the target band $b$. This might correct the optimization in scenarios where no empirical observation is found from the SN-rise and SN-peak regions in the current band. The optimal SPM parameters from the companion band are found using Maximum Likelihood Estimation (MLE) over the empirical observations (see Appendix \refsec{sec:spm_bounds} for details). This prior formulation could be extended, for more than two bands, by using a Gaussian Mixture Model (GMM) as the prior distribution.
\end{enumerate}

\subsection{Sampling Time Window} 
To evaluate the SPM analytical function at different observation-times, a Sampling Time Window (STW), consisting of a collection of $L$ time values, is defined as ${\Delta t^\ob_\text{stw}}_{i}=\set{t|t\sim \uniformdis{{t^\ob_\text{init}}_{i}, {t^\ob_\text{final}}_{i}}}{j\seq 1}{L}$, where the STW consists of a temporal grid with $L$ time values sampled from a uniform distribution. The size of the STW is defined by the number of empirical observations from the current single-band light-curve $\Phi_{i}^\ob$ ($L=L_{i}^\ob$).

The STW uniform distribution lower bound is defined as follows:
\begin{align}
{t^\ob_\text{init}}_{i} &= \begin{cases}
t_{i,1^\ob}, & \text{if }t_{i,1^\ob}<t^\ob_\text{max}{}_{i},\\
t_{i,1^\ob}-\Delta t, & \text{otherwise},\label{eq:tinit}
\end{cases}
\end{align}
where $t_{i,1^\ob}$ represents the first empirical observation-time from the single-band light-curve $\Phi_{i}^\ob$. The term $t^\ob_\text{max}{}_{i}$ is the time value associated with the global maximum of the optimized SPM function evaluated using the optimal SPM parameters $\thetabf_{i}^{\ob*}$. By setting $\Delta t=10\tunits{days}$, the STW can be used to sample observation-times before the first empirical observation, e.g., when no observation is available close to the SN-peak (according to the SPM function maximum). This extension allows generating plausible observations over the poorly represented SN-rise and SN-peak regions, originally observed from the dataset $\dataset{}$.

The uniform distribution upper bound is defined as ${t_\text{final}}_{i}=t_{i,-1^\ob}$, where $t_{i,-1^\ob}$ represents the last empirical observation-time from the single-band light-curve $\Phi_{i}^\ob$. This bound ensures that the sampled observation-times are bounded by the last empirical observation-time.

\subsection{Generation of Synthetic Observations}\label{sec:parametric} 
\begin{figure*}[!t]
\centering
\minipage{0.5\textwidth}
\includegraphics[trim={0} {\yoffsetb} {0} {0}, clip=true, width=.9999\linewidth]{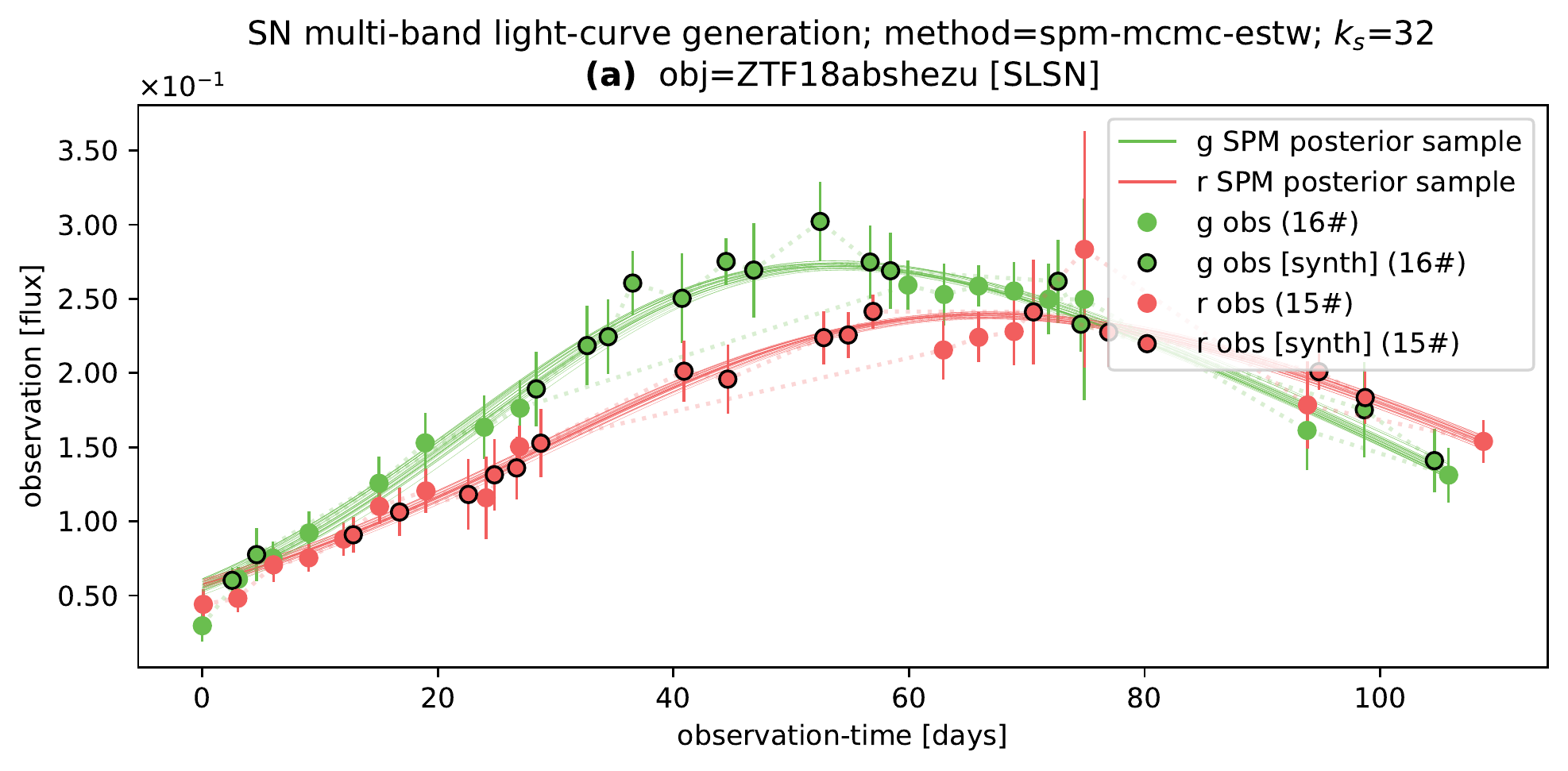}
\endminipage
\minipage{0.5\textwidth}
\includegraphics[trim={\xoffset} {\yoffsetb} {-\xoffset} {0}, clip=true, width=.9999\linewidth]{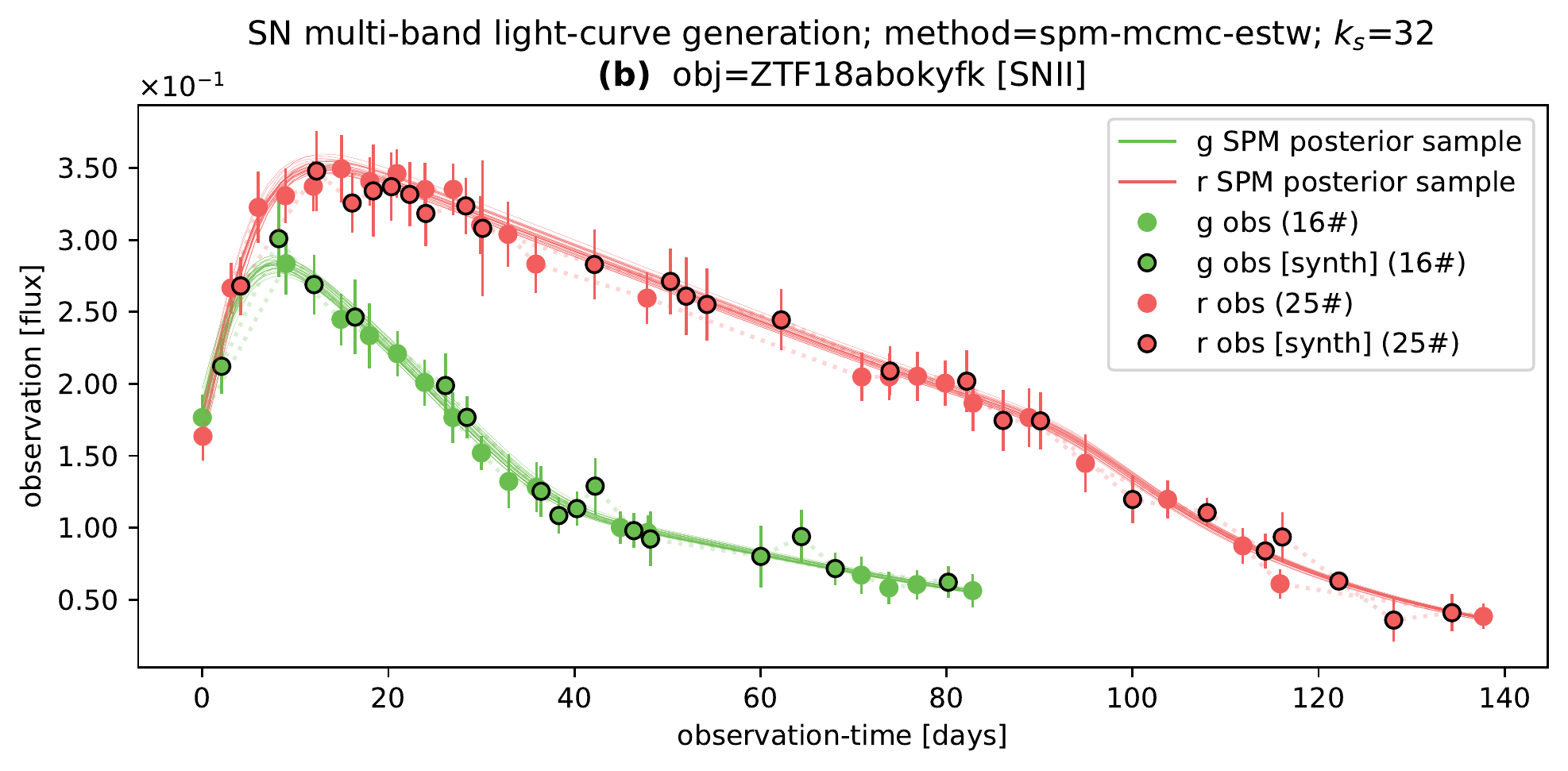}
\endminipage
\\
\minipage{0.5\textwidth}
\includegraphics[trim={0} {0} {0} {\yoffsett}, clip=true, width=.9999\linewidth]{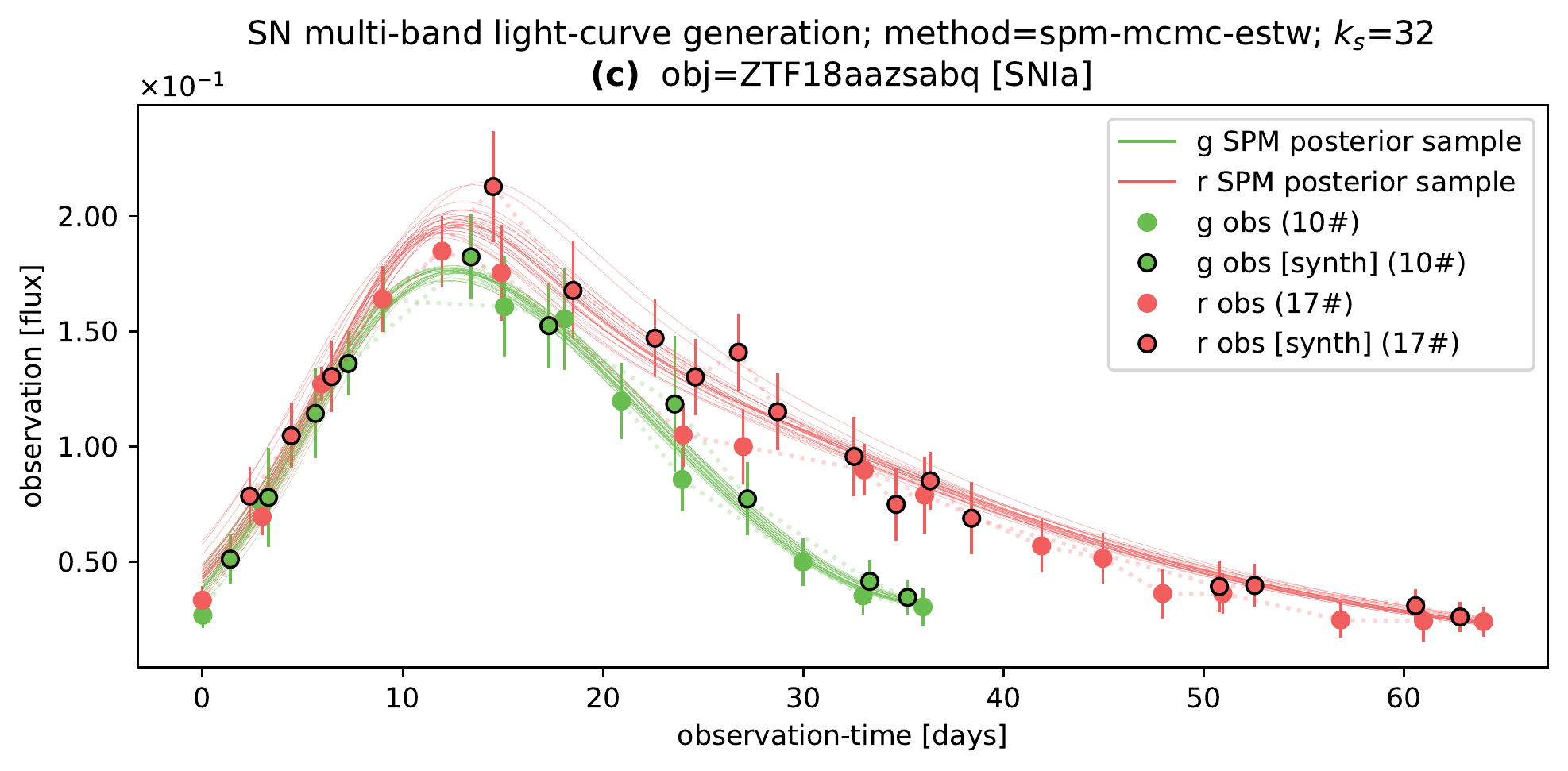}
\endminipage
\minipage{0.5\textwidth}
\includegraphics[trim={\xoffset} {0} {-\xoffset} {\yoffsett}, clip=true, width=.9999\linewidth]{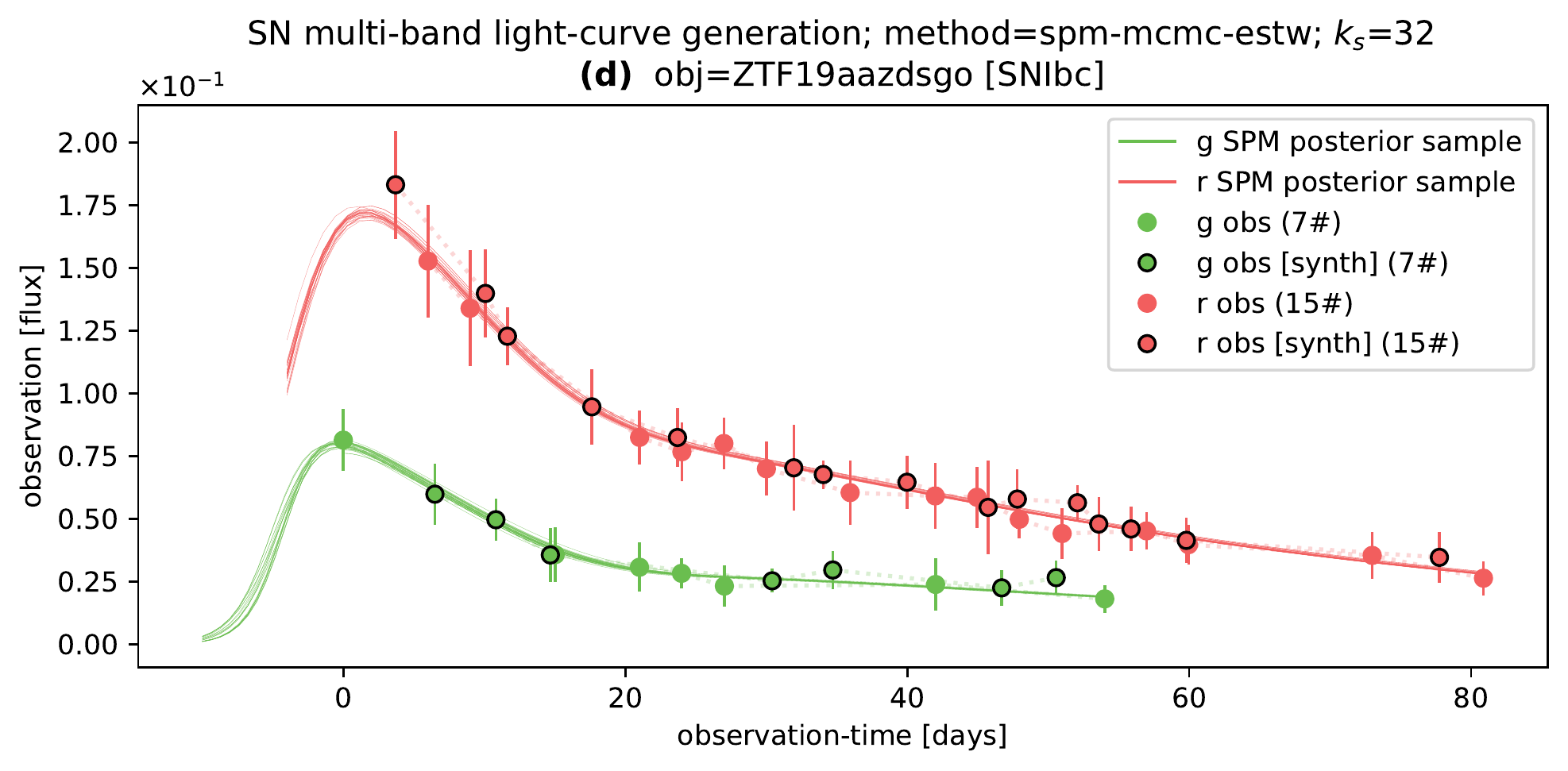}
\endminipage
\caption{
Examples of synthetic SN multi-band light-curves (before the observation-time re-offset). The SPM posterior samples ($k_s=32$) from MCMC are shown as continuous-time curves for each band. Empirical observation-fluxes are shown as color circles with observation-error bars. Synthetic observation-fluxes, using a random SPM posterior sample, are shown as black outlined circles. (a) SLSN type. (b) SNII type. (c) SNIa type. (d) SNIbc type.
}
\label{fig:synth_examples}
\end{figure*}

Given an arbitrary SN multi-band light-curve $\Phi_{i}$, the process to generate synthetic light-curves is shown in algorithm \ref{alg:parametric}. In addition, Fig. \ref{fig:synth_examples} shows several examples of synthetic generation of multi-band light-curves for each SN type.

\SetInd{0.5em}{0.2em}
\begin{algorithm}[t]
\caption{
SN multi-band synthetic light-curve generation based on SPM.
}
\label{alg:parametric}
  \Repeat{A number of $k_s$ new light-curves are generated from $\Phi_{i}$}{
  \For{$b\in\keys{1,\dots,B}$}{
  $\thetabf_{i}^{\ob*}\sim p\tuple{\Phi_{i}^\ob|\thetabf_{i}^\ob}  p\tuple{\thetabf_{i}^\ob}$\;
  \For{$t_{i,\job} \in {\Delta t^\ob_\text{stw}}_{i}$}{
$\hatobs_{i,\job} = \fsne\tuple{t_{i,\job};\thetabf_{i}^{\ob*}}$\;
$\hatobse_{i,\job} \sim p(\obse|\hatobs_{i,\job},b)$\;
$\hatobs_{i,\job}\gets \hatobs_{i,\job}+ k \cdot \hatobse_{i,\job} \cdot \varepsilon, \varepsilon \sim \tdist{\nu}$\;
  }
  }
 \For{$b\in \keys{1,\dots,B}$}{
 $t_{i,\job}\gets t_{i,\job}-t_{i,1},\forall \job$\tcp{\scriptsize Observation-time re-offset}
 }
  }
\end{algorithm}

First, a set of optimal SPM parameters $\thetabf_{i}^{\ob*}$ are sampled by using an MCMC posterior distribution estimation from the empirical observations. The MCMC sampling procedure allows us to introduce a moderate diversity of SPM parameters when generating a new light-curve\footnote{\highlighttext{Possible MCMC algorithm exceptions and divergences are handled by replacing the SPM analytical function ${\fsne}$ with a linear interpolation between empirical observations. This strategy is also used when there is no other band information.}}. Next, the STW is generated, and the sampled time values are evaluated using the SPM analytical function ${\fsne}$ and a set of optimal SPM parameters $\thetabf_{i}^{\ob*}$. In this way, multiple synthetic observation-fluxes $\hatobs_{i,\job}$ are generated.

Next, the synthetic observation-errors are sampled from a conditional distribution $p(\obse|\hatobs_{i,\job},b)$, which describes the observation-error distribution $\obse$, given an observation-flux $\hatobs$ and a band $b$. Sampling from this distribution allows generating plausible observation-errors given an arbitrary observation-flux. To estimate this distribution, a collection of Gaussian distributions are fitted using a variable binning strategy over a transformed observation-flux versus observation-error distribution from the training-set $\dataset{train}$ (see Appendix \refsec{sec:obse_estimation} for details).

A new synthetic observation-flux is then generated by re-sampling the synthetic observation-flux $\hatobs_{i,\job}$ using a clipped t-student distribution scaled by $k \cdot \hatobse_{i,\job}$. Several methods of light-curve observation-flux re-sampling have been explored in the literature using the Gaussian distribution \citep{Moss2018, Naul2018, Gomez2020, Hosenie2020}, but in this work, we use the t-student distribution as it has a higher dispersion over the distribution tails, generating a higher proportion of outlier observation-fluxes along the light-curve. The Gaussian distribution can be recovered by increasing $\nu\to\infty$, where $\nu$ is the degree of freedom of the t-student distribution. For the dataset $\dataset{}$, we set $k=\kexp{5}{-1}$ and $\nu=2$ to obtain a general well-behaved re-sampling dispersion.

\subsection{Synthetic Training-Set Generation} 
Given an empirical SN multi-band light-curve, the proposed method can generate an arbitrary number $k_s$ of new synthetic light-curves. Thus, an augmented synthetic training-set $\dataset{train[s]}$ is built by generating a number of $k_s=32$ new synthetic light-curves for each empirical light-curve from the original training-set $\dataset{train}$. The construction of this new synthetic training-set $\dataset{train[s]}$ is an effort to increase both, the total number of samples and the diversity of the irregular cadence population observed in the original training-set $\dataset{train}$.
\section{BRF Baseline Classifier}\label{sec:brf_baseline} 
In this section, the Balanced Random Forest (BRF) model is described as a baseline classifier. This model uses a set of features extracted from the light-curves to classify different types of SNe.

\subsection{Photometric and Astrophysical Features} 
Given a multi-band light-curve $\Phi_i$, several features can be extracted using irregular time-series related methods and astrophysical knowledge. These features aim to characterize the general behavior of a variable-length multi-band light-curve into a fixed-length feature vector. For the feature extraction, photometric and astrophysical features implemented by the ALeRCE broker\footnote{\url{https://github.com/alercebroker/lc_classifier}.} \citep{Sanchez-Saez2021} are used, which consist of a collection of 152 photometric features that are computed from light-curves. The ALeRCE broker proposed a vast set of novel features, but it also collected features from previous works \citep[][]{Nun2015}. For instance, the ALeRCE broker proposed to fit the SPM parameters, based on an MLE estimation, as a novel approach to characterize SN light-curves.

In this work, we exclude some metadata-based features such as the ALLWISE colors or the galactic coordinates features, as these features do not influence the performance of the transient classifier as reported by the ALeRCE team \citep{Sanchez-Saez2021}. Other transient features, such as the Star Galaxy Separation score (SGS score) metadata or non-detection features, are not used because we aim to classify SNe based solely on the detected photometric information, i.e., the light-curves. This gives us a total of 144 features computed from each SN multi-band light-curve. The observation-fluxes, used for our methods, are consequently transformed to apparent magnitude to correctly compute the features.

\subsection{Balanced Random Forest} 
For the baseline classifier, the Balanced Random Forest model \citep[BRF;][]{Chen2004} is used, which is a variation of the original Random Forest model \citep[RF;][]{Breiman2001}. The main advantage of the BRF algorithm is that it can deal with the high class imbalance of the training-set $\dataset{train}$. To deal with the imbalance, the BRF train each decision tree with a bootstrapped sample that is balanced in class samples, where the minority class is potentially well-represented.

To train and test the BRF model, the photometric features are computed from all the SN multi-band light-curves. Infinite and NaN features, which are produced due to incorrect feature extraction and programming code exceptions, are replaced with a special value of $-999$ \citep{Sanchez-Saez2021}. In addition, the best hyperparameter configuration is found by using a grid search over different values for the split quality criterion (e.g., gini, entropy) and tree maximum depth. The best selected configuration is the one associated with the best performance reported over the validation-set $\dataset{val}$ by monitoring the maximum value of the balanced b-$F_1$score metric (see Appendix \refsec{sec:metrics} for details).
\section{TimeModAttn Model}\label{sec:deep} 
\begin{figure}[!t]
\centering
\includegraphics[width=\onecolumnscale]{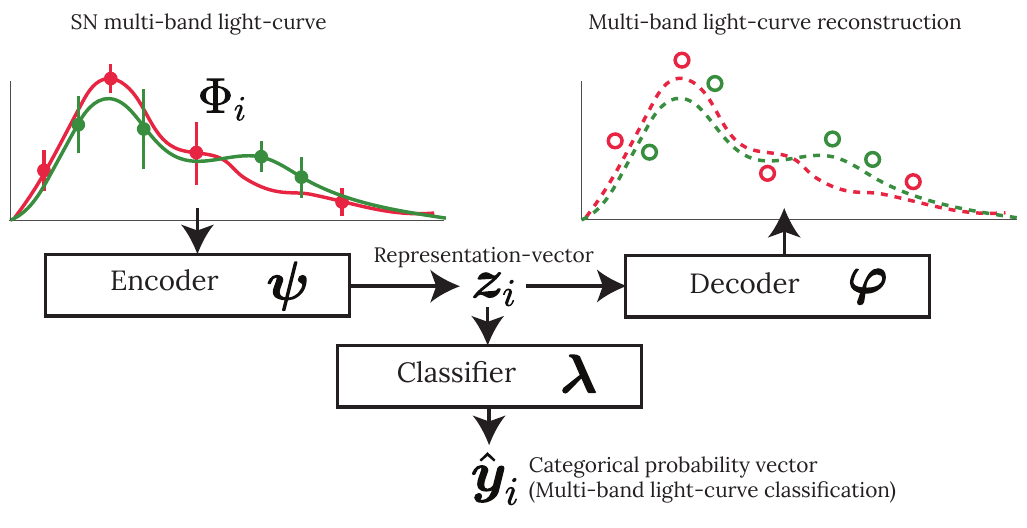}
\caption{
Proposed model architecture based on an autoencoder $\keys{\psibf, \varphibf}$, which is composed of an encoder $\psibf$ and a decoder $\varphibf$. The representation-vector $\z_i$ serves as input to a classifier $\lambdabf$.
}
\label{fig:arch_simple}
\end{figure}

\begin{figure*}[!t]
\centering
\includegraphics[width=1\linewidth]{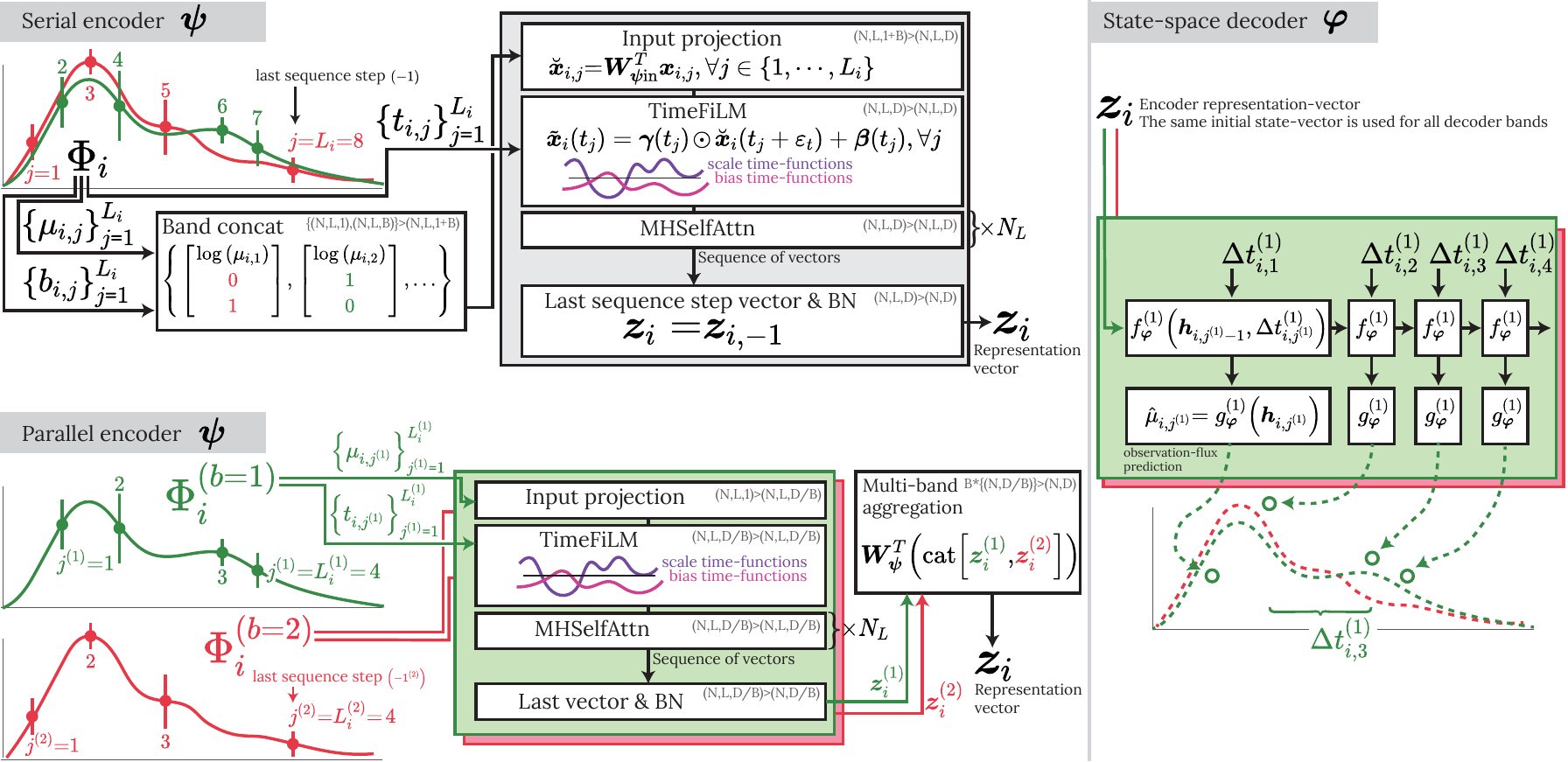}
\caption{
Diagram for the proposed autoencoder model (TimeModAttn), which is composed of an encoder $\psibf$ (shown on the left side of the diagram) and a decoder $\varphibf$ (shown on the right side of the diagram). There are two variations for the encoder: serial and parallel. For the band operator, the notation $\Phi^{(b=1)}$ is used for the band g, and $\Phi^{(b=2)}$ is used for the band r. An arbitrary number of $N_L$ stacked layers is shown for the MHSelfAttn multi-layer sequence processing. Optional tensor dimensional transformations, related with the model implementation, are also shown (e.g., \texttt{(N,L,1+B)>(N,L,D)}). $N$ stands for the mini-batch size, $L$ for the sequence steps tensor dimension, $B$ for the number of bands, and $D$ for the representation-vector dimension.
}
\label{fig:proposed_model_arch}
\end{figure*}

In this section, the proposed TimeModAttn model for the SN multi-band light-curve classification is described. As shown in Fig. \ref{fig:arch_simple}, this model is based on an autoencoder (encoder-decoder) and a classifier. A more detailed diagram is illustrated in Fig. \ref{fig:proposed_model_arch}. Note that this model can be used to process arbitrary multi-band light-curves; therefore, it is not limited to SN light-curves.

\subsection{Loss Functions}\label{sec:losses} 
\highlighttext{First, given a multi-band light-curve $\Phi_i$, a Mean Squared Error (MSE) reconstruction loss function is defined as follows:
\begin{align}
    \loss_{\text{rec}_i}&=\frac{1}{B} \sum_{b\seq 1}^{B} \frac{1}{L_i^\ob} \sum_{\job\seq 1}^{L_i^\ob} \tuple{\obs_{i,\job}-\hatobs_{i,\job}}^2,\label{eq:mse}
\end{align}
where $\obs_{i,\job}$ are the observation-fluxes from the single-band light-curve $\Phi_i^\ob$ (with variable-length $L_i^\ob$). The observation-fluxes predicted by the decoder, for the single-band light-curve $\Phi_i^\ob$, are denoted as $\hatobs_{i,\job}$. Note that the final value of the loss is the average of the reconstruction losses from all the $B$ bands\customfootnote{During the preparation of this \docname, we thought that we were using another definition for the reconstruction loss that was based on the Weighted MSE (WMSE) loss. After an inspection of our implementation, we finally decided that the shown loss (MSE) was the most theoretically correct way to define the used reconstruction loss. For more details regarding this issue, see Appendix \refsec{sec:loss_issue}}.}

Second, given a multi-band light-curve $\Phi_i$ and based on the Cross-Entropy (CE) $H(p|q)$, a categorical loss function is defined as follows:
\begin{align}
    \loss_{\text{cat}_i}&=H(p(y_i|\Phi_i),q(y_i|\Phi_i)),\nonumber\\
    &=\sum_{c\seq 1}^{C} -p_{c}(y_i|\Phi_i) \cdot\logfun{\hat{y}_{i,c}},
\end{align}
where $C$ is the total number of classes presented in the dataset, $p(y_i|\Phi_i)$ is the true class distribution, $q(y_i|\Phi_i)$ is the model estimated class distribution, and $\hat{y}_{i,c}$ is the model estimated probability for class $c$.

\subsection{Proposed Model Formulation} 
To model the loss functions defined above, we use an autoencoder model (encoder-decoder). Following \citea{Jamal2020}, two main architectures are implemented to deal with the multi-band light-curve processing: serial encoder and parallel encoder.

\subsection{Serial Encoder} 
The main goal of the encoder is to generate an automatic fixed-length representation-vector $\z_i$ from the variable-length multi-band light-curve $\Phi_i$ as shown in Fig. \ref{fig:arch_simple}. For the serial encoder $\psibf$, the formulation is as follows:
\begin{align}
\breve{\x}_{i,j}&=\mtw^T_{\psibf\text{in}} \x_{i,j},\forall j \in \keys{1,\dots,L_i}  ,&&\text{}\label{eq:s_inproj}\\
\z_{i,j}&=f_\psibf\tuple{\set{\tuple{\breve{\x}_{i,j'}, t_{i,j'}}}{j'\leq j}{}},\forall j \in \keys{1,\dots,L_i} ,&&\text{}\label{eq:s_sec}\\
\z_i&= \z_{i,-1},&&\text{}\label{eq:s_aggr}
\end{align}
where $\x_{i,j}$ is the encoder input vector, which is composed of photometric observations\footnote{All model input vectors are normalized using a standardization method with statistics computed from synthetic training-set $\dataset{train[s]}$ (see Appendix \refsec{sec:normalization} for details).} at the sequence step $j$. For the model input, the logarithm of the observation-flux is used as $\x_{i,j}=\vect{\logfun{\obs_{i,j}+\varepsilon}}$, \highlighttext{to attenuate large observation-flux values from the SN-peak, helping in the normalization of the input distribution for the Deep Learning models}\footnote{\highlighttext{An inverse hyperbolic sine (asinh) function can be used if a forced photometry scenario is presented (with possible negative observation-fluxes).}}. Note that no explicit time information is included in the input vector as the TimeModAttn model uses a temporal modulation strategy (see details in section \refsec{sec:timefilm}).

For the serial encoder, a one-hot vector, which is defined as $\b_{i,j}=\vect{0,\dots,1_{b_{i,j}\seq b},\dots,0}\in\real{B}$, is used as a band indicator and concatenated with the encoder input. This operation produces a new encoder input vector $\x_{i,j}\in\real{(1+B)}$. Then, the input vector is projected into a higher $D=128$ dimensional space $\breve{\x}_{i,j}\in\real{D}$ by using the linear projection $\mtw_{\psibf\text{in}}\in\real{(1+B)\times D}$, with shared parameters over all sequence steps.

In eq. \refeq{eq:s_sec}, a causal sequence processing formulation is given. For an arbitrary sequence step $j$, a representation-vector $\z_{i,j}$ is computed using the current and previous vectors and observation-times $\keys{\tuple{\breve{\x}_{i,1},t_{i,1}},\dots,\tuple{\breve{\x}_{i,j}, t_{i,j}}}$. We propose a temporal modulation (section \refsec{sec:timefilm}) followed by an attention mechanism (section \refsec{sec:attn}) to solve this formulation. The idea of this setting is to process a time-modulated sequence of representation-vectors by using the attention mechanism.

In eq. \refeq{eq:s_aggr}, the final representation-vector from the encoder is defined as the last representation-vector $\z_{i,-1}$ from the sequence $\set{\z_{i,j}}{j\seq1}{L_i}$. This vector is associated with the last sequence step $j=L_i$, where $L_i$ is the length of the multi-band light-curve $\Phi_i$. Additionally, a Batch Normalization \citep[BN;][]{Ioffe2015} operation is performed over the final representation-vector $\z_{i,-1}$.

\subsubsection{Temporal Modulation (TimeFiLM)}\label{sec:timefilm} 
A method is required to induce the sequential information in attention mechanisms, as well as the observation-time information, to correctly process and characterize the SN light-curves with highly irregular cadence.

Inspired by the idea of the Feature-wise Linear Modulation operation \citep[FiLM;][]{Perez2018}, we propose a temporal modulation (TimeFiLM) to induce the time information in the input sequence that is processed by the attention mechanism (see Fig. \ref{fig:proposed_model_arch}). The formulation of the proposed temporal modulation is as follows:
\begin{align}
&\tilde{\x}_{i}(t_{i,j})=\nonlinearfun{tanh}{\gammabf(t_{i,j})}\odot \breve{\x}_{i}(t_{i,j}+\varepsilon_t)+\betabf(t_{i,j}),&&\text{}\label{eq:mod_mod}\raisetag{20pt}\\
&\tilde{\x}'_{i}(t_{i,j})=\nonlinearfun{ReLU}{\mtw^T \tilde{\x}_{i}(t_{i,j})+\b},&&\text{}\label{eq:mod_out}\\
&\gamma_k(t)=\sum_{m\seq1}^{M} a'_{k,m} \sin\tuple{\frac{2\pi m}{T_\text{max}}t}+b'_{k,m} \cos\tuple{\frac{2\pi m}{T_\text{max}}t} ,&&\text{}\label{eq:mod_scale}\\
&\beta_k(t)=\sum_{m\seq1}^{M} v'_{k,m}  \sin\tuple{\frac{2\pi m}{T_\text{max}}t}+w'_{k,m} \cos\tuple{\frac{2\pi m}{T_\text{max}}t} ,&&\text{}\label{eq:mod_bias}
\end{align}
where, given an arbitrary input vector time function $\breve{\x}_{i}(t_{i,j}+\varepsilon_t):\funmap{\real{}}{\real{K}}$, the result of the modulation operation $\tilde{\x}_{i}(t_{i,j}):\funmap{\real{}}{\real{K}}$, in eq. \refeq{eq:mod_mod}, is defined as the element-wise product ($\odot$) followed by the element-wise addition $(+)$ (FiLM operation) using the vector time functions $\gammabf(t_{i,j})$ and $\betabf(t_{i,j})$, respectively. An optional hyperbolic tangent function $\nonlinear{tanh}$ is used to prevent explosive product values. Note that the vector $\breve{\x}_{i,j}$ (shown in eqs. \ref{eq:s_inproj} and \ref{eq:s_sec}), associated with the observation-time $t_{i,j}$, stands for the vector time function evaluated at time $t_{i,j}$, i.e., $\breve{\x}_{i,j}\newdef\breve{\x}_{i}(t_{i,j})$.

\begin{figure}[!t]
\centering
\includegraphics[width=\onecolumnscale]{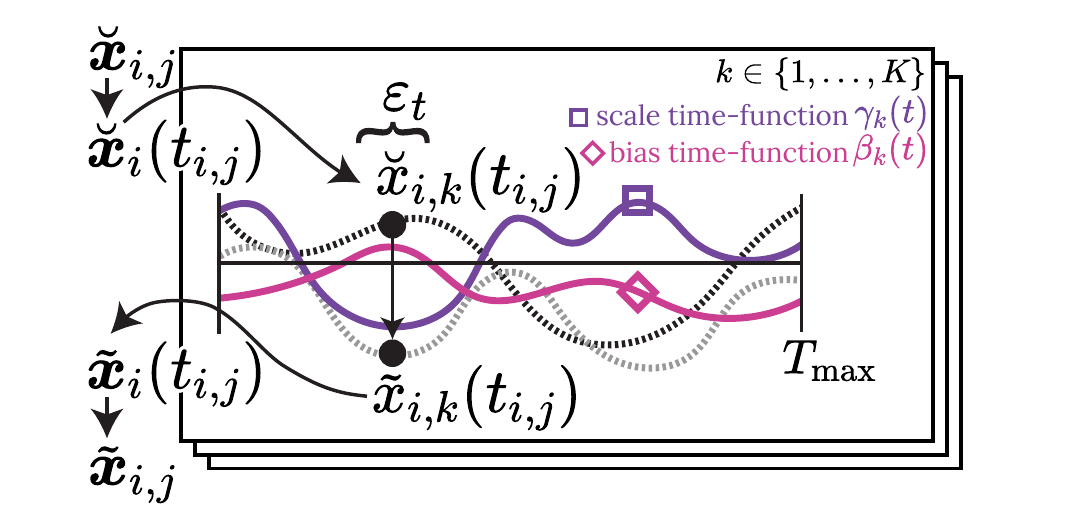}
\caption{
Proposed temporal modulation $\tilde{\x}_{i,j}$, where $\breve{x}_{i,k}(t_{i,j})$ is the $k$-th component of the vector time function $\breve{\x}_{i}(t_{i,j})$, which represents the vector $\breve{\x}_{i,j}$ associated with the time value $t_{i,j}$. The scale $\gamma_k(t)$ and bias $\beta_k(t)$ functions, represented with solid lines, can be evaluated at any arbitrary continuous-time value, giving a non-redundant and continuous-time modulation-range $[0,T_\text{max}]$. Dashed lines represent empirical unobserved time functions, which are associated with the model input.
}
\label{fig:timefilm}
\end{figure}

The vector time functions are constructed as $\gammabf(t)=\vect{\gamma_1(t),\dots,\gamma_K(t)}$ and $\betabf(t)=\vect{\beta_1(t),\dots,\beta_K(t)}$, where $\gamma_k(t):\funmap{\real{}}{\real{}}$ and $\beta_k(t):\funmap{\real{}}{\real{}}$ are the scale and bias time functions, respectively. We assume that these functions are continuously defined and can be evaluated at any time value. In eq. \refeq{eq:mod_mod}, a new time-modulated vector function $\tilde{\x}_{i}(t_{i,j})$ is generated from the input vector function by using $K$ different scale and bias time functions, each one associated with one dimension component of the modulator input vector $\breve{\x}_{i,j}$ (see Fig. \ref{fig:timefilm}).

The construction of both time functions, the scale, in eq. \refeq{eq:mod_scale}, and bias, in eq. \refeq{eq:mod_bias}, is based on a Fourier decomposition with $M$ harmonic components. The term $m=0$ is not included to attenuate the risk of constructing time functions that are invariant in time, which may be produced by significantly high linear term values. Then, the $4 M K$ learnable parameters of the modulation are $\set{\set{a'_{k,m},b'_{k,m},v'_{k,m},w'_{k,m}}{k\seq 1}{K}}{m\seq 1}{M}$\footnote{The parameters are randomly initialized with uniform Kaiming initialization \citep[][]{He2015}. Also, higher harmonics are attenuated by using an exponential decay $e^{-k \cdot (m-1)}, k=.5$, stabilizing the early training epochs by starting with smooth and amplitude controlled modulation curves.}, where $K=D$ is used, corresponding to the dimensionality of the vector $\breve{\x}_{i,j}$.

As an optional and direct regularization technique, a noise term $\varepsilon_t$ is added to each evaluated time value $t_{i,j}$ only during the training process. For the SNe, this noise can be dynamically sampled from a uniform distribution with values between -6 and 6 hours, which can induce a dynamic and moderate disturbance over the original light-curves.

This formulation allows the model to learn any smooth and continuous-time functions, inducing a non-redundant temporal modulation over a finite time interval $[0, T_\text{max}]$, where $T_\text{max}=k_T \cdot \maxx{}\set{t_{i,-1}}{i\seq 1}{N}$ is defined as the maximum period, associated with the minimum harmonic frequency of the Fourier decomposition. This maximum period is arbitrarily defined such that it exceeds the maximum empirical last observation-time $t_{i,-1}$, found among the light-curves from the training-set $\dataset{train}$, by setting $k_T=1.5$.

The flexibility to learn any arbitrary time function\footnote{Maximum flexibility and smoothness are constrained by the selected number of $M$ harmonics components used.}, and not just a collection of periodic functions, as done in \citep[][]{Vaswani2017, Kazemi2019, Sousa2020}, might be especially beneficial for non-periodic transient events such as SNe, as there could be potentially more informative time regions in the early explosion days: earlier than and close to the SN-peak, instead of periodically spaced informative zones. Thus, the TimeModAttn model might learn an adequate modulation modulation over those SN time regions to correctly extract useful information. The learning of periodic functions was tested in preliminary experiments, but the collapse of some learned periods was observed, increasing the risk of constructing time-invariant functions. This may be because no hidden or intrinsic periodic behavior is expected in the SNe dataset.

In eq. \refeq{eq:mod_out}, the final modulated vector function $\tilde{\x}'_{i}(t_{i,j})$ is computed using a linear projection $\mtw \in \real{K\times K}$, plus a bias vector $\b$, and the ReLU function $\nonlinear{ReLU}$, with shared parameters over all sequence steps. This operation allows the model to perform nonlinear interactions among the components of the time-modulated vector.

In summary, the proposed temporal modulation allows the model to properly capture the highly irregular cadence of light-curves by directly using the observation-times to induce a smooth and non-redundant modulation over the time interval $[0,T_\text{max}]$. This allows us to avoid using missing-value assumptions, as well as any explicit imputation and interpolation methods. The latter methods might be detrimental because they can introduce artifacts and anomaly light-curve behaviors. Additionally, the construction of continuous-time defined functions using Fourier decomposition allows us to directly inspect the learned functions, exploring for possible and interpretable temporal modulation tendencies produced by the processing of transient events.

\subsubsection{Multi-Head Self-Attention Mechanism (MHSelfAttn)}\label{sec:attn} 
To complete the formulation of a solution for the causal sequence processing described in eq. \refeq{eq:s_sec}, we propose the use of a multi-head self-attention (MHSelfAttn) mechanism\footnote{For simplicity, a single MHSelfAttn's layer is used ($N_L=1$). Also, the number of units in the MHSelfAttn's MLP hidden-layer is reduced ($k_\text{mlp}=1$). Note that more MHSelfAttn's layers could be used if required.}. This operation is performed over the output sequence vectors $\set{\tilde{\x}'_{i,j}}{j\seq 1}{L_i}$ obtained from the temporal modulation method (TimeFiLM). The goal of the attention mechanism is to construct informative and meaningful context vectors given the query, key, and value vectors as explained in section \refsec{sec:background/attn}.

We highlight that the context vectors are computed over a sequence of vectors with induced time information from the proposed temporal modulation (TimeFiLM). Thus, a meaningful context vector could contain automatic time representations computed along the light-curve, such as time differences, elapsed times, short and long-range trends and time dependencies, among others. This can be achieved by computing the correlations between the time-modulated versions of the query and key vectors, but also with the final vector aggregation using the time-modulated value vectors.

\subsection{Parallel Encoder} 
The formulation for the parallel encoder $\psibf$ model is as follows:
\begin{align}
\breve{\x}_{i,\job}=&\mtw^{\ob ^T}_{\psibf\text{in}} \x_{i,\job},\forall \job \in \keys{1,\dots,L_i^\ob} ,&&\text{}\label{eq:p_inproj}\\
\z_{i,\job}=&f_\psibf^\ob\tuple{\set{\tuple{\breve{\x}_{i,\job'}, t_{i,\job'}}}{\job'\leq \job}{}},&&\text{}\label{eq:p_sec}\\
&\forall\job\in\keys{1,\dots,L_i^\ob},\nonumber\\
\z^\ob_i=& \z_{i,-1^\ob},&&\text{}\label{eq:p_aggr}\\
\z_i=&\mtw^T_{\psibf}\tuple{\cat{\z_i^{(1)},\dots,\z_i^{(B)}}},&&\text{}\label{eq:mb_aggr}
\end{align}
where eqs. \refeq{eq:p_inproj}-\refeq{eq:p_aggr} follow similar ideas as those from the serial encoder. In general, in the parallel case, the sequence processing is performed independently for each of the $B$ bands in the multi-band light-curve $\Phi_i$.

In contrast to the serial encoder, in eq. \refeq{eq:p_inproj} the one-hot vector $\b_{i,j}$ is not included in the encoder input vector $\x_{i,\job}=\vect{\logfun{\obs_{i,\job}+\varepsilon}}$, as it is not required to include the band information for the parallel encoder. The high-dimensional space of the encoder is decreased, from $D$ to $(D/B)$, by using a linear projection $\mtw^{\ob}_{\psibf\text{in}}\in \real{1\times (D/B)}$. This dimensionality reduction is performed to have a similar total number of learnable parameters for both, the serial and parallel encoders. Additionally, to keep an equal attention head dimensional space ($D_q$, $D_k$, $D_v$), we decrease the number of attention heads from $H$ to $H/B$.

For the final representation-vector note that, in eq. \refeq{eq:p_aggr}, the final representation-vector $\z_{i,-1^\ob}$ is defined as the last representation-vector from the sequence $\set{\z_{i,\job}}{\job\seq1}{L_i^\ob}$. This vector is associated with the last sequence step $\job=L_i^\ob$, where $L_i^\ob$ is the length of the single-band light-curve $\Phi_i^\ob$.

In eq. \refeq{eq:mb_aggr}, the final representation-vector $\z_i$ is projected by using the linear projection $\mtw_{\psibf}\in\real{D \times D}$ from the joint space constructed with the concatenation of each of the $B$ band representation-vectors: $\keys{\z_i^{(1)},\dots,\z_i^{(B)}}$. This operation allows the model to capture relevant information, from each band representation-vector, into a final representation-vector with the same number of dimensions $D$ as those of the serial encoder.

\subsection{Parallel Decoder}\label{sec:p_dec} 
The parallel decoder $\varphibf$ formulation is based on a state-space model as follows:
\begin{align}
& \h_{i,\job} =
\begin{cases}
f_\varphibf^\ob\tuple{\z_i, \Delta t^\ob_{i,1^{(b)}}} & \text{if }\job=1,\\
f_\varphibf^\ob\tuple{\h_{i,\job-1}, \Delta t^\ob_{i,\job}}, & \text{otherwise},\\
\end{cases},\label{eq:trans_fun}\\
&\hatobs_{i,\job} = g_\varphibf^\ob\tuple{\h_{i,\job}},&&\text{}\label{eq:obs_fun}
\end{align}
In eq. \refeq{eq:trans_fun}, the transfer function $f_\varphibf^\ob$ is defined to model the evolution dynamic for the current sequence step state-vector $\h_{i,\job}$ given both, the previous state-vector $\h_{i,\job-1}$ and the time difference $\Delta t^\ob_{i,\job}$ between both observations. Note that, when decoding the first sequence step ($\job=1$), the initial state-vector $\h_{i,1^\ob}$ is defined as the representation-vector $\z_i$ computed by the encoder $\psibf$, i.e., $\h_{i,1^\ob}=\z_i$. In eq. \refeq{eq:obs_fun}, the function $g_\varphibf^\ob$ is defined to generate the single-band light-curve observation-flux estimations $\hatobs_{i,\job}$ given the current state-vector $\h_{i,\job}$.

For simplicity and because our major research interest is the expressive capacity of the encoder $\psibf$, only the parallel approach is used for the decoder. This approach is also shared with all the encoder variations and baselines implemented in this work. As a remark, the same initial state-vector $\z_i$ is used for each of the $B$ parallel decoders, where each decoder is associated with a particular band.

The function ${f_\varphibf^\ob}$ is modeled with a Gated Recurrent Unit \citep[GRU;][]{Chung2014}, having a 1-dimensional input space for the time difference input and $D$ dimensions for the hidden state-vector. The function ${g_\varphibf^\ob}$ is modeled with a 1-hidden-layer Multi-Layer Perceptron \citep[MLP;][]{Rumelhart1986}, with a 1-dimensional output space and a linear activation function at the output. This MLP model shared parameters over all sequence steps.

\subsection{Classifier} 
The proposed formulation for the classifier $\lambdabf$ (see Fig. \ref{fig:arch_simple}) is as follows:
\begin{align}
\hat{\y}_{i}&=\nonlinearfun{softmax}{f_\lambdabf\tuple{\z_i}},&&\text{}
\end{align}
where the function $f_\lambdabf$ is modeled using a 2-hidden-layer MLP. The output dimension is set as the number of classes $C$. The softmax function $\nonlinear{softmax}$ is used to generate a final categorical probability vector for the discrete class prediction distribution $\hat{\y}_i = \vect{\hat{y}_{i,1},\dots,\hat{y}_{i,C}}$. A dropout probability \citep[][]{Srivastava2014} of $50\%$ is used for the MLP model.

\subsection{Optimization Problem} 
Given the aforementioned loss functions and the TimeModAttn model formulation, the complete optimization problem is defined as follows:
\begin{align}
    \loss_\text{pre-training} &= \frac{1}{N}\sum_{i\seq 1}^{N}{  \underbrace{k_0\cdot\loss_{\text{rec}_i}}_{\mathclap{\substack{\text{multi-band}\\
    \text{light-curve}\\ \text{reconstruction}}}}+ \underbrace{k_1\cdot\loss^{}_{\text{cat}_i}}_{\mathclap{\substack{\text{cross-entropy}\\ \text{regularization}}}}},&&\text{}\label{eq.pre_training}\\
    \loss_\text{fine-tuning} &= \frac{1}{N}\sum_{i\seq 1}^{N}\underbrace{\loss_{\text{cat}_i}}_{\mathrlap{\text{cross-entropy}}},&&\text{}
\end{align}
where $k_0=\kexp{1}{4}$ and $k_1=1$. This setting implies a higher relevance for the reconstruction loss term, i.e., the unsupervised learning term $\loss_{\text{rec}_i}$. Here, $N$ is the number of light-curves used to compute the loss functions (mini-batch size). The minimization problem is split into the following two main steps:

\begin{enumerate}
\item \textbf{Pre-training}: first, the autoencoder $\keys{\psibf, \varphibf}$ is trained to solve an auxiliary multi-band light-curve reconstruction task using a training-set composed of synthetic light-curves from $\dataset{train[s]}$. The encoder $\psibf$ computes a fixed-length representation-vector $\z_i$ from the variable-length multi-band light-curve $\Phi_i$ (see Fig. \ref{fig:arch_simple}). This representation-vector $\z_i$ automatically summarizes all the relevant aspects of the dynamics of the light-curve in order to estimate the correct light-curve reconstruction using the decoder $\varphibf$. The dynamics might include short and long-range trends and time dependencies; significant information about the first days of observations, SN-peak and SN-plateau regions; correlations and differences between bands; among others.

The representation-vector $\z_i$ serves as input to the classifier $\lambdabf$ to discriminate among SN types. Additionally, in eq. \refeq{eq.pre_training}, the labels of the synthetic light-curves are included on a cross-entropy regularization term over the representation-vector from the encoder. This regularization aims to improve the degree of nonlinear separation of the SN types over the representation space. The encoder aims to automatically generate an informative representation-vector $\z_i$ useful for both, a multi-band reconstruction task and a categorical discrimination task.

The pre-training optimization step is performed using the Adam optimizer \citep[][]{Kingma2014} with the following setting: \{\verb+params+=$\keys{\psibf, \varphibf, \lambdabf}$, \verb+batch_size+$=200$, \verb+betas+=$(.9, .999)$, \verb+weight_decay+=$\kexp{2}{-4}$\}. A linear learning rate warm-up schedule is implemented, increasing the learning rate $\text{lr}$, from $\text{lr}_\text{min}=\kexp{1}{-10}$ to $\text{lr}_\text{max}=\kexp{1.1}{-3}$, in $\Delta_\text{epoch}=10$ train epochs \citep[][]{Vaswani2017, Lee2021}.

\item \textbf{Fine-tuning}: after the pre-training process, a classification task is solved using a training-set composed only of empirical light-curves from $\dataset{train}$. In this step, no synthetic light-curves are used. The fine-tuning process is performed as a domain adaptation technique, aiming to minimize the model's gaps and discrepancies between the synthetic distribution and the empirical distribution of light-curves. Discrepancies may arise as the synthetic light-curves behavior could be biased towards the SPM's overly smooth behavior, general incorrect SPM parameters fit, inter-band peak time shifts and time differences, among others.

The fine-tuning optimization step is trained using the Stochastic Gradient Descent optimizer (SGD) with the following setting: \{\verb+params+$=\keys{\lambdabf}$, \verb+batch_size+$=50$, \verb+lr+$=\kexp{1}{-3}$, \verb+momentum+$=.9$\}. In the fine-tuning process only the parameters of the classifier $\lambdabf$ are re-optimized, while freezing the autoencoder parameters $\keys{\psibf, \varphibf}$. Thus, the encoder is used as a fixed-length representation-vector extractor from the multi-band light-curves.
\end{enumerate}

\subsubsection{Imbalance Learn and Regularization} 
To deal with class imbalance, the model is trained using mini-batches sampled from balanced auxiliary training-sets, which are dynamically and randomly constructed at each new training epoch using a stratified bootstrap strategy. This balancing strategy is applied during both optimization steps (see Appendix \refsec{sec:balance_learning} for details).

In addition, as a regularization technique, an early stopping routine is executed during both optimization steps. This technique is performed by evaluating the losses over the imbalanced validation-sets $\dataset{val}$. The losses of each light-curve $\Phi_i$, of class $c$, are weighted according to the factor $k_c=\frac{1}{N_c \cdot C}$, where $N_c$ is the number of samples from class $c$ and $C$ is the total number of classes. This procedure allows us to evaluate a kind of balanced loss function over an imbalanced validation-set.

Finally, during both optimization steps, a dynamical data-augmentation procedure is also implemented by introducing observation-flux noise and removing random observations along the multi-band light-curves. For further comparison purposes, we implement three levels of data-augmentation: zero, weak, and strong data-augmentation levels (see Appendix \refsec{sec:dataaugmentation} for details).

\subsection{RNN Baseline Models} 
In addition to the proposed attention-based encoder, baseline models based on Recurrent Neural Networks \citep[RNNs;][]{Rumelhart1986, Zimmermann2012} were implemented.

For the serial encoder, two different RNN models are tested: the Long Short-Term Memory \citep[LSTM;][]{Hochreiter1997} and the Gated Recurrent Unit \citep[GRU;][]{Chung2014}, which results in an alternative formulation for eq. \refeq{eq:s_sec}. Similar to previous works \citep[][]{Naul2018, Carrasco-Davis2019, Tsang2019, Gomez2020, Becker2020, Moller2020, Tachibana2020}, the encoder input vector $\x_{i,j}$, shown in eq. \refeq{eq:s_inproj}, is re-defined as $\x_{i,j}=\vect{\logfun{\obs_{i,j}+\varepsilon},\Delta t_{i,j}}$, where the time difference term $\Delta t_{i,j}$, for a multi-band light-curve $\Phi_i$, is included. This term aims to describe the irregular cadence information for the encoder to correctly capture relevant time dependencies.

Likewise, for the parallel encoder, eq. \refeq{eq:p_sec} is modeled with LSTM and GRU models. Additionally, the encoder input vector, described in eq. \refeq{eq:p_inproj}, is re-defined to include the time information as $\x_{i,\job}=\vect{\logfun{\obs_{i,\job}+\varepsilon},\Delta t^\ob_{i,\job}}$, where $\Delta t^\ob_{i,\job}$ is the time difference computed for the single-band light-curve $\Phi_i^\ob$.
\section{Results and Analyses}\label{sec:results} 
In this section, the experimental results are presented and the main analyses are performed. Due to computational cost limitations, all SN light-curve observations beyond a threshold-day of $100\tunits{days}$ were removed from all sets: $\dataset{train[s]}$, $\dataset{train}$, $\dataset{val}$, and $\dataset{test}$. We found this threshold-day representative enough to effectively study the SN events from the ZTF survey\footnote{Note that, under this consideration, the extended non-redundant modulation-range is $[0,150]\tunits{days}$.}.

If not specified otherwise, all results are reported by evaluating the models over the test-set $\dataset{test}$, which is composed of empirical light-curves only. The reported results consist of the aggregation of the results reported from all folds in the stratified 5-fold cross-validation. In addition, six random model's initializations (runs) per fold were performed, giving a total number of $N_\text{runs}=30$ runs per model implementation.

Due to the general high variance found in the results, the p-values ($\pvalue$) from significance statistical tests are also included when required, complementing the analysis of results. Given the non-Gaussian distribution observed in the test-set results\footnote{The non-Gaussianity (and high variance) of the aggregated 5-fold cross-validation results is produced due to differences in the reported classification performance among the test-set folds. These performance differences are usually influenced by the high class imbalance, the small number of samples, and the general quality of the light-curves presented in each test-set fold.}, a non-parametric statistical test is used: the permutation test\footnote{\url{http://rasbt.github.io/mlxtend/user_guide/evaluate/permutation_test/}.}. We use a threshold of $\pvalue<.05$ to denote a statistical significance when comparing differences ($\Delta$) between population means.

The notations \texttt{S-model} and \texttt{P-model} stand for the serial and parallel encoders used in the Deep Learning models, respectively. \highlighttext{Additionally, to further study the performance of the BRF classifier baseline, we propose two BRF settings: 1) Training with only empirical data (\texttt{training-set=[r]}): it is the original setting described in section \refsec{sec:brf_baseline} where only empirical light-curves, from the training-set $\dataset{train}$, are used to optimize the BRF baseline. 2) Training with only synthetic data (\texttt{training-set=spm-mcmc-estw[s]}): in this setting, only the synthetic light-curves, from the training-set $\dataset{train[s]}$, are used to optimize the BRF baseline.}

All Deep Learning models were implemented in Pytorch 1.8.1 \citep{Paszke2019}, using a GeForce GTX 1080 Ti GPU\footnote{\url{https://github.com/oscarpimentel/astro-lightcurves-classifier}.}.

\subsection{Late-Classification Scenario} 
We report the late-classification performance for all models using multi-band light-curves with a maximum threshold-day of $t_\text{th}=100\tunits{days}$, which are called \cdays light-curves in what follows. These light-curves are still of variable-length, where the maximum light-curve length found in this scenario is close to $L_i=150$.

Table \ref{tab:late_multiclass_metrics} shows the balanced metrics (\verb+b-metric+) for the multi-class classification scenario: Precision, Recall, F$_1$score, Area Under the Receiver Operating Characteristic Curve (AUCROC), and Area Under the Precision-Recall Curve (AUCPR). These balanced metrics assume that each class is equally important despite the high class imbalance (see Appendix \refsec{sec:metrics} for details).

For comparison purposes, the results for the zero, weak, and strong data-augmentation levels are presented and analyzed. In the zero data-augmentation level, none of the data-augmentation procedures are used. In the weak level, a probability of $10\%$ is used to randomly remove observations in the light-curves during training. Consequently, this probability value produces a moderate disturbance in the irregular cadence observed during training. In contrast, in the strong data-augmentation level, a probability of $50\%$ is used, heavily affecting the cadence observed during training (see Appendix \refsec{sec:dataaugmentation} for details). Additionally, the results of the pre-training, using empirical light-curves from the training-set $\dataset{train}$, are shown for comparison purposes (empirical pre-training).

%
\def\tabsrule{\rule{0pt}{0pt}\rule[0pt]{0pt}{0pt}}
\def\tabbline{\Xcline{1-6}{1.5pt}\tabsrule}
\begin{table*}[!t]
\centering
\caption{
Late-classification performances for the BRF baselines, RNN baselines, and attention-based models (TimeModAttn) using \cdays multi-band light-curves. Both, the serial (\texttt{S-model}) and parallel (\texttt{P-model}) encoders are reported along with several pre-training and data-augmentation schemes (mean$\pm$std from 5-fold cross-validation).
}
\label{tab:late_multiclass_metrics}\vspace{.1cm}
\tiny\scriptsize\footnotesize\small\normalsize
\footnotesize
\begin{tabular}{lccccc}
\tabbline
Feature-based models & b-Precision$_{ }^{ }$ & b-Recall$_{ }^{ }$ & b-$F_1$score$_{ }^{ }$ & b-AUCROC$_{ }^{ }$ & b-AUCPR$_{ }^{ }$ \tabsrule\\
\cmidrule{2-6}
BRF (fmode=all; training-set=[r]) & .527$\pm$.030 & .687$\pm$.052 & .525$\pm$.039 & .866$\pm$.020 & .602$\pm$.051  \tabsrule\\
BRF (fmode=all; training-set=spm-mcmc-estw[s]) & \textbf{.592$\pm$.032} & \textbf{.719$\pm$.048} & \textbf{.594$\pm$.047} & \textbf{.890$\pm$.018} & \textbf{.654$\pm$.053}  \tabsrule\\
\tabbline
Serial Deep Learning models\\
&\multicolumn{5}{c}{Empirical pre-training (zero data-augmentation)}\tabsrule\\
\cmidrule{2-6}
S-RNN+$\Delta t$ (cell=GRU) & .520$\pm$.043 & .626$\pm$.050 & .528$\pm$.039 & .852$\pm$.021 & .577$\pm$.049  \tabsrule\\
S-RNN+$\Delta t$ (cell=LSTM) & .497$\pm$.030 & .602$\pm$.044 & .502$\pm$.034 & .840$\pm$.019 & .568$\pm$.031  \tabsrule\\
\rowcolor{clr:light_gray}S-TimeModAttn (M=12; H=8; $\varepsilon_t$=6/24) & .551$\pm$.034 & .664$\pm$.058 & .565$\pm$.040 & .874$\pm$.024 & .597$\pm$.036  \tabsrule\\
&\multicolumn{5}{c}{Synthetic pre-training (zero data-augmentation)}\tabsrule\\
\cmidrule{2-6}
S-RNN+$\Delta t$ (cell=GRU) & .562$\pm$.051 & .688$\pm$.058 & .579$\pm$.049 & .885$\pm$.036 & .627$\pm$.062  \tabsrule\\
S-RNN+$\Delta t$ (cell=LSTM) & .561$\pm$.035 & .680$\pm$.053 & .578$\pm$.040 & .884$\pm$.028 & .619$\pm$.046  \tabsrule\\
\rowcolor{clr:light_gray}S-TimeModAttn (M=12; H=8; $\varepsilon_t$=6/24) & \textbf{.598$\pm$.030} & .736$\pm$.056 & \textbf{.614$\pm$.036} & .904$\pm$.029 & .665$\pm$.060  \tabsrule\\
&\multicolumn{5}{c}{Synthetic pre-training (weak data-augmentation)}\tabsrule\\
\cmidrule{2-6}
S-RNN+$\Delta t$ (cell=GRU) & .545$\pm$.034 & .706$\pm$.070 & .556$\pm$.045 & .879$\pm$.034 & .610$\pm$.066  \tabsrule\\
S-RNN+$\Delta t$ (cell=LSTM) & .550$\pm$.031 & .711$\pm$.070 & .558$\pm$.040 & .887$\pm$.033 & .621$\pm$.070  \tabsrule\\
\rowcolor{clr:light_gray}S-TimeModAttn (M=12; H=8; $\varepsilon_t$=6/24) & .588$\pm$.023 & \textbf{.759$\pm$.040} & .596$\pm$.033 & .910$\pm$.020 & \textbf{.671$\pm$.056}  \tabsrule\\
&\multicolumn{5}{c}{Synthetic pre-training (strong data-augmentation)}\tabsrule\\
\cmidrule{2-6}
S-RNN+$\Delta t$ (cell=GRU) & .491$\pm$.024 & .649$\pm$.068 & .496$\pm$.036 & .860$\pm$.032 & .561$\pm$.063  \tabsrule\\
S-RNN+$\Delta t$ (cell=LSTM) & .497$\pm$.021 & .657$\pm$.066 & .494$\pm$.028 & .864$\pm$.031 & .565$\pm$.055  \tabsrule\\
\rowcolor{clr:light_gray}S-TimeModAttn (M=12; H=8; $\varepsilon_t$=6/24) & .582$\pm$.017 & .754$\pm$.039 & .584$\pm$.031 & \textbf{.911$\pm$.019} & .665$\pm$.053  \tabsrule\\
\tabbline
Parallel Deep Learning models\\
&\multicolumn{5}{c}{Empirical pre-training (zero data-augmentation)}\tabsrule\\
\cmidrule{2-6}
P-RNN+$\Delta t$ (cell=GRU) & .521$\pm$.042 & .613$\pm$.042 & .527$\pm$.044 & .849$\pm$.013 & .561$\pm$.032  \tabsrule\\
P-RNN+$\Delta t$ (cell=LSTM) & .497$\pm$.034 & .604$\pm$.049 & .500$\pm$.041 & .834$\pm$.016 & .548$\pm$.027  \tabsrule\\
\rowcolor{clr:light_gray}P-TimeModAttn (M=12; H=4; $\varepsilon_t$=6/24) & .543$\pm$.026 & .671$\pm$.053 & .562$\pm$.029 & .865$\pm$.022 & .599$\pm$.038  \tabsrule\\
&\multicolumn{5}{c}{Synthetic pre-training (zero data-augmentation)}\tabsrule\\
\cmidrule{2-6}
P-RNN+$\Delta t$ (cell=GRU) & .566$\pm$.038 & .685$\pm$.056 & .582$\pm$.040 & .883$\pm$.027 & .624$\pm$.047  \tabsrule\\
P-RNN+$\Delta t$ (cell=LSTM) & .567$\pm$.029 & .683$\pm$.041 & .580$\pm$.036 & .881$\pm$.027 & .645$\pm$.048  \tabsrule\\
\rowcolor{clr:light_gray}P-TimeModAttn (M=12; H=4; $\varepsilon_t$=6/24) & \textbf{.591$\pm$.021} & .729$\pm$.038 & \textbf{.610$\pm$.026} & .897$\pm$.023 & .676$\pm$.059  \tabsrule\\
&\multicolumn{5}{c}{Synthetic pre-training (weak data-augmentation)}\tabsrule\\
\cmidrule{2-6}
P-RNN+$\Delta t$ (cell=GRU) & .547$\pm$.030 & .697$\pm$.070 & .552$\pm$.041 & .879$\pm$.031 & .610$\pm$.055  \tabsrule\\
P-RNN+$\Delta t$ (cell=LSTM) & .541$\pm$.022 & .704$\pm$.061 & .540$\pm$.032 & .876$\pm$.029 & .606$\pm$.051  \tabsrule\\
\rowcolor{clr:light_gray}P-TimeModAttn (M=12; H=4; $\varepsilon_t$=6/24) & .580$\pm$.020 & \textbf{.753$\pm$.044} & .594$\pm$.035 & \textbf{.911$\pm$.017} & \textbf{.689$\pm$.047}  \tabsrule\\
&\multicolumn{5}{c}{Synthetic pre-training (strong data-augmentation)}\tabsrule\\
\cmidrule{2-6}
P-RNN+$\Delta t$ (cell=GRU) & .490$\pm$.020 & .645$\pm$.057 & .482$\pm$.024 & .856$\pm$.032 & .577$\pm$.064  \tabsrule\\
P-RNN+$\Delta t$ (cell=LSTM) & .499$\pm$.020 & .660$\pm$.061 & .484$\pm$.031 & .857$\pm$.031 & .573$\pm$.053  \tabsrule\\
\rowcolor{clr:light_gray}P-TimeModAttn (M=12; H=4; $\varepsilon_t$=6/24) & .581$\pm$.019 & .750$\pm$.039 & .585$\pm$.036 & .907$\pm$.016 & .679$\pm$.043  \tabsrule\\
\tabbline
\end{tabular}
\end{table*}

A significant performance improvement, w.r.t. the empirical pre-training, can be observed (all metrics) when using the synthetic pre-training settings, i.e., when synthetic light-curves are used to perform the pre-training optimization step. This improvement is achieved by all the tested Deep Learning models for both, the serial and parallel encoders. In the case of the BRF baseline, a significant performance improvement is also achieved when training with synthetic data w.r.t. the case of training with real data. \highlighttext{The results obtained with a BRF model trained with synthetic data are analyzed in section \refsec{sec:brf_synth}}. These results confirm that the use of synthetic light-curves is effectively beneficial to support the optimization of both, the Deep Learning models (attention-based and RNN models) and also the BRF model.

For all the synthetic pre-training settings, it can be observed (all metrics) that the proposed TimeModAttn model outperformed the BRF baseline trained with real data (\texttt{training-set=[r]}). In particular, we highlight the weak data-augmentation level, where the following are the metric mean’s differences, w.r.t. the BRF baseline, for the serial encoder: $\Delta$b-Precision=$\diffpv{.0611}{***}$, $\Delta$b-Recall=$\diffpv{.0719}{***}$, $\Delta$b-$F_1$score$=\diffpv{.0703}{***}$, $\Delta$b-AUCROC=$\diffpv{.0437}{***}$, and $\Delta$b-AUCPR$=\diffpv{.0691}{***}$\footnote{The statistical significance notation used is as follows:\\$^\text{***}p\leq.001$, $^\text{**}p\leq.01$, $^\text{*}p\leq.05$, and $^\text{+}p\leq.1$.\label{foot:pvalues}}. For the parallel encoder, the metric differences are as follows: $\Delta$b-Precision=$\diffpv{.0530}{***}$, $\Delta$b-Recall=$\diffpv{.0654}{***}$, $\Delta$b-$F_1$score$=\diffpv{.0687}{***}$, $\Delta$b-AUCROC=$\diffpv{.0446}{***}$, and $\Delta$b-AUCPR$=\diffpv{.0865}{***}$. \highlighttext{The comparison w.r.t. the BRF model trained with synthetic data is analyzed in section \refsec{sec:brf_synth}}. Additionally, no strong or consistent statistical evidence was found to conclude that either, the serial encoder or the parallel encoder, is the best alternative ($\pvalue\in\boxx{.021, .403}$\footnote{$\pvalue\in\keys{.021, .206, .403, .391, .036}$.}), implying that the type of encoder may be irrelevant in terms of general performance for the TimeModAttn model.

\subsubsection{TimeModAttn Model Versus RNN Baselines}
It can be observed that the level of the data-augmentation affected the performance of the RNN baselines for both the serial and parallel encoders. The strong data-augmentation level was detrimental to the performance of the RNN baselines. This effect may be explained because the proposed data-augmentation dynamically influences the number of observations in the light-curves during the model optimization, directly affecting the values of the computed time differences $\Delta t_{i,j}$: the higher the probability of removing observations, the longer the computed time differences. Thus, the data-augmentation produces a discrepancy between the time difference distributions of the training-set and the test-set, which may lead to a poor model generalization for unobserved light-curves presented in the test-set. Note that the maximum discrepancy between these distributions arises in the strong data-augmentation level, where the worst performance was reported for the RNN baselines.

In contrast, high robustness against different data-augmentation levels can be observed for the TimeModAttn model. This could be explained because the encoder in the TimeModAttn model (TimeFiLM) directly uses the observation-times, without just relying on the time difference values. Thus, the time representation used in the TimeModAttn model might be less sensitive to the general irregularity of the cadence, achieving a higher degree of model generalization for the unobserved light-curves in the test-set. This effect can be observed in the reported performances in Table \ref{tab:late_multiclass_metrics} where, for all the tested data-augmentation levels (all metrics) the TimeModAttn model achieved higher performance than the RNN baselines. Note that the TimeModAttn model achieved high classification performances even in the strong data-augmentation level.

For further comparisons, we analyze the zero data-augmentation level because the RNN baselines achieved the best general performance in that setting. The TimeModAttn model outperforms the GRU baseline (all metrics) for both, the serial encoder ($\pvalue\leq\rightonly{...}{.001},\forall\pvalue$) and the parallel encoder ($\pvalue\leq\rightonly{...}{.001},\forall\pvalue$). Similar statistical evidence was also found when comparing the TimeModAttn model w.r.t. the LSTM baseline (all metrics) for both, the serial encoder ($\pvalue\leq\rightonly{...}{.001},\forall\pvalue$) and the parallel encoder ($\pvalue\leq\rightonly{...}{.001},\forall\pvalue$). Additionally, no strong or consistent statistical significance was found when comparing the GRU and LSTM baselines for both, the serial encoder ($\pvalue\in\boxx{.212, .459}$\footnote{$p\in\keys{.453, .212, .459, .440, .215}$.}) and the parallel encoder ($\pvalue\in\boxx{.009, .482}$\footnote{$p\in\keys{.482, .447, .380, .370, .009}$.}). Thus, no major difference exists between both RNN models in the context of this work.

\subsection{Early-Classification Scenario}\label{sec:early} 
Herein we study the case when a higher number of observations is gradually available in the test-set $\dataset{test}$. With this aim, a moving threshold-day $t_\text{th}\in [1,100]\tunits{days}$ is used in order to remove all observations, from test-set $\dataset{test}$, above a given threshold. Note that if $t_\text{th}=100\tunits{days}$; then, the test-set $\dataset{test}$ is equivalent to the set used in the late-classification scenario. For the feature extraction, algorithm instabilities arise when just a single observation is used to fit the SPM model. Therefore, for the BRF baseline, the results start being reported only after a minimum number of observations is reached: when all light-curves in the test-set $\dataset{test}$ have at least one band with a number equal or higher than $L_i^\ob\geq 2$ observations.

\begin{figure*}[!t]
\centering
\includegraphics[width=1\linewidth]{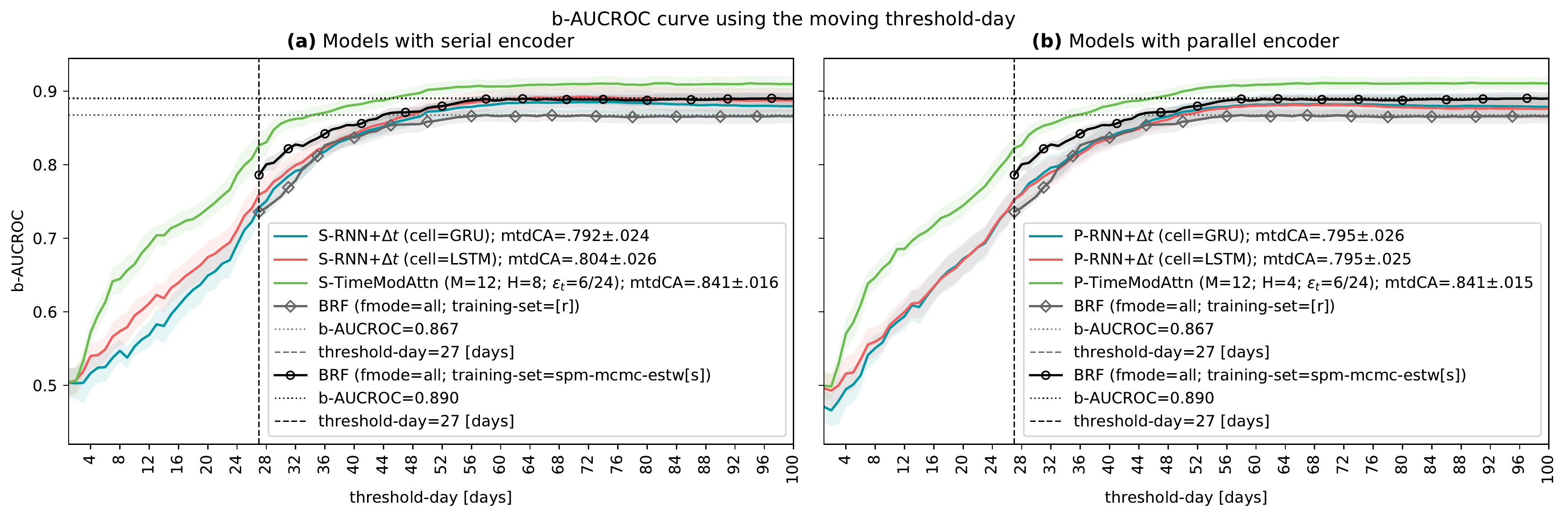}
\caption{
b-AUCROC metric-curve v/s moving threshold-day for the attention-based models, BRF baselines, and RNN baselines in the weak data-augmentation level (for the sake of better visualization, the mean$\pm\frac{1}{2}$std curve is shown from 5-fold cross-validation). The moving threshold-day Curve Average (mtdCA) is shown for the TimeModAttn model and RNN baselines. \highlighttext{Note that the horizontal axis (evolution of the threshold-day) is related with the observation-time since the first observation}. (a) Models with serial encoder. (b) Models with parallel encoder.
}
\label{fig:dl_models_and_brf_b-aucroc}
\end{figure*}

For example, Fig. \ref{fig:dl_models_and_brf_b-aucroc} shows the evolution of the b-AUCROC metric as a function of a moving threshold-day $t_\text{th}$ in the weak data-augmentation level. As expected, the general performance of the b-AUCROC increased with larger threshold-days. This is because the models have access to longer light-curves; hence, more information about the evolution of the SN transient event. From the b-AUCROC curves, we can observe that the performance of the TimeModAttn model tended to be higher than the rest of the tested baselines for most of the operation points. In particular, the TimeModAttn model achieved the maximum performance of b-AUCROC of the BRF baseline trained with real data (\texttt{training-set=[r]}) several days earlier ($t_\text{th}\in(32, 40)\tunits{days}$) than this baseline ($t_\text{th}\in(52,60)\tunits{days}$). This result indicates that the TimeModAttn model can discriminate between SN types using light-curves with fewer observations. The results obtained with a BRF model trained with synthetic data are analyzed in section \refsec{sec:brf_synth}.

To summarize the early-classification results, the Curve Average (CA) is computed for the curves constructed by using the moving threshold-day (mtdCA). High values of the mtdCA are achieved if the performance of the model is consistently high along all the operation points defined by the moving threshold-day. Table \ref{tab:early_multiclass_metrics} shows the computed mtdCA for all the reported balanced metrics.

%
\def\tabsrule{\rule{0pt}{0pt}\rule[0pt]{0pt}{0pt}}
\def\tabbline{\Xcline{1-6}{1.5pt}\tabsrule}
\begin{table*}[!t]
\centering
\caption{
Early-classification performances for the RNN baselines and attention-based models (TimeModAttn). The moving threshold-day Curve Average (mtdCA) is used ($\ddag$). Both, the serial (\texttt{S-model}) and parallel (\texttt{P-model}) encoders are reported along with several pre-training and data-augmentation schemes (mean$\pm$std from 5-fold cross-validation).
}
\label{tab:early_multiclass_metrics}\vspace{.1cm}
\scriptsize\footnotesize\small\normalsize
\footnotesize
\begin{tabular}{lccccc}
\tabbline
Serial Deep Learning models & b-Precision$_\text{ }^\text{\ddag}$ & b-Recall$_\text{ }^\text{\ddag}$ & b-$F_1$score$_\text{ }^\text{\ddag}$ & b-AUCROC$_\text{ }^\text{\ddag}$ & b-AUCPR$_\text{ }^\text{\ddag}$ \tabsrule\\
\cmidrule{2-6}
&\multicolumn{5}{c}{Empirical pre-training (zero data-augmentation)}\tabsrule\\
\cmidrule{2-6}
S-RNN+$\Delta t$ (cell=GRU) & .427$\pm$.025 & .491$\pm$.029 & .404$\pm$.034 & .745$\pm$.020 & .461$\pm$.022  \tabsrule\\
S-RNN+$\Delta t$ (cell=LSTM) & .412$\pm$.021 & .489$\pm$.028 & .395$\pm$.026 & .749$\pm$.018 & .450$\pm$.028  \tabsrule\\
\rowcolor{clr:light_gray}S-TimeModAttn (M=12; H=8; $\varepsilon_t$=6/24) & .474$\pm$.025 & .535$\pm$.033 & .448$\pm$.030 & .806$\pm$.023 & .509$\pm$.028  \tabsrule\\
&\multicolumn{5}{c}{Synthetic pre-training (zero data-augmentation)}\tabsrule\\
\cmidrule{2-6}
S-RNN+$\Delta t$ (cell=GRU) & .471$\pm$.033 & .536$\pm$.036 & .440$\pm$.036 & .780$\pm$.029 & .505$\pm$.038  \tabsrule\\
S-RNN+$\Delta t$ (cell=LSTM) & .481$\pm$.035 & .559$\pm$.035 & .459$\pm$.042 & .797$\pm$.024 & .518$\pm$.038  \tabsrule\\
\rowcolor{clr:light_gray}S-TimeModAttn (M=12; H=8; $\varepsilon_t$=6/24) & .516$\pm$.022 & .608$\pm$.033 & \textbf{.497$\pm$.022} & .832$\pm$.024 & .562$\pm$.038  \tabsrule\\
&\multicolumn{5}{c}{Synthetic pre-training (weak data-augmentation)}\tabsrule\\
\cmidrule{2-6}
S-RNN+$\Delta t$ (cell=GRU) & .481$\pm$.030 & .577$\pm$.044 & .454$\pm$.031 & .792$\pm$.024 & .520$\pm$.039  \tabsrule\\
S-RNN+$\Delta t$ (cell=LSTM) & .480$\pm$.023 & .590$\pm$.036 & .457$\pm$.027 & .804$\pm$.026 & .527$\pm$.043  \tabsrule\\
\rowcolor{clr:light_gray}S-TimeModAttn (M=12; H=8; $\varepsilon_t$=6/24) & \textbf{.522$\pm$.022} & .630$\pm$.026 & .495$\pm$.020 & .841$\pm$.016 & \textbf{.580$\pm$.040}  \tabsrule\\
&\multicolumn{5}{c}{Synthetic pre-training (strong data-augmentation)}\tabsrule\\
\cmidrule{2-6}
S-RNN+$\Delta t$ (cell=GRU) & .447$\pm$.023 & .556$\pm$.036 & .430$\pm$.026 & .783$\pm$.020 & .491$\pm$.033  \tabsrule\\
S-RNN+$\Delta t$ (cell=LSTM) & .445$\pm$.019 & .567$\pm$.035 & .422$\pm$.021 & .790$\pm$.021 & .492$\pm$.031  \tabsrule\\
\rowcolor{clr:light_gray}S-TimeModAttn (M=12; H=8; $\varepsilon_t$=6/24) & \textbf{.522$\pm$.018} & \textbf{.632$\pm$.023} & .493$\pm$.020 & \textbf{.845$\pm$.012} & .579$\pm$.039  \tabsrule\\
\tabbline
Parallel Deep Learning models\\
&\multicolumn{5}{c}{Empirical pre-training (zero data-augmentation)}\tabsrule\\
\cmidrule{2-6}
P-RNN+$\Delta t$ (cell=GRU) & .447$\pm$.028 & .501$\pm$.029 & .420$\pm$.030 & .762$\pm$.018 & .471$\pm$.018  \tabsrule\\
P-RNN+$\Delta t$ (cell=LSTM) & .422$\pm$.026 & .494$\pm$.024 & .400$\pm$.033 & .749$\pm$.026 & .458$\pm$.026  \tabsrule\\
\rowcolor{clr:light_gray}P-TimeModAttn (M=12; H=4; $\varepsilon_t$=6/24) & .461$\pm$.023 & .516$\pm$.026 & .438$\pm$.022 & .789$\pm$.021 & .495$\pm$.023  \tabsrule\\
&\multicolumn{5}{c}{Synthetic pre-training (zero data-augmentation)}\tabsrule\\
\cmidrule{2-6}
P-RNN+$\Delta t$ (cell=GRU) & .492$\pm$.032 & .549$\pm$.034 & .464$\pm$.033 & .788$\pm$.026 & .519$\pm$.033  \tabsrule\\
P-RNN+$\Delta t$ (cell=LSTM) & .488$\pm$.025 & .552$\pm$.024 & .461$\pm$.029 & .791$\pm$.030 & .530$\pm$.034  \tabsrule\\
\rowcolor{clr:light_gray}P-TimeModAttn (M=12; H=4; $\varepsilon_t$=6/24) & \textbf{.516$\pm$.017} & .597$\pm$.018 & \textbf{.499$\pm$.018} & .826$\pm$.020 & .565$\pm$.027  \tabsrule\\
&\multicolumn{5}{c}{Synthetic pre-training (weak data-augmentation)}\tabsrule\\
\cmidrule{2-6}
P-RNN+$\Delta t$ (cell=GRU) & .485$\pm$.023 & .580$\pm$.041 & .462$\pm$.028 & .795$\pm$.026 & .524$\pm$.040  \tabsrule\\
P-RNN+$\Delta t$ (cell=LSTM) & .476$\pm$.018 & .586$\pm$.036 & .451$\pm$.024 & .795$\pm$.025 & .516$\pm$.034  \tabsrule\\
\rowcolor{clr:light_gray}P-TimeModAttn (M=12; H=4; $\varepsilon_t$=6/24) & .514$\pm$.018 & .621$\pm$.027 & \textbf{.499$\pm$.019} & \textbf{.841$\pm$.015} & \textbf{.587$\pm$.029}  \tabsrule\\
&\multicolumn{5}{c}{Synthetic pre-training (strong data-augmentation)}\tabsrule\\
\cmidrule{2-6}
P-RNN+$\Delta t$ (cell=GRU) & .440$\pm$.016 & .561$\pm$.045 & .422$\pm$.022 & .777$\pm$.023 & .496$\pm$.039  \tabsrule\\
P-RNN+$\Delta t$ (cell=LSTM) & .442$\pm$.016 & .564$\pm$.043 & .416$\pm$.018 & .782$\pm$.020 & .493$\pm$.032  \tabsrule\\
\rowcolor{clr:light_gray}P-TimeModAttn (M=12; H=4; $\varepsilon_t$=6/24) & .515$\pm$.015 & \textbf{.623$\pm$.022} & .495$\pm$.020 & \textbf{.841$\pm$.014} & .582$\pm$.034  \tabsrule\\
\tabbline
\end{tabular}
\end{table*}

Our findings in the early-classification scenario follow a similar trend as those previously reported in the late-classification scenario. As before, pre-training with synthetic light-curves was beneficial for all the tested Deep Learning models. No strong or consistent statistical evidence was found to conclude which encoder alternative is the best (serial or parallel) for the TimeModAttn model ($\pvalue\in\boxx{.022, .458}$\footnote{$p\in\keys{.022, .034, .164, .458, .118}$.}) in the weak data-augmentation level.

\subsubsection{TimeModAttn Model Versus RNN Baselines}
Table \ref{tab:early_multiclass_metrics} shows that for all the different pre-training settings (all metrics) the TimeModAttn model achieved higher performance than the RNN baselines. When comparing performances of the TimeModAttn model against the RNN baselines (GRU and LSTM), in the zero data-augmentation level (best late-classification setting for the RNN baselines), a significant difference was found (all metrics) for both, the serial encoder ($\pvalue \leq \rightonly{...}{.001},\forall \pvalue$) and the parallel encoder ($\pvalue \leq \rightonly{...}{.001},\forall \pvalue$). These results, along with Fig. \ref{fig:dl_models_and_brf_b-aucroc}, indicate that the TimeModAttn model outperforms the RNN baselines in the early-classification scenario of light-curves with few observations.

As before, general robustness against the level of the data-augmentation was again observed for the TimeModAttn model. Furthermore, the results shown in Table \ref{tab:early_multiclass_metrics} suggest that the use of data-augmentation could be beneficial in the early-classification performance for the TimeModAttn model.

The above results show that the performance of the TimeModAttn model is not only higher in the late-classification scenario, but it is also consistently higher along with different early-classification operation points that are defined by changing the moving threshold-day. Moreover, the early-classification performance of the TimeModAttn model was higher than all the other tested baselines.

If not specified otherwise, the weak data-augmentation level is selected and explored as the main pre-training setting for the following experiments in this work. Additionally, examples of SN multi-band light-curve reconstructions can be found in Appendix \refsec{sec:reconstruction} for the weak data-augmentation level.

\subsection{Confusion Matrices, Misclassifications, and Operational Curves}\label{sec:cms_and_roccs} 
\begin{figure*}[!t]
\centering
\includegraphics[width=1\linewidth]{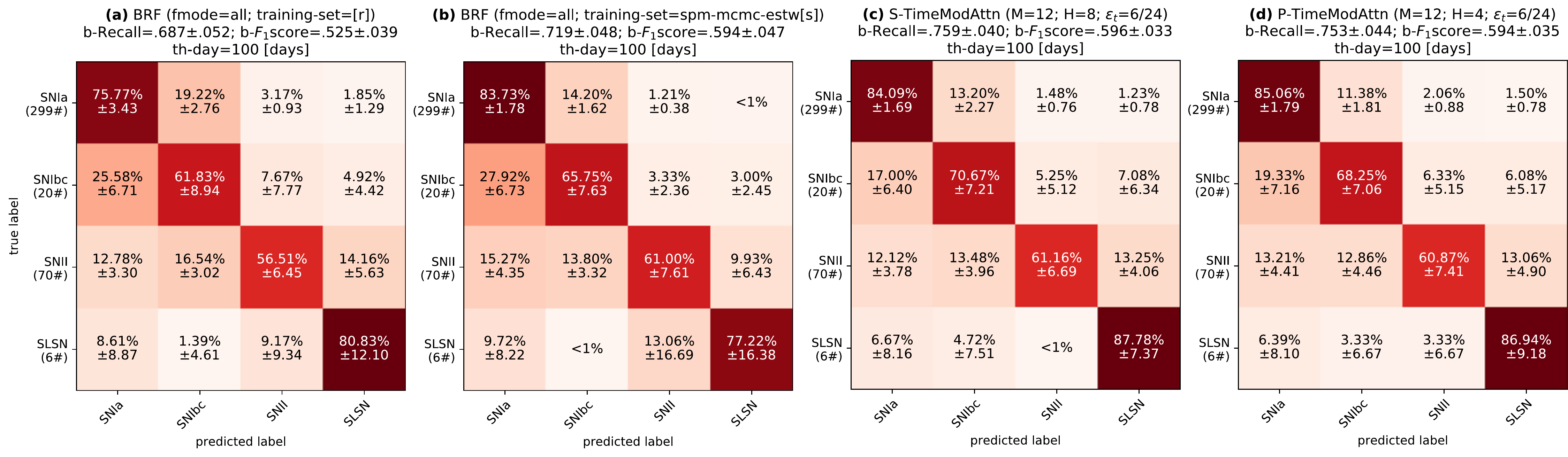}
\caption{
Confusion matrices for the SNe classification task using \cdays multi-band light-curves (mean$\pm$std from 5-fold cross-validation). The corresponding b-Recall and b-$F_1$score metrics are given on the top of each matrix. (a) BRF baseline trained with real data. (b) BRF baseline trained with synthetic data. (c) S-TimeModAttn model. (d) P-TimeModAttn model.
}
\label{fig:cms}
\end{figure*}

Fig. \ref{fig:cms} shows confusion matrices for the classification of SN multi-band light-curves. As previously reported in the literature \citep{Moss2018, Villar2019, Sanchez-Saez2021}, we can observe a common confusion between the SNIa and SNIbc types, in all confusion matrices, which may be related with intrinsic similarities of the mechanisms that cause the SN-peak: the diffusion of energy deposited by radioactive $^{56}$Ni \citep{Arnett2008}. The TimeModAttn model decreased the confusion between the SNIa and SNIbc types w.r.t. the BRF baseline trained with real data (\texttt{training-set=[r]}) for both, the serial and parallel encoders. The TimeModAttn model achieved a maximum increment of the True Positive (TP) percentage for the SNIa type of $\Delta \text{TP}_\text{SNIa}=\diffpv{9.2939}{***}$, for the SNIbc type of $\Delta \text{TP}_\text{SNIbc}=\diffpv{8.8333}{***}$, for the SNII type of $\Delta \text{TP}_\text{SNII}=\diffpv{4.6453}{***}$, and for the SLSN type of $\Delta \text{TP}_\text{SLSN}=\diffpv{6.9444}{***}$\footref{foot:pvalues}.

\begin{figure*}[!t]
\centering
\includegraphics[width=1\linewidth]{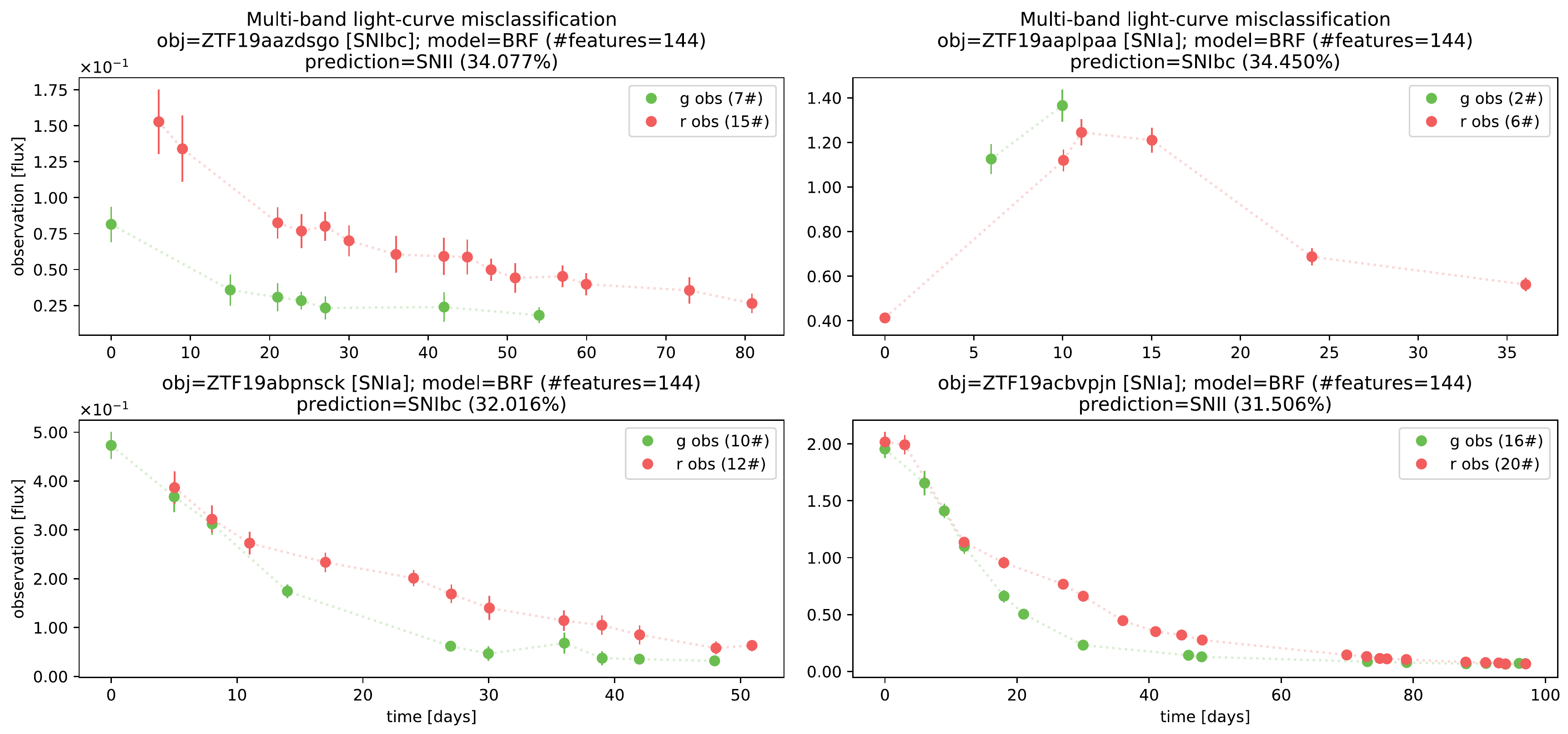}
\caption{
Four examples of misclassified SN multi-band light-curves by the BRF baseline trained with real data (\texttt{training-set=[r]}) that were correctly classified by the P-TimeModAttn model.
}
\label{fig:only_brf_misclassifications}
\end{figure*}

Fig. \ref{fig:only_brf_misclassifications} shows four light-curve examples that were correctly classified by the TimeModAttn model but incorrectly classified by the BRF baseline trained with real data (\texttt{training-set=[r]}). Several misclassification errors arise among curves that, due to the irregular cadence, do not present observations in the SN-rise and/or SN-peak regions. This might cause instabilities in the SPM fitting, producing misleading features for the BRF baseline. For instance, features related with the rising time or the maximum brightness could be incorrectly estimated. Consequently, this may be especially detrimental for the discrimination between the SNIa and SNIbc types. In addition, multi-band light-curves having zero or few observations in one band tended to be misclassified by the BRF baseline too. As stated before, the scarcity of observations could lead to highly unstable SPM fittings and misleading features.

\begin{figure*}[!t]
\centering
\includegraphics[width=1\linewidth]{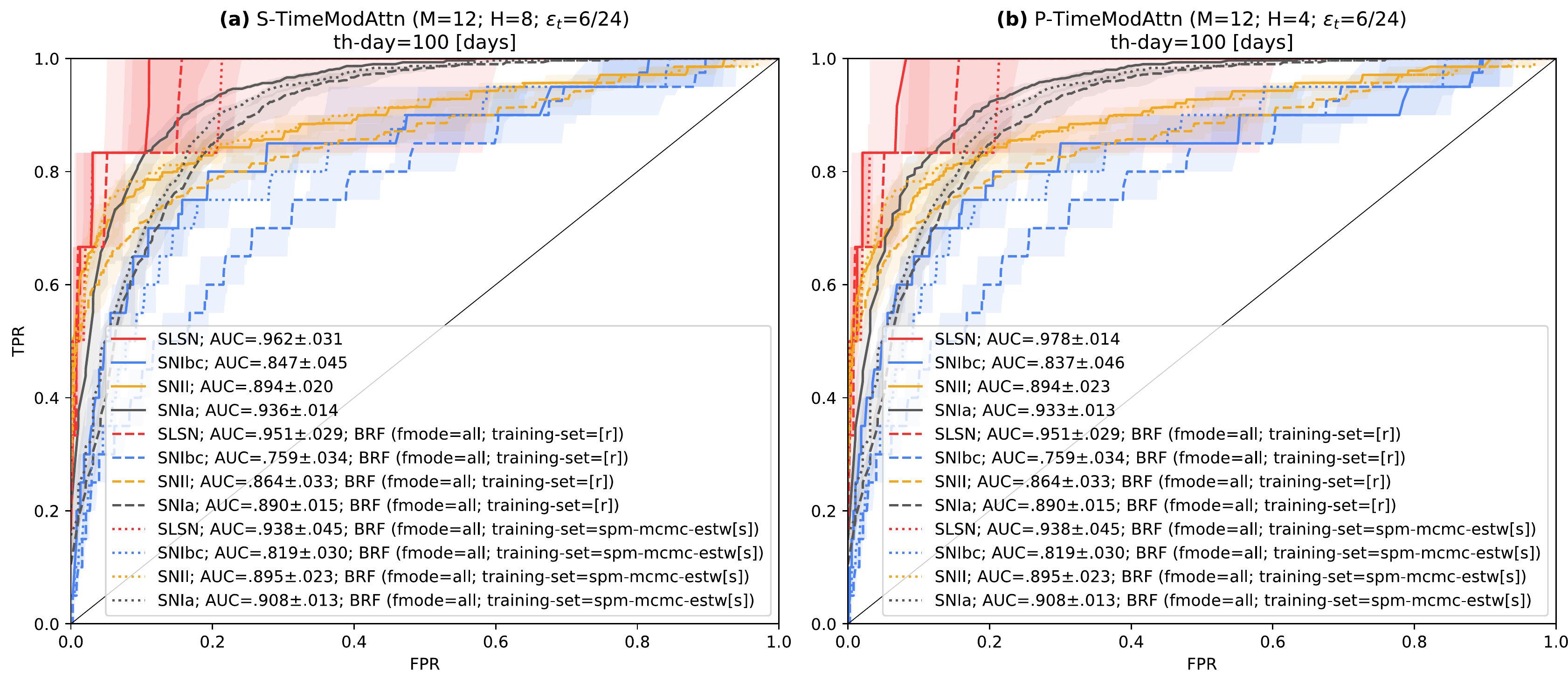}
\caption{
ROC curves for the SNe classification task using \cdays light-curves, where TPR and FPR stand for the True Positive Rate and False Positive Rate, respectively (50-percentile curve and 30-70-percentile are shown from 5-fold cross-validation). In both plots, the BRF ROC curves are shown as dashed lines (both: trained with real data and trained with synthetic data). (a) S-TimeModAttn. (b) P-TimeModAttn model.
}
\label{fig:roccs}
\end{figure*}

Fig. \ref{fig:roccs} shows the ROC operational curves for each SN type. In general, it can be observed that, for all the SN types, the ROC curves reported from the TimeModAttn models are above those from the BRF baseline trained with real data (\texttt{training-set=[r]}), leading to higher AUCROC scores per SN type for the TimeModAttn model. We highlight the ROC curves separation for the SNIa and SNIbc types, with maximum AUCROC differences of $\Delta \text{AUCROC}_\text{SNIa}=\diffpv{.0457}{***}$ and $\Delta \text{AUCROC}_\text{SNIbc}=\diffpv{.0877}{***}$\footref{foot:pvalues}, respectively. This fact correlates with the decrease of confusion errors found in the confusion matrices for these SN types. The results obtained with a BRF model trained with synthetic data are analyzed in section \refsec{sec:brf_synth}.

\subsection{BRF Baseline Trained With Synthetic Data}\label{sec:brf_synth}
\highlighttext{To further study the late-classification scenario performance of the BRF baseline settings w.r.t. the TimeModAttn model, Table \ref{tab:late_multiclass_metrics} includes the additional BRF baseline setting of training with synthetic data (\texttt{training-set=spm-mcmc-estw[s]}). Additionally, Fig. \ref{fig:cms} also shows the confusion matrix associated with this setting for the BRF baseline. We can observe that the confusion matrices of the TimeModAttn models show a general higher and well-distributed TP percentage along the diagonal, ensuring good performance for all classes. This is important because all the classes must be, in principle, equally relevant in the classification task.}

\highlighttext{We also studied the evolution of the performance w.r.t. the general number of observations in the test-set (by using the moving threshold-day). Fig. \ref{fig:dl_models_and_brf_b-aucroc} shows the evolution of the performance for the b-AUCROC metric, of both BRF baseline settings, as a function of the moving threshold-day $t_\text{th}$. It can be observed that the TimeModAttn model still achieved higher performance w.r.t. both BRF baseline settings in the early-classification and the late-classification.}

\highlighttext{In general, we highlight the fact that using synthetic data for training helps in the performance for both, the TimeModAttn model and the BRF baseline. In fact, although the TimeModAttn model shows a general better overall performance than the tested BRF settings (early and late-classification), it is not possible to
firmly conclude that the proposed model is always better when using synthetic data for training: no strong statistically significance was found for all metrics. In any case, the proposed TimeModAttn model still has advantages in this comparison scenario:
\begin{enumerate}
\item It is worth remembering that the computation and extraction of features, from the light-curves, is mandatory when using a feature-based model. Thus, to optimize the BRF using synthetic light-curves, the features of the latter are strictly required. Because the computation of features is a costly process, the optimization of the BRF model using synthetic light-curves incurs in an even higher computational cost w.r.t. the use of only empirical light-curves.

\item Note that including light-curves obtained with data-augmentation methods (such as the one described in Appendix \refsec{sec:dataaugmentation}) may be overly demanding, incurring in an even higher computational cost since the computation of the features of each possible augmented light-curve must be carried out.

\item It should be remembered that the inference of the TimeModAttn model is straightforward for short light-curves with one or few observations. This is not the case for the feature-based BRF model as the features extracted with one or few observations are highly unstable. This effect can not be solved by using synthetic light-curves.

\item The use of features still heavily depends on expert knowledge which is a non-trivial and costly task (and may even induce undesired expert biases). This task, moreover, should always be
subject to revision, especially if new types of astronomical objects need to be studied in the future.
\end{enumerate}}

\subsection{Multi-band Effect in Classification} 
%
To study the effect of the multi-band information, we designed a setting to train and evaluate the TimeModAttn model using only a single-band $b^*$. With this aim, we use the parallel encoder where, given a target preserved single-band $b^*$, all the representation-vectors associated with the rest of the bands are replaced with zero vectors as $\z_i^{(b')} = \vec{0},\forall b'\in\keys{1,\dots,B}-\keys{b^*}$. Moreover, extra considerations were implemented to properly test this experiment, e.g., a zero reconstruction loss, in eq. \refeq{eq.pre_training}, is used for all bands except for the target preserved single-band $b^*$; all information related with the observation-times is properly adjusted. Table \ref{tab:preserved} shows the reported performance for the TimeModAttn model with different target preserved single-bands.

%
\def\tabsrule{\rule{0pt}{0pt}\rule[0pt]{0pt}{0pt}}
\def\tabbline{\Xcline{1-6}{1.5pt}\tabsrule}
\begin{table*}[!t]
\centering
\caption{
Late-classification performances for the BRF baselines and attention-based models (TimeModAttn) using \cdays multi-band light-curves and different target preserved single-bands. The value pb=\{g, r\} indicates the target preserved single-band $b^*$. To avoid empty light-curve evaluations in the preserved single-band scenarios, light-curves with zero observations, in any of the $B$ bands, were removed from the test-set $\dataset{test}$. Both, the serial (\texttt{S-model}) and parallel (\texttt{P-model}) encoders are reported (mean$\pm$std from 5-fold cross-validation).
}
\label{tab:preserved}\vspace{.1cm}
\scriptsize\footnotesize\small\normalsize
\footnotesize
\begin{tabular}{lccccc}
\tabbline
Multi-band models& b-Precision$_\text{ }^\text{ }$ & b-Recall$_\text{ }^\text{ }$ & b-$F_1$score$_\text{ }^\text{ }$ & b-AUCROC$_\text{ }^\text{ }$ & b-AUCPR$_\text{ }^\text{ }$ \tabsrule\\
\cmidrule{2-6}
BRF (fmode=all; training-set=[r]) & .530$\pm$.032 & .692$\pm$.055 & .530$\pm$.042 & .867$\pm$.020 & .605$\pm$.052  \tabsrule\\
BRF (fmode=all; training-set=spm-mcmc-estw[s]) & \textbf{.593$\pm$.033} & .722$\pm$.052 & .596$\pm$.050 & .892$\pm$.017 & .657$\pm$.055  \tabsrule\\
\rowcolor{clr:light_gray}S-TimeModAttn (M=12; H=8; $\varepsilon_t$=6/24) & .592$\pm$.022 & \textbf{.762$\pm$.039} & \textbf{.600$\pm$.033} & .911$\pm$.019 & .675$\pm$.055  \tabsrule\\
\rowcolor{clr:light_gray}P-TimeModAttn (M=12; H=4; $\varepsilon_t$=6/24) & .586$\pm$.018 & .757$\pm$.044 & .599$\pm$.036 & \textbf{.914$\pm$.017} & \textbf{.692$\pm$.047}  \tabsrule\\
\tabbline
Single-band models\\
\rowcolor{clr:light_gray}P-TimeModAttn (M=12; H=4; $\varepsilon_t$=6/24; pb=g) & .518$\pm$.019 & \textbf{.661$\pm$.044} & \textbf{.508$\pm$.027} & \textbf{.848$\pm$.026} & \textbf{.584$\pm$.074}  \tabsrule\\
\rowcolor{clr:light_gray}P-TimeModAttn (M=12; H=4; $\varepsilon_t$=6/24; pb=r) & \textbf{.525$\pm$.019} & .625$\pm$.057 & .495$\pm$.036 & .846$\pm$.024 & .582$\pm$.031  \tabsrule\\
\tabbline
\end{tabular}
\end{table*}

The reported metrics show that the exclusive use of a single-band information (g or r) is significantly detrimental for the classification performance of the TimeModAttn model; therefore, using both bands is required to correctly characterize the SN transient events. Using all the available bands is especially beneficial when just a few observations are available in a particular band, where the model requires to support the classification task by using observations from the rest of the bands. Consequently, the use of serial or parallel encoders is recommended to properly capture all the information from a multi-band light-curve.

\subsection{Interpretability Experiments} 
To explore, evaluate, and validate the automatic decisions of the TimeModAttn model, several experiments on interpretability are presented in this section. These experiments are based on the parallel encoder formulation, allowing us to explore the attention scores and the learned temporal modulation in each band.

\subsubsection{Attention Scores}\label{sec:exp_attn} 
Given a single-band light-curve $\Phi_i^\ob$, the attention scores $\set{s_{i,\job}}{\job=1}{L_i^\ob}$ are collected from the last MHSelfAttn's layer. In the multi-head attention scenario, the average score among the $H$ heads is used: $s_{i,\job}=\frac{1}{H}\sum_{h\seq 1}^{H}s^{(h)}_{i,\job},\forall \job$. Then, these average attention scores are normalized as follows:
\begin{align}
\bar{s}_{i,\job} &= \fractuple{s_{i,\job}-s^\ob_{i_\text{min}}}{s^\ob_{i_\text{max}}-s^\ob_{i_\text{min}}},&&\text{}\label{eq:norm_attn}\raisetag{20pt}
\end{align}
where $\bar{s}_{i,\job} \in [0,1]$ is the normalized attention score given the average attention score $s_{i,\job}\in \real{+}$. The maximum and minimum attention scores, found in the band $b$, are denoted as $s^\ob_{i_\text{max}}=\maxx{}\set{s_{i,\job}}{\job\seq1}{L_i^\ob}$ and $s^\ob_{i_\text{min}}=\minn{}\set{s_{i,\job}}{\job\seq1}{L_i^\ob}$, respectively. This procedure generates a maximum normalized attention score value of $\bar{s}_{i,\job}=1$ in the observation with the highest original attention score and a value of $\bar{s}_{i,\job}=0$ at the lowest.

\begin{figure*}[!t]
\centering
\includegraphics[width=1\linewidth]{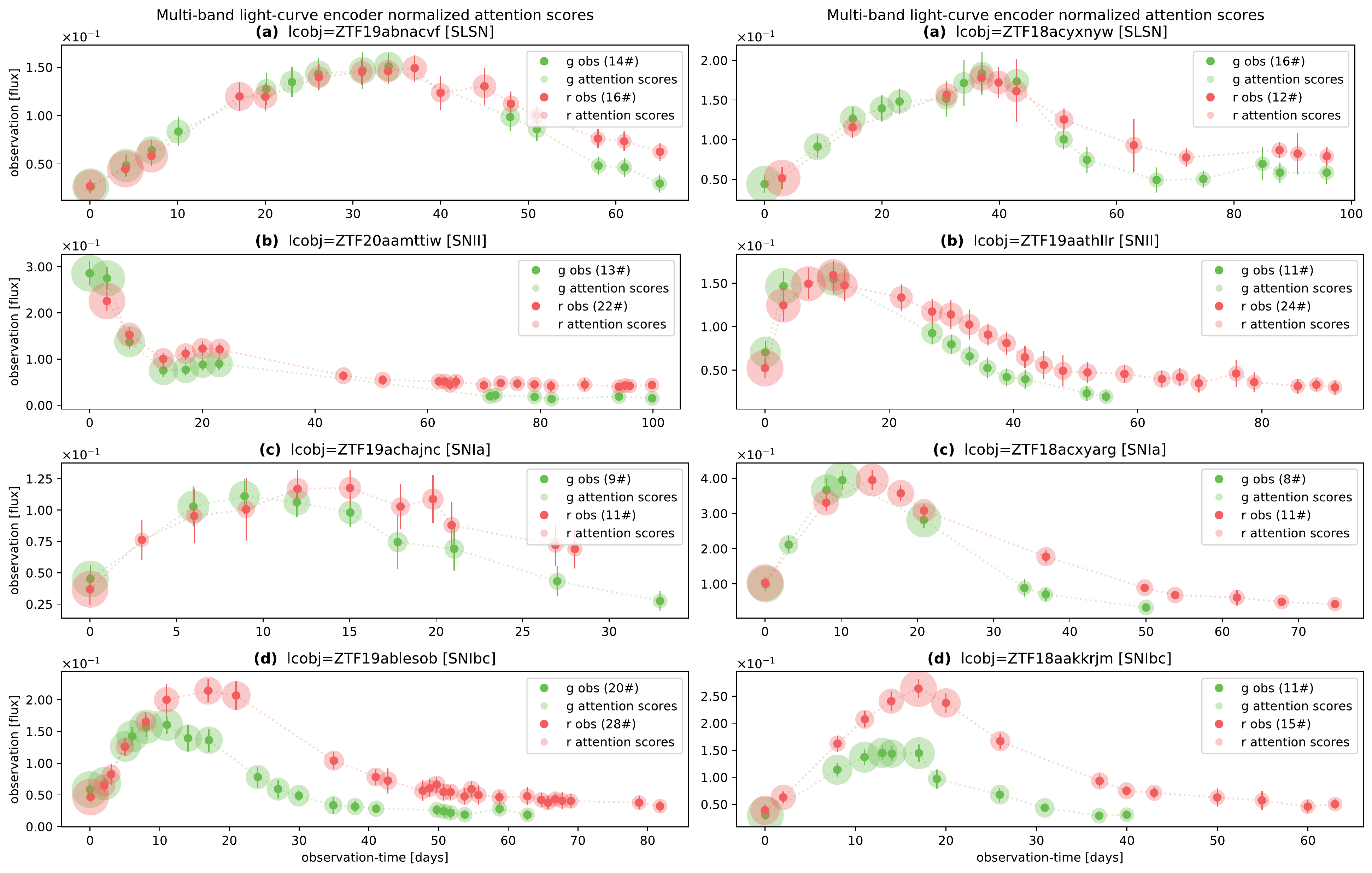}
\caption{
Examples of the normalized attention scores for the P-TimeModAttn model after pre-training. The bigger the shadow circle, the higher the attention score of an observation. (a) SLSN type. (b) SNII type. (c) SNIa type. (d) SNIbc type.
}
\label{fig:attnscores_test}
\end{figure*}

Fig. \ref{fig:attnscores_test} shows examples of the normalized attention scores for different SN multi-band light-curves. In general, the model tended to assign high normalized attention scores to early observations from the SN events, i.e., paying more attention to observations earlier than and close to the SN-peak.

We hypothesize that this attention behavior is produced because the first observations seem to contain highly valuable information about the following evolution of the SN transient events. For example, the first observations can help the model to construct inner representations of the SN rising time (SN-rise region) and maximum brightness (SN-peak) that are relevant to discriminate among SNe, such as the SNIa and SNIbc types. In addition, the early SN observations (before the SN-peak) can be scarce due to the limited duration of the SN events and irregular cadence, which could also explain how the model handles the inner attention resources, prioritizing high attention scores on the early time region.

We highlight the high attention scores for the very first observations of a light-curve. For example, by using just the first observation, the model could construct an initial slope representation to distinguish if an SN light-curve started to be detected before the SN-peak (in the SN-rise region) or after the SN-peak (in the SN-fall region) due to the irregular cadence. The first observation could be used also as an observation-time offset, allowing the model to construct a representation of the elapsed time for each relevant and posterior observation. The first observation-time differences, among the bands, may offer relevant information of the multi-band behavior, which could be especially important when a specific band is started to be followed-up a long time after the rest of the bands.

\subsubsection{Attention-Based Statistics}\label{sec:attnstats} 
The main limitation of the attention score visualization presented above is that it heavily relies on a visual interpretation over a set of examples from a dataset $\dataset{}$. Based on a statistical approach, a new interpretability experiment is conducted to explore the attention behavior over a complete dataset $\dataset{}$. With this aim, we use two simple and interpretable local features for a SN light-curve. Given a single-band light-curve $\Phi_i^\ob$, the following local SN-features are defined:
\begin{align}
&m^*_{i,\job},n^*_{i,\job}=\argminn{m_{i,\job},n_{i,\job}} \sum_{\job'\in \Delta J_i^\ob}^{}\tuple{\obs_{i,\job'}-\hatobs_{i,\job'}}^2, && \text{}\label{eq:sne_local_slope}\\
&\Delta t^*_{i,\job} = \frac{1}{L}\sum_{\job'\in \Delta J_i^\ob}^{} \tuple{ t_{i,\job'}-t_{i,\jmaxob}}, &&\text{}\label{eq:sne_peak_distance}
\end{align}
where the explanation is the following:
\begin{enumerate}
    \item \textbf{SN-local-slope}: Given a linear function $\hatobs_{i,\job'}=m_{i,\job}\cdot t_{i,\job'}+n_{i,\job}$, in eq. \refeq{eq:sne_local_slope}, a local slope value $m_{i,\job}$ and an offset value $n_{i,\job}$ are computed. The optimal values are estimated using a Mean Square Error (MSE) optimization, fitting the linear function over a group of empirical observation-fluxes defined by a window of local sequence steps $\Delta J_i^\ob$ centered in the sequence step $\job$. A window size of $L=\cardi{\Delta J_i^\ob}=3$ is used, i.e., the slope values are fitted using the observation-times $\keys{t_{i,\job-1},t_{i,\job},t_{i,\job+1}}$ and the observation-fluxes $\keys{\obs_{i,\job-1},\obs_{i,\job},\obs_{i,\job+1}}$. Positive values of $m^*_{i,\job}$ are associated with observations with a local increase in brightness, while negative values are associated with observations with a local decrease in brightness.
    
    \item \textbf{SN-peak-distance}: eq. \refeq{eq:sne_peak_distance} represents the average time difference (days) between the observation-times used to fit the SN-local-slope and the SN-peak time: the observation-time associated with the empirical maximum brightness found along the light-curve. The maximum brightness observation-time is denoted as $t_{i,\jmaxob}$, where $\jmaxob=\argmaxx{\job}\allowbreak\set{\obs_{i,\job}}{\job=1}{L_i^\ob}$ corresponds to the sequence step with the maximum observation-flux. Negative values of $\Delta t^*_{i,\job}$ are associated with observations detected earlier than the SN-peak, while positive values are associated with observations detected after the SN-peak.
\end{enumerate}

Given a dataset $\dataset{}$ with $N$ light-curves $\Phi_i$, we can gather a collection of local SN-features, for each observation, as $\set{\set{\tuple{m^*_{i,\job},\Delta t^*_{i,\job},b_{i,\job},\bar{s}_{i,\job}}}{\job\seq1}{L_i^\ob}}{i\seq 1}{N}$, where $m^*_{i,\job}$, $\Delta t^*_{i,\job}$, $b_{i,\job}$ and $\bar{s}_{i,\job}$, are the SN-local-slope, the SN-peak-distance, the band, and the normalized attention score, respectively.

\begin{figure*}[!t]
\centering
\includegraphics[width=1\linewidth]{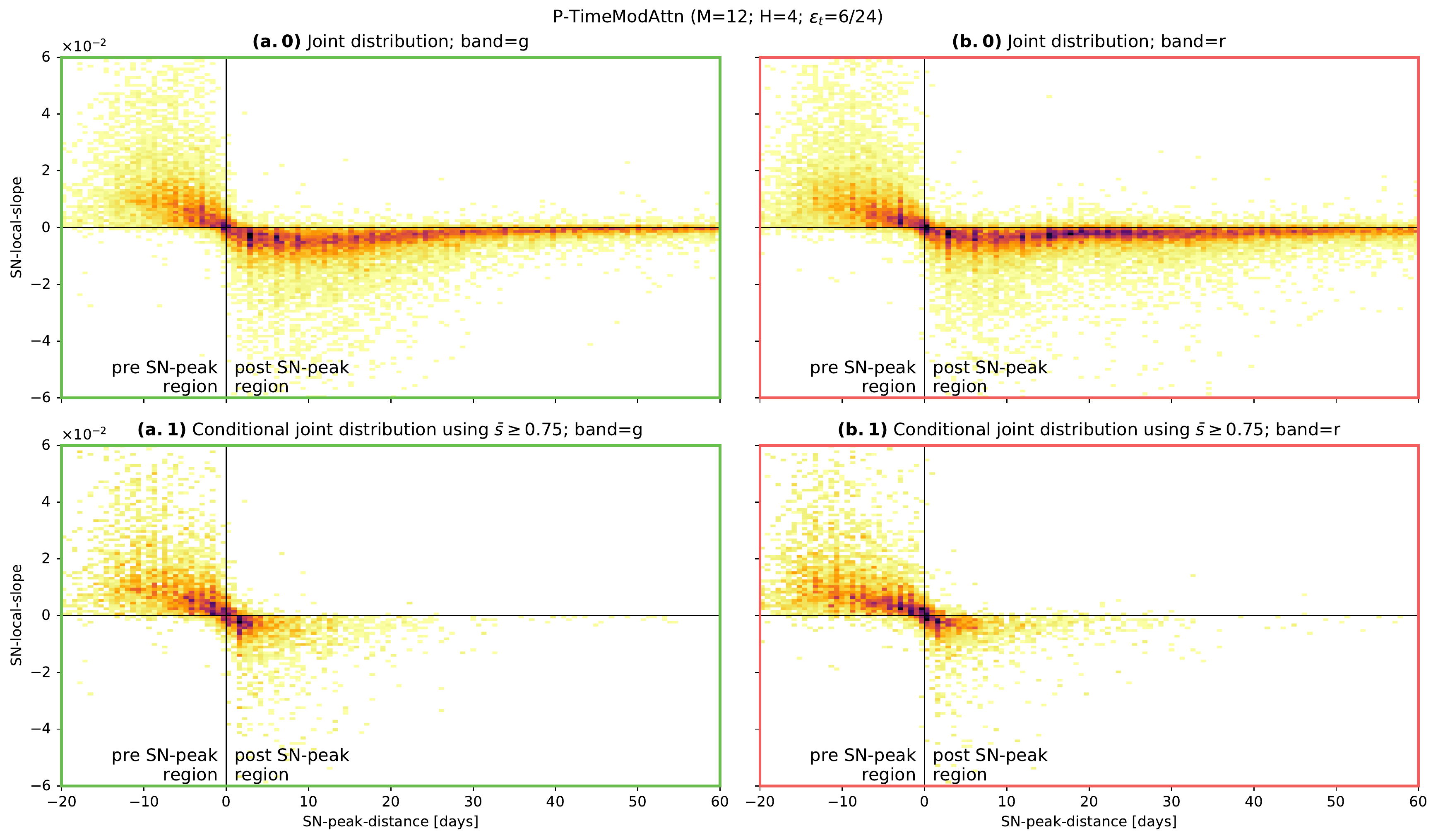}
\caption{
Attention-based statistics for the P-TimeModAttn model after pre-training. In the top row (plots (a.0) and (b.0) corresponding to the g and r band, respectively), the joint distribution is shown for both local SN-features: the SN-local-slope and the SN-peak-distance. In the bottom row (plots (a.1) and (b.1)), the conditional joint distribution is shown using a threshold for normalized attention scores of $\bar{s}_\text{th}=.75$. This allows highlighting the regions with higher attention scores. \highlighttext{Visual guides, for zero SN-local-slope and zero SN-peak-distance, are shown as black lines}. Green borders (plots (a.0) and (a.1)) correspond to the band g. Red borders (plots (b.0) and (b.1)) correspond to the band r.
}
\label{fig:attnstats}
\end{figure*}

Using a probabilistic framework, let $m^*$, $\Delta t^*$, $b$, and $\bar{s}$ be discrete random variables. Fig. \ref{fig:attnstats} shows the joint distribution $p(m^*,\Delta t^*,b)=\sum_{\bar{s}} p(m^*,\Delta t^*,b,\bar{s})$ (marginalizing over the normalized attention score $\bar{s}$), in plots (a.0) and (b.0), for the bands g and r, respectively. An expected SN behavior is observed as the positive SN-local-slope values are distributed earlier than the SN-peak (SN-rise region), while negative SN-local-slope values are distributed after the SN-peak (SN-fall region). Small and zero values of the SN-local-slope are found in two scenarios: close to the SN-peak and in the SN-dimming region.

Fig. \ref{fig:attnstats} shows the conditional joint distribution $p(m^*,\Delta t^*,b|\bar{s}\geq \bar{s}_\text{th})$, in plots (a.1) and (b.1), for the bands g and r, respectively. These distributions show the local SN-features that are related with high normalized attention scores using an attention threshold of $\bar{s}_\text{th}=.75$. When comparing the high distribution density w.r.t. the joint distribution, it can be observed that the high attention region is correlated with the attention score exploration shown in section \refsec{sec:exp_attn}, i.e., the model tended to pay more attention over observations earlier than and close to the SN-peak. The region of high attention was up to several days after the SN-peak, which might be an informative region to characterize the SN-plateau slope and duration\footnote{In general, this behavior is similar for both bands. A similar tendency was observed when using more MHSelfAttn's layers ($N_L=2$) or a different number of $H$ attention heads.}.

\highlighttext{Our findings suggest that early SN observations are the most important observations for the TimeModAttn model. Moreover, these observations can be directly captured by the attention-based models even in the case of long light-curves. We think this is important during training as early observations may be always available regarding the length of the light-curve, helping in the generalization of incomplete light-curves. In contrast, the RNN models may be biased to complete light-curves during training as their processing is forced to be sequential through all the sequence steps, which may hurt the generalization of incomplete light-curves. Moreover, RNN models may have difficulties when capturing information from early observations because the maximum path length could be large and composed of uninformative observations (SN-dimming region), especially in long-duration SN light-curves. This may explain why the TimeModAttn model achieved a general higher performance in early-classification than the RNN baselines.}

\subsubsection{Temporal Modulation Variability}\label{sec:timemod_var} 
\begin{figure*}[!t]
\centering
\includegraphics[width=1\linewidth]{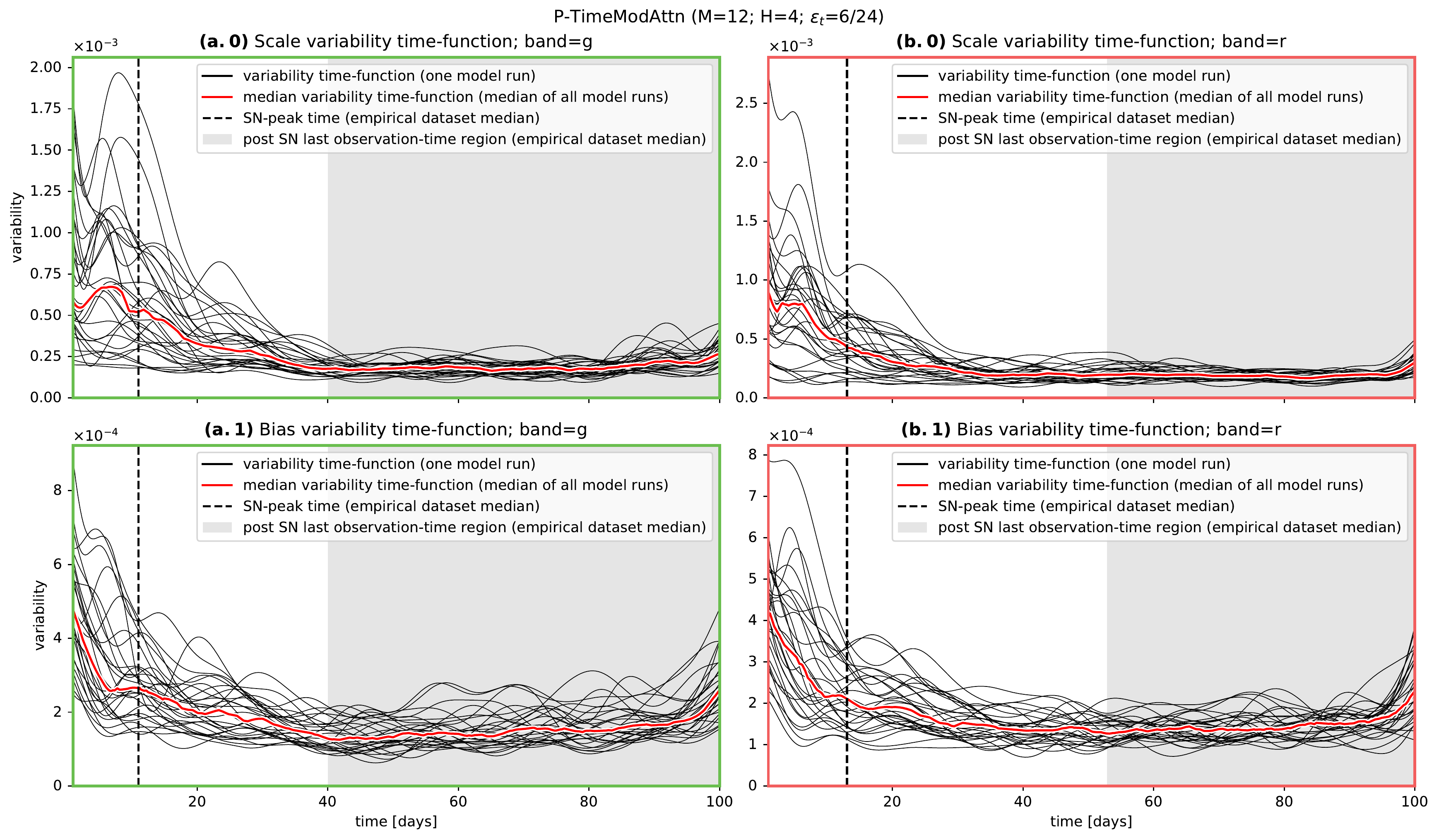}
\caption{
Variability functions for the scale time-function (plots (a.0) and (b.0)) and bias time-function (plots (a.1) and (b.1)) computed from the temporal modulations learned by the P-TimeModAttn model after the pre-training process as a function of time in the range $[0,100]\tunits{days}$. A high variability can be observed before the SN-peak time (dotted line). Each model iteration (total of $N_\text{runs}$) is represented with a black curve. The red curves are computed as the median curve using all the $N_\text{runs}$ model iterations. The SN-peak time is computed as the median empirical maximum brightness time from the original dataset $\dataset{}$. The gray region starts at the empirical median SN last observation-time computed from the dataset $\dataset{}$. Green borders (plots (a.0) and (a.1)) correspond to the band g. Red borders (plots (b.0) and (b.1)) correspond to the band r.
}
\label{fig:variability}
\end{figure*}

In this section, the scale and bias variability time-functions of the proposed temporal modulation (TimeFiLM) are further analyzed. Given an arbitrary encoder associated with the band $b$, the temporal modulation variability time-functions are defined as follows:
\begin{align}
\bar{\gamma}^\ob(t)&=\frac{1}{K}\sum_{k\seq 1}^{K}\tuple{\frac{\partial\gamma^\ob_k(t)}{\partial t}}^2,&&\text{}\label{eq:scale_variability}\\
\bar{\beta}^\ob(t)&=\frac{1}{K}\sum_{k\seq 1}^{K}\tuple{\frac{\partial\beta^\ob_k(t)}{\partial t}}^2,&&\text{}\label{eq:bias_variability2}
\end{align}
where the functions $\bar{\gamma}^\ob(t)$ and $\bar{\beta}^\ob(t)$ are the variability time-functions for the scale and bias, respectively. These time-functions are defined as the average variability of the $K$ modulation time-functions learned by the model. The variability is defined as the squared derivative of the modulation functions w.r.t. the time value $t$. Thus, high values of the variability time-functions indicate a high average variability of the modulation over time. Fig. \ref{fig:variability} shows the learned scale and bias variability time-functions for each run of the model.

For both, the scale and bias variability time-functions, a general high variability over the early time range can be observed, i.e., earlier than the empirical median of the SN-peak time. This high variability could help the model to properly capture and differentiate small changes of the time values. This effect can be seen as a high temporal modulation resolution over the early time range\footnote{Note that the scale function is the modulation operation that can offer a high degree of change when applying the temporal modulation, with direct attenuations and sign inversions.}. This behavior is also correlated with the early high attention scores previously explored. We hypothesize that a high resolution is required to support the attention mechanisms, over the early time range, to correctly characterize the SN most important time regions. This high resolution can be beneficial when computing correlations between the time-modulated query and key vectors, as well as the final vector aggregation using the time-modulated value vectors.

The observed high variability, for both modulation functions, starts decreasing after passing the empirical median of the SN-peak time. A lower modulation resolution could be required in this time range, as the observation distribution becomes scarcer and sparser because the majority of light-curves have stopped being observed. Moreover, a large population of low attention score observations can be found in this time range, suggesting the presence of irrelevant observations. A final high variability when reaching the time $t=100\tunits{days}$ can be observed. This final variability rise might be influenced by long-duration light-curves (e.g., SNII, SLSN) or by modulation instabilities when reaching the time range $t>100\tunits{days}$, where no empirical observations were found during training\footnote{In general, this variability behavior was similar for both bands. A similar tendency was observed when using more MHSelfAttn's layers ($N_L=2$) or a different number of $M$ harmonic components for the temporal modulation.}.

\subsection{Empirical Computational Cost} 
Table \ref{tab:times} shows statistics of the optimization for the RNN baselines and TimeModAttn models. In an effort to fairly compare different models capacities, we also include the metric $\text{mbIT}/\#\text{p}$, representing the training time per total number of model learnable parameters. We do not include the BRF baseline because its optimization does not rely on GPU usage and it heavily depends on the CPU specifications and multi-threading strategies.

%
\def\tabsrule{\rule{0pt}{0pt}\rule[0pt]{0pt}{0pt}}
\def\tabbline{\Xcline{1-5}{1.5pt}\tabsrule}
\begin{table*}[!t]
\centering
\caption{
\highlighttext{Total number of learnable parameters (\#p), best epoch, and empirical training times for RNN and attention-based models during the pre-training process (using GPU). We denote mbIT as the mini-batch Iteration Time measured for a complete forward and backward operation by each model.}
}
\label{tab:times}\vspace{.1cm}
\tiny\scriptsize\footnotesize\small\normalsize
\footnotesize
\begin{tabular}{lcccc}
\tabbline
Serial DL models & \#p & best epoch & mbIT [s]  & mbIT/\#p [\SI{}{\micro\relax}s] \tabsrule\\
\cmidrule{2-5}
S-RNN+$\Delta t$ (cell=GRU) & \textbf{267,270} &  \textbf{44.233$\pm$24.052} & .040$\pm$.007 & .151$\pm$.026  \tabsrule\\
S-RNN+$\Delta t$ (cell=LSTM) & 300,294 & 61.967$\pm$31.818 & .041$\pm$.007 & .135$\pm$.023  \tabsrule\\
\rowcolor{clr:light_gray}S-TimeModAttn (M=12; H=8; $\varepsilon_t$=6/24) & 290,310 & 47.100$\pm$23.604 & \textbf{.038$\pm$.006} & \textbf{.132$\pm$.020}  \tabsrule\\
\tabbline
Parallel DL models\\
P-RNN+$\Delta t$ (cell=GRU) & \textbf{234,246} & 65.767$\pm$34.951 & .044$\pm$.008 & .187$\pm$.033  \tabsrule\\
P-RNN+$\Delta t$ (cell=LSTM) & 250,886 & 74.200$\pm$23.484 & .045$\pm$.007 & .179$\pm$.029  \tabsrule\\
\rowcolor{clr:light_gray}P-TimeModAttn (M=12; H=4; $\varepsilon_t$=6/24) & 249,094 & \textbf{52.333$\pm$25.971} & \textbf{.043$\pm$.007} & \textbf{.174$\pm$.027}  \tabsrule\\
\tabbline
\end{tabular}
\end{table*}

For a similar number of parameters, it can be observed that the TimeModAttn models achieved highly competitive empirical training times w.r.t. the RNN baselines. Note that the computational cost of the time modulation is also captured in the reported training times. From our experiments, the use of the serial encoder leads to lower values of mbIT/\#p for each of the tested Deep Learning models, suggesting that the serial encoder is more efficient in terms of training time per total number of parameters\footnote{Even though the parallel encoder is parallel in our formulation, the optimization procedure was sequentially implemented in this work. This implementation handles the computation of all $B$ parallel representation-vectors $\keys{\z_1,\dots,\z_B}$ one after another, which may not be optimal in terms of GPU usage.}.

\highlighttext{Table \ref{tab:times} shows the convergence of the pre-training in terms of the best epochs (early stopping). We only show the convergence of the pre-training as this is notoriously more time-consuming than the fine-tuning process. In general, it is hard to conclude which model converges faster due to the high variance related with the best epoch. In any case, we can observe that the TimeModAttn and GRU models obtain the lowest best epochs, i.e., faster convergence. Also, these results suggest that the use of the serial encoder may lead to a faster convergence for all these models. As a rough estimate, the complete convergence of each run of the TimeModAttn model empirically took between 90 and 120 minutes. This estimation contemplates both, pre-training and fine-tuning, as well as additional optimization routines such as preparation of mini-batches, validation for early stopping, etc. As future work, the convergence of these models could be further and finely measured using special datasets created for this type of experiment.}
\section{Conclusions}\label{sec:conclusions} 
In this work, a Deep Learning model (TimeModAttn), based on attention mechanisms (MHSelfAttn) with temporal modulation (TimeFiLM), was proposed to process and classify multi-band light-curves for different SN types. The proposed model avoids the requirement for hand-crafted feature computations, missing-value assumptions, and explicit light-curve imputation and interpolation methods. The training process was performed in two sequential steps. First, a pre-training process was performed, using synthetic SN multi-band light-curves with two simultaneous goals to solve: a multi-band light-curve reconstruction task and an SN type classification task. Second, a domain adaptation fine-tuning process was performed, using empirical multi-band light-curves, in a supervised learning scheme, to solve a classification task. Moreover, we proposed a method for the generation of synthetic SN multi-band light-curves, which is based on the SPM function. This helps to increase both, the number of samples and the diversity of the irregular cadence population.

Using SN multi-band light-curves from the ZTF survey, we first tested the proposed TimeModAttn model in the late-classification scenario using different performance metrics. From our experiments, we found that the TimeModAttn model outperformed the feature-based BRF baseline trained with real data. By comparing the confusion matrices of the TimeModAttn model, w.r.t. the BRF baseline, it was observed that the TimeModAttn model obtained fewer confusions between the SN types, with maximum and significant increments of the True Positive (TP) percentages for the SNIa, SNIbc, SNII, and SLSN types. These results are correlated with the ROC curves separation between the TimeModAttn model and the BRF baseline. We highlight the confusion reduction obtained between the SNIa and SNIbc types, especially recalling the importance of the SNIa type for cosmology.

In the early-classification scenario, we found that the TimeModAttn model achieved the maximum BRF's trained with real data reported b-AUCROC performance several days earlier. This indicates that the TimeModAttn model can correctly discriminate between SN types using fewer observations, i.e., shorter SN multi-band light-curves. \highlighttext{This early-classification capability of the TimeModAttn model could be especially useful for future high-volume data surveys such as the LSST survey, where a fast and accurate classification of astronomical events must be performed with the fewest number of observations as possible}. These findings show us that the TimeModAttn model can be effectively used to process and classify multi-band light-curves from different SN types, without relying on any costly hand-crafted feature computation.

\highlighttext{Next, we compared the TimeModAttn model w.r.t. the BRF baseline trained with synthetic data. In general, we highlight the fact that using synthetic data for training improved the performance of both, the TimeModAttn model and the BRF baseline. The TimeModAttn model shows a similar overall performance than the BRF in both tested settings (early and late-classification), but there are no statistical significant differences in the context of this work. In any case, the TimeModAttn model has other advantages over the feature-based models (BRF baseline). We highlight, for example: 1) Although the performance of both models improves when using synthetic data, the use of synthetic data is only scalable when using raw light-curves directly because the computation of features becomes extremely costly when including a large number of synthetic light-curves. 2) Possible useful data-augmentation techniques are only really scalable when using raw light-curves. 3) In contrast to the BRF baseline, the inference of the TimeModAttn model is straightforward for short light-curves with one or few observations. This issue is not solved by using synthetic light-curves. 4) The design of features still heavily depends on expert knowledge which is a non-trivial and costly task that should always be in constant revision, especially if new types of astronomical objects need to be studied.}

\highlighttext{The TimeModAttn model outperformed the tested RNN baselines (LSTM and GRU models). The proposed model obtained a higher performance, which is statistically significant, in both scenarios: the late-classification and early-classification w.r.t. the RNN baselines. For both types of Deep Learning models, we found that training with synthetic light-curves allowed to increase their general performance. We conjecture that attention-based models have the advantage of being able to access any observation from a light-curve regarding the total length or current sequence step. In contrast, in RNN models the processing is forced to be sequential through all the sequence steps (although many of them could be uninformative). Thus, we conjecture that the proposed model performs better as it can focus and pay attention to the relevant observations given the SN context, i.e., the early observations from the SN light-curves.}

On the other hand, by testing different levels of data-augmentation, we observed a high sensitivity of the RNN baselines against the level of data-augmentation. Specifically, a strong data-augmentation was highly detrimental for the RNN baselines. This effect may be due to discrepancies in the time difference distributions between the training-set and test-set when using the proposed data-augmentation procedures. In contrast, the TimeModAttn model showed high robustness against the data-augmentation levels. This robustness could be helpful when a significant discrepancy in the irregular cadence, between the training-set and test-set, may be expected due to survey conditions. Additionally, we found that the TimeModAttn model was highly competitive w.r.t. the RNN baselines in terms of the reported empirical training computational cost.

From our experiments, we found that using all the available band information (multi-band light-curve) resulted in a significant benefit for the TimeModAttn model performance w.r.t. the use of a single-band information. Thus, the serial or parallel encoder should be used to address the SN classification task by capturing all the information from the multi-band light-curves. Additionally, no consistent or strong statistical evidence was found to conclude which encoder alternative is the best for the SN classification task. In this work, we used the parallel encoder to conduct interpretability experiments; however, by taking into account the reported empirical computational cost and convergence of the serial encoder, the latter encoder could be more suitable when faster and more efficient models (and no explicit multi-band interpretability) are required. The parallel encoder could be further explored when a higher number of bands will be available with surveys such as the LSST survey, which will use six bands \citep{Ivezic2019}.

\highlighttext{Since the proposed model processes raw light-curves, it should not be a major problem to apply it to other ongoing and upcoming astronomical surveys based on light-curves. We expect that the proposed model will be scalable to surveys where a large volume of data is expected per night (e.g., the current ZTF or the future LSST surveys), mainly because it is not necessary to perform a continuous and costly computation of features from light-curves. On the other hand, the proposed model can be easily parallelized given that its architecture is based on the multi-head dot-attention mechanism. This can be a great advantage when optimizing pipelines for fast real-time inference of a large volume of data. Additionally, for class inference, the decoder can be discarded from the model to further reduce the inference time for real-time classification.}

Finally, we conducted several experiments on interpretability to explore the automatic decisions of the TimeModAttn model. We observed that the model tended to pay more attention to the first observations of the SN light-curves, i.e., the observations earlier than and close to the SN-peak. This behavior might be because the first observations offer highly valuable information about the evolution of SNe. This early attention behavior could allow the model to construct meaningful inner representations to characterize a SN light-curve, e.g., the initial brightness slope; the SN brightness rising time, maximum peak, and early decay region; the elapsed time between the very first observation and the following observations. We found that the early high attention is correlated with a higher temporal modulation variability or \quo{resolution} over the early time range. This increase in the variability could be beneficial to correctly induce the time information in the attention mechanisms operations: the computation of correlations between the time-modulated query and key vectors, as well as the final vector aggregation using the time-modulated value vectors.

\subsection{Future Work} 
No astrophysical external metadata (e.g., ALLWISE colors, galactic coordinates, SGS score, redshift) was used in this work. As feature work, we propose to extend our model to include metadata values using an extra modulation process. Given that metadata values could exhibit non-Gaussian distribution behaviors (e.g., multi-modal, clipped range, sparse distributions), we can directly use our proposed time modulation as a new \quo{metadata modulation} over the corresponding non-redundant metadata-range associated with the handled metadata values. This modulation could be used, over the sequence input or directly over the encoder representation-vector, including the metadata information in the representation-vectors. This could be extended to multiple metadata values in a multi-layer metadata modulation architecture.

Given the space-state model formulation presented in this work (decoder), the forecasting of SN light-curves could be further explored as this model can evolve arbitrarily over unobserved and future time values. Neural ODE decoders \citep{Chen2018, Rubanova2019} could be also tested to perform continuous-time forecasting, avoiding the use of any explicit time differences information. Attention-based decoders could be also explored to implement an autoencoder model based solely on attention mechanisms. Moreover, a direct projection of the representation-vector (from the encoder) could be used as a decoding strategy, where the temporal modulation could be used to induce the information of the time values.

\highlighttext{We believe that one of the major difficulties to be faced in a new survey may be the nature of its irregular cadence as well as the class imbalance. To further validate our proposed methodology, it would be ideal to process data from other astronomical surveys and, eventually, test it on the future LSST survey.} Alternatively, we propose to classify other astronomical objects, such as stochastic events and periodic stars (using unfolded or folded light-curves). Given that periodic stars usually have a larger number of observations than SNe (thus, longer light-curves), we expect that the use of attention mechanisms could be beneficial given its natural long-term time dependencies learning capability. In addition, new interpretability experiments could be proposed for periodic stars, e.g., by exploring periodicities in the attention scores that could be better exposed by using folded light-curves.

\highlighttext{Through the use of new earth and space facilities thousands of new objects will be detected every night in the near future, allowing the scientific community to discover unexpected and rare events. Thus, it is critically important to design new algorithms that can process photometric information to detect anomalous light-curves. Several works have aimed to tackle anomaly detection \citep[][]{Webb2020, Villar2021, Malanchev2021, Muthukrishna2021, Sanchez-Saez2021-agn}. For example, \citea{Villar2021} proposed the use of a Variational AutoEncoder (VAE) to collapse the information from a light-curve into a latent space. Then, anomalous events can be detected using an Isolation Forest. The encoder proposed in this \docname could be used as an alternative when implementing an unsupervised VAE architecture, exploring anomalous light-curves along with methods such as Isolation Forest. The exploration of the attention scores could help us to understand local behaviors that may cause a light-curve to be anomalous.}

\section{Acknowledgments}
The authors acknowledge support from the National Agency of Research and Development's Millennium Science Initiative through grant IC12009, awarded to the Millennium Institute of Astrophysics (OP, PE, FF) and from the National Agency for Research and Development (ANID) grants: BASAL Center of Mathematical Modelling AFP-170001, ACE210010, FB210005 (FF), and FONDECYT Regular \#1200710 (FF) and \#1220829 (PE). We thank the ALeRCE broker for collecting and providing the data used in this \docname. We also thank Pablo Montero and Nicolás Astorga for the useful discussions.

\textbf{Software}: Pytorch \citep{Paszke2019}, Jupyter\footnote{\url{https://jupyter.org/}.}, Dask \citep{Rocklin2015}, Matplotlib \citep{Hunter2007}, Pandas \citep{Mckinney2011}, Python\footnote{\url{https://www.python.org/}.}, Scikit-learn \citep{Pedregosa2011}, Emcee \citep{Foreman-Mackey2013}.

\appendix
\section{Balanced Multi-Class Performance Metrics}\label{sec:metrics} 
Given a multi-class dataset $\dataset{}$ and an arbitrary target class $c\in\keys{1,\dots,C}$, where $C$ is the total number of classes in $\dataset{}$, a new binary class dataset $\dataset{c}$ is constructed, where $\dataset{c}$ has $C=2$ classes: the positive class \quo{$c$} and the negative class \quo{$\bar{c}$}. The new binary classes are assigned according to the original true class label $c$, and assigning the auxiliary negative class $\bar{c}$ to every sample from any other class different than the positive class $c$. Similarly, the new model binary class predictions are assigned according to the original model class prediction $c_i=\argmaxx{c}\vect{\hat{y}_{i,1},\dots,\hat{y}_{i,C}}$, where $c_i$ is the class associated with the highest predicted probability \footnote{\url{https://scikit-learn.org/stable/modules/generated/sklearn.metrics.precision_recall_fscore_support.html}.}.

Given a binary class dataset $\dataset{c}$, the Precision, Recall, and $F_1$score metrics are defined as follows:
\begin{align}
\precision_{c} &= \frac{TP_{c}}{TP_{c}+FP_{c}},\\
\recall_{c} &= \frac{TP_{c}}{TP_{c}+FN_{c}},\\
F_1\text{score}_{c} &= 2\cdot \frac{\precision_{c}\cdot\recall_{c}}{\precision_{c}+\recall_{c}},
\end{align}
where $TP_{c}$, $FP_{c}$, and $FN_{c}$ stand for the True Positive, False Positive, and False Negative binary class prediction scenarios given the dataset $\dataset{c}$, respectively.

The Receiver Operating Characteristic (ROC)\footnote{\url{https://scikit-learn.org/stable/modules/generated/sklearn.metrics.roc_curve.html}.} and the Precision-Recall (PR)\footnote{\url{https://scikit-learn.org/stable/modules/generated/sklearn.metrics.precision_recall_curve.html}.} curves are constructed by using the predicted probability $\hat{y}_{i,c}$ for the positive class \quo{$c$} and the probability $1-\hat{y}_{i,c}$ for the negative class \quo{$\bar{c}$}. For the experiments, the Area Under the Curve (AUC) is reported for both, the ROC curve (AUCROC) and the PR curve (AUCPR).

All kind of metrics computed for the binary datasets $\dataset{c}$ can be aggregated into a new balanced metric. For example, the balanced $F_1$score is computed as follows:
\begin{align}
    \text{b-$F_1$score} &= \frac{1}{C}\sum_{c=1}^C \text{$F_1$score}_{c},
\end{align}
where each class performance is equally important in the final balanced metric. The same process can be applied to each metric used in this work, obtaining the following balanced metrics: b-Precision, b-Recall, b-$F_1$score, b-AUCROC, and b-AUCPR.
\section{SPM Bounds and MCMC Prior Distribution}\label{sec:spm_bounds} 
The Maximum Likelihood Estimation (MLE) optimization is performed using the curve-fit algorithm\footnote{\url{https://docs.scipy.org/doc/scipy/reference/generated/scipy.optimize.curve_fit.html}.}. To ensure positive flux values and a general MLE fit stabilization, we impose valid bounds over the SPM parameter values, as shown in Table \ref{tab:spm_bounds}. The initial MLE parameter guesses $p_0$ are also shown.

%
\def\tabsrule{\rule{0pt}{10pt}\rule[0pt]{0pt}{0pt}}
\def\tabbline{\Xcline{1-3}{1.5pt}\tabsrule}
\begin{table}[!t]
\centering
\caption{
SPM bounds and MLE initial guesses $p_0$, given an arbitrary single-band light-curve $\Phi_i^\ob$, for different SPM parameters (P). The sequence step $\jmaxob=\argmaxx{\job}\allowbreak\set{\obs_{i,\job}}{\job\seq1}{L_i^\ob}$ corresponds to the sequence step with the maximum observation-flux (maximum brightness). The observation-times $\mathcal{T}_i^\ob=\set{t_{i,\job}}{\forall \job|\obs_{i,\job} \geq \frac{1}{3} \obs_{i,\jmaxob}}{}$ are the observation-times at which the observation-fluxes are higher than the brightness threshold $\frac{1}{3} \obs_{i,\jmaxob}$, where $\obs_{i,\jmaxob}$ is the maximum observation-flux. The observation-time $t_{i,1^\ob}$ is the first observation-time from the single-band light-curve $\Phi_i^\ob$.
}
\label{tab:spm_bounds}\vspace{.1cm}
\scriptsize\footnotesize\small\normalsize\small
\scriptsize
\begin{tabular}{lcc}
\tabbline
P & Lower \& upper SPM bounds & $p_0$ \tabsrule\\
\cmidrule{2-3}
$A_i^\ob$                     &  $\tuple{\frac{1}{5} \obs_{i,\jmaxob}, 5 \obs_{i,\jmaxob}}$ & $ 1.2 \obs_{i,\jmaxob}$  \tabsrule\\
$\tzeroob_i$      &   $\tuple{t_{i,1^\ob}-10,t_{i,\jmaxob}+50}$  &  $t_{i,\jmaxob}$ \tabsrule\\ 
$\gamma^\ob_i$ &  $\tuple{1,120}$ & $\maxx{}\mathcal{T}_i^\ob -\minn{}\mathcal{T}_i^\ob$ \tabsrule\\
$\spmbeta_i^\ob$                     & $\tuple{0,1}$  &  $.5$ \tabsrule\\
$\triseob_i$       & $\tuple{1,50}$  & $\frac{1}{2}\tuple{t_{i,\jmaxob}-t_{i,1^\ob}}$ \tabsrule\\
$\tfallob_i$        & $\tuple{1,130}$  & $40$ \tabsrule\\
\tabbline
\end{tabular}
\end{table}

As mentioned in section \refsec{sec:synth.mcmc}, an isotropic multivariate Gaussian distribution is used for the MCMC prior distribution with a diagonal standard deviation matrix $\Sigmabf\in\real{6 \times 6}$. Each diagonal entry is proportional to the associated SPM bound range. For example, for the SPM parameter $A$, we define the standard deviation entry $\Sigma_{1,1}$ as $\sigma_{A_i^\ob}=k\tuple{\text{sup}\tuple{A_i^\ob}-\text{inf}\tuple{A_i^\ob}}$, where $k=.1$ is a scaling factor. Additionally, all Gaussian distributions used in this work are truncated by the SPM bounds to avoid sampling any invalid SPM parameter value during the MCMC optimization.

\section{Conditional Observation-Error Distribution Estimation}\label{sec:obse_estimation} 
\begin{figure*}[!t]
\centering
\includegraphics[width=1\linewidth]{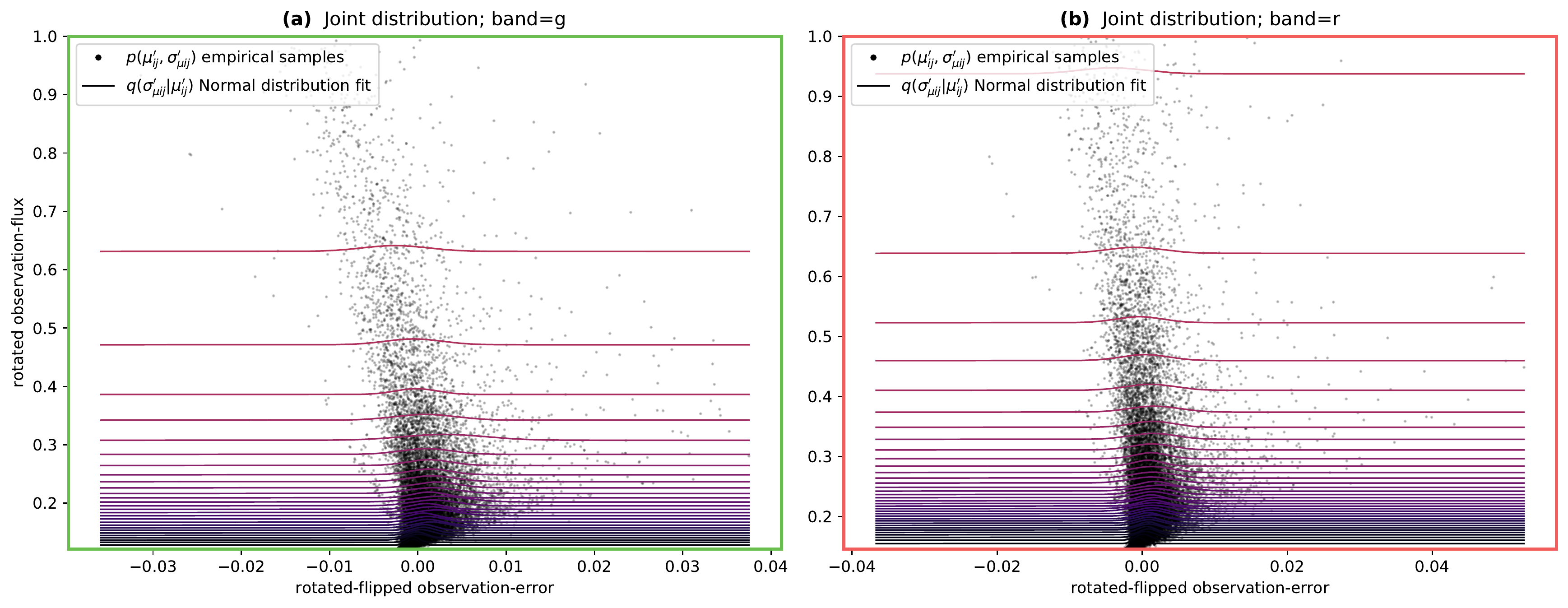}
\caption{
Gaussian distribution fits for the conditional observation-error distribution (from the training-set $\dataset{train}$ of an arbitrary fold split). Green border (a) corresponds to the band g. Red border (b) corresponds to the band r.
}
\label{fig:cond_distr}
\end{figure*}

To estimate the observation-error versus the observation-flux conditional distribution, we use empirical samples from the training-set $\dataset{train}$. We compute the maximum dispersion axis over the joint distribution $p(\obs,\obse,b)$ using the principal component from a Principal Component Analysis (PCA) reduction\footnote{\url{https://scikit-learn.org/stable/modules/generated/sklearn.decomposition.PCA.html}.}. Then, a rotation operation of the original space $p(\obs,\obse,b)$ is performed by using the maximum dispersion axis slope for the construction of a rotation linear projection $\mtw\in\real{2\times 2}$ (see Fig. \ref{fig:cond_distr} for an example of the rotated space samples). Next, a collection of Gaussian distributions are fitted by MLE over the rotated space $p(\obse'|\obs',b)$ by using several binned regions, where each bin has at least 50 empirical samples (see Fig. \ref{fig:cond_distr} for examples of the Gaussian distribution fits).

To generate a new observation-error sample $\hatobse$, given the observation-flux $\obs$ and the band $b$, we first find the corresponding bin (target bin) associated with the observation-flux $\obs$ in the rotated space. Then, we sample a rotated observation-error $\hatobse'$ from the fitted Gaussian distribution associated with the target bin. Finally, the rotated observation-error $\hatobse'$ is rotated back using the inverse rotation linear projection $\mtw^{-1}$ to obtain the required observation-error sample $\hatobse$.
\section{Model Input Normalization}\label{sec:normalization} 
For the serial formulation, the input vectors are normalized as follows:
\begin{align}
\x_{i,j} &\gets \frac{\x_{i,j} -\text{mean}\tuple{\x}}{\text{std}\tuple{\x}+\varepsilon},
\end{align}
where the mean and standard deviation vectors are computed over the auxiliary vector set $\x = \keys{\set{\x_{1,j}}{j\seq1}{L_1},\dots,\set{\x_{N,j}}{j\seq1}{L_N}}$, which is a variable-length vector collection consisting of all the samples from the synthetic training-set $\dataset{train[s]}$.

For the parallel formulation, the normalization follows as follows:
\begin{align}
\x_{i,\job} &\gets \frac{\x_{i,\job} -\text{mean}\tuple{\x^\ob}}{\text{std}\tuple{\x^\ob}+\varepsilon},
\end{align}
where an auxiliary vector set is collected, given the band $b$, as $\x^\ob = \keys{\set{\x_{1,\job}}{\job\seq1}{L_1^\ob},\dots,\set{\x_{N,\job}}{\job\seq1}{L^\ob_N}}$.

This method is used to normalize the input for both, the encoder model and the decoder model (time difference values). It is also used over the observation-fluxes evaluated in the reconstruction loss shown in section \refsec{sec:losses}.

\section{Class Balance Strategy}\label{sec:balance_learning} 
Algorithm \ref{alg:balance} describes the strategy used to dynamically construct an auxiliary and balanced training-set $\dataset{train}^\text{balanced}$. This strategy is used in both, the pre-training and fine-tuning processes to deal with the class imbalance.

\SetInd{0.5em}{0.2em}
\begin{algorithm}[!t]
\caption{
Dynamic training-set class balancing strategy.
}
\label{alg:balance}
$\data_{c} = \set{\Phi_i|y_i=c}{i\seq 1}{N} \subset \dataset{train}$\tcp{\scriptsize Collect the light-curves, of class $c$, into a new auxiliar subset $\dataset{(c)}$}
$N_\text{max}=\maxx{}\set{N_{c}= \cardi{\data_{c}}}{c\seq 1}{C}$\tcp{\scriptsize Find the maximum population $N_\text{max}$ associated with the most populated class $c$}
\For{$epoch \in epochs$}{
 $\dataset{train}^\text{balanced}=\keys{\varnothing}$\tcp{\scriptsize Init a new empty balanced training-set}
  \For{$c \in \keys{1,\dots,C}$}{
  \For{$i \in \keys{1,\dots,N_\text{max}}$}{
  $\Phi_i'\sim\dataset{(c)}$\tcp{\scriptsize Randomly choose a light-curve of class $c$ with probability $p=\frac{1}{N_{c}}$}
   $\dataset{train}^\text{balanced}\overset{\cup}{\gets}\keys{\Phi_i'}$\tcp{\scriptsize Append the selected light-curve into the balanced training-set}
  }
 }
 }
\end{algorithm}

\section{Multi-Band Light-Curve Data-Augmentation}\label{sec:dataaugmentation} 
Algorithm \ref{alg:dataaugmentation} describes the data-augmentation strategy used to dynamically construct new multi-band light-curves during the training processes, allowing us to induce a degree of variability in the light-curves processed by the models. This strategy is used in both, the pre-training and fine-tuning processes.

\SetInd{0.5em}{0.2em}
\begin{algorithm}[!t]
\caption{
Dynamic data-augmentation strategy for a multi-band light-curve $\Phi_i$.
}
\label{alg:dataaugmentation}
 \If{model is training}{
   \For{$b\in \keys{1,\dots,B}$}{
  $\Phi_i^\ob\gets f_\text{lcrss}\tuple{\Phi_i^\ob}$\tcp{\scriptsize LCRSS}
  $\Phi_i^\ob\gets f_\text{lcrod}\tuple{\Phi_i^\ob}$\tcp{\scriptsize LCROD}
  \For{$\job\in \keys{1,\dots,L_i^\ob}$}{
  $\obs_{i,\job} \gets \obs_{i,\job} + k\cdot\obse_{i,\job}\cdot\varepsilon, \varepsilon \sim\tdist{\nu}$\tcp{\scriptsize LCORE}
 }
 }
$\Phi_i\gets f\tuple{\keys{\Phi_i^{(1)},\dots,\Phi_i^{(B)}}}$\tcp{\scriptsize Re-define the multi-band light-curve using the new single-band light-curves}
 \For{$b\in \keys{1,\dots,B}$}{
 $t_{i,\job}\gets t_{i,\job}-t_{i,1},\forall \job$\tcp{\scriptsize Observation-time re-offset}
 }
 }
\end{algorithm}

The data-augmentation explanation is as follows:
\begin{enumerate}
\item \textbf{Light-Curve Random Sub-Slide (LCRSS)}: given a single-band light-curve $\Phi_i^\ob$, a random light-curve sub-slide is selected from $\Phi_i^\ob$, re-defining the original single-band light-curve. The sub-slide is performed by randomly sampling both, an initial sequence step and a new variable-length $L_i^\ob$.

\item \textbf{Light-Curve Random Observation Dropout (LCROD)}: given a single-band light-curve $\Phi_i^\ob$, random individual observations are removed from the $\Phi_i^\ob$, re-defining the original single-band light-curve. A dropout probability of $p_\text{lcrod}$ is used.

\item \textbf{Light-Curve Observation-flux Re-Estimation (LCORE)}: given a single-band light-curve $\Phi_i^\ob$, and following the same method shown in section \refsec{sec:parametric}, a clipped t-student distribution is used to re-sample each observation-flux $\obs_{i,\job}$.

\item \textbf{Observation-time re-offset}: to avoid ill-defined multi-band light-curves $\Phi_i$, the first observation-time $t_{i,1}$, from the resulting multi-band light-curve, is subtracted from all the observation-times. This is performed to construct new multi-band light-curves where the first observation-time is zero: $t_{i,1}=0$. This procedure is also performed whenever changes in the observation-times occur. For example, when applying pre-processing methods (section \refsec{sec:preprocessing}) and generating synthetic light-curves (section \refsec{sec:synth}).
\end{enumerate}

Three main levels of data-augmentation are used in this work: zero, weak, and strong data-augmentation levels. In the zero data-augmentation level, all the aforementioned procedures are ignored, returning the original multi-band light-curves. In the weak data-augmentation level, a dropout probability of $p_\text{lcrod}=.1$ is used. In the strong data-augmentation level, a probability of $p_\text{lcrod}=.5$ is used. Because some data-augmentation procedures remove observations, a minimum single-band light-curve length threshold of $L_i^\ob \geq 5$ is imposed as a new augmented light-curve requirement to avoid problems of short or empty light-curves.
\section{Multi-band Light-Curve Reconstruction}\label{sec:reconstruction} 
\begin{figure*}[!t]
\centering
\includegraphics[width=1\linewidth]{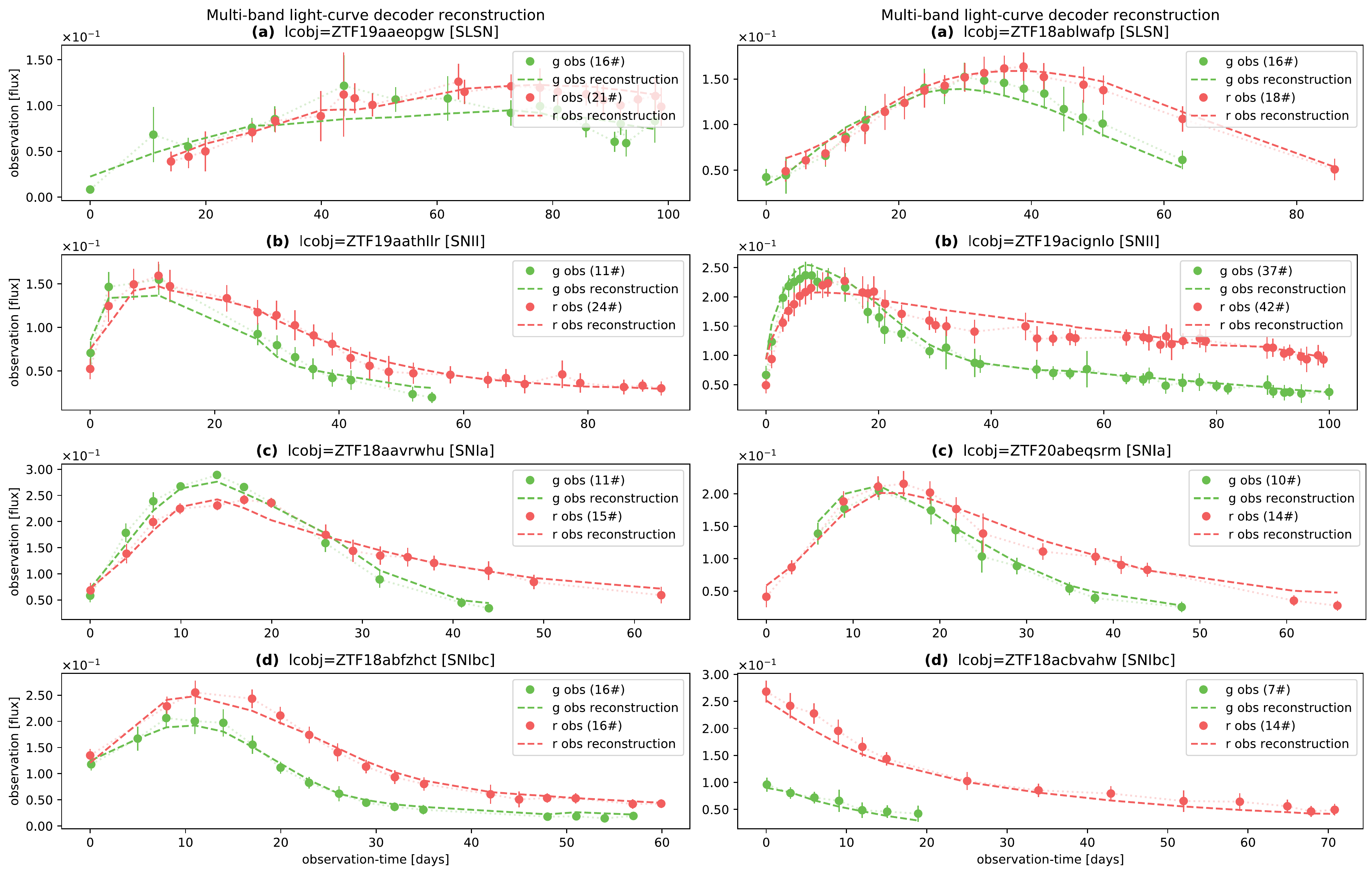}
\caption{
Examples of multi-band light-curve reconstructions for the P-TimeModAttn model after the pre-training process. Dashed lines are used for the reconstructed light-curves. (a) SLSN type. (b) SNII type. (c) SNIa type. (d) SNIbc type.
}
\label{fig:recs}
\end{figure*}

Fig. \ref{fig:recs} shows examples of multi-band light-curve reconstructions for different SN types using the TimeModAttn model. These examples show that the decoder can correctly estimate, given the representation-vector $\z_i$ generated by the encoder, the observation-fluxes for the reconstruction of SN multi-band light-curves. This reconstruction is well-performed despite the existence of long time gaps without any observation due to the irregular cadence.

It is worth noticing that the representation of the time information, handled by the TimeModAttn model, is different for the encoder and decoder. For the encoder, a temporal modulation based on the raw time values is induced in the input; while for the decoder, the time difference values are used as the input. In general, this dual representation of the time values produces a highly challenging scenario for the learning of time dependencies, forcing the autoencoder model to transform raw time value representations into time difference representations. Given a correct optimization process, this dual time representation may ultimately lead to a highly meaningful representation space for the reconstruction task and the classification task. This could explain the high classification performance obtained by the TimeModAttn model, but further analyses and experiments must be performed, which are left for future work.

\section{Ablation Studies and Additional Models}\label{sec:ablations} 
\subsection{Number of Fourier Decomposition Harmonics and Attention Heads}\label{sec:m_and_h} 
\highlighttext{To study the influence and relevance of the components of the TimeModAttn model, the following settings were tested:}

\highlighttext{
\begin{enumerate}
\item \textbf{Case} $M=0$: a fully invariant temporal modulation is directly induced by setting invariant scale and bias time-functions in eq. \refeq{eq:mod_mod}: $\gamma_k(t)=1\land\beta_k(t)=0,\forall k\in\keys{1,\dots,K},\forall t$. This setting allows us to simulate the effect of bypassing the temporal modulation: the encoder can access the information of the observation-flux but not any meaningful information about the observation-time.

\item \textbf{Case} $H=0$: the self-attention mechanism, in the MHSelfAttn layers, is bypassed by imposing $\vtc_{i,j}=\vtzero$ in eq. \refeq{eq:attn_res1}. Thus, in this case, the encoder can only access to the last observation-flux and the last observation-time of the light-curves, as the last observation is still modulated by the TimeFiLM.

\item \textbf{Case} $M=0$; $H=0$: in this extreme case, the encoder can only access the last observation-flux information from the light-curves, but not the last observation-time information.
\end{enumerate}
}

\highlighttext{Table \ref{tab:ablations} shows the results associated with the aforementioned architecture settings. Note that, in all settings, the achieved performances are notably lower w.r.t. the BRF baseline performance, suggesting that the temporal modulation and the attention mechanism are required components for the TimeModAttn model, as expected.}

%
\def\tabsrule{\rule{0pt}{10pt}\rule[0pt]{0pt}{0pt}}
\def\tabbline{\Xcline{1-6}{1.5pt}\tabsrule}
\begin{table*}[!t]
\centering
\caption{
Late-classification performances for the BRF baselines and several ablation case studies for the attention-based models (TimeModAttn) using \cdays multi-band light-curves. Both, the serial (\texttt{S-model}) and parallel (\texttt{P-model}) encoders are reported (mean$\pm$std from 5-fold cross-validation).
}
\label{tab:ablations}\vspace{.1cm}
\tiny\scriptsize\footnotesize\small\normalsize
\footnotesize
\begin{tabular}{lccccc}
\tabbline
Feature-based models & b-Precision$_{ }^{ }$ & b-Recall$_{ }^{ }$ & b-$F_1$score$_{ }^{ }$ & b-AUCROC$_{ }^{ }$ & b-AUCPR$_{ }^{ }$ \tabsrule\\
\cmidrule{2-6}
BRF (fmode=all; training-set=[r]) & .527$\pm$.030 & .687$\pm$.052 & .525$\pm$.039 & .866$\pm$.020 & .602$\pm$.051  \tabsrule\\
BRF (fmode=all; training-set=spm-mcmc-estw[s]) & \textbf{.592$\pm$.032} & \textbf{.719$\pm$.048} & \textbf{.594$\pm$.047} & \textbf{.890$\pm$.018} & \textbf{.654$\pm$.053}  \tabsrule\\
\tabbline
Serial Deep Learning models\tabsrule\\
\rowcolor{clr:light_gray}S-TimeModAttn (M=0; H=0; $\varepsilon_t$=6/24) & .291$\pm$.046 & .333$\pm$.043 & .194$\pm$.052 & .575$\pm$.034 & .305$\pm$.020  \tabsrule\\
\rowcolor{clr:light_gray}S-TimeModAttn (M=0; H=8; $\varepsilon_t$=6/24) & \textbf{.410$\pm$.027} & \textbf{.561$\pm$.058} & \textbf{.386$\pm$.024} & \textbf{.787$\pm$.036} & \textbf{.480$\pm$.032}  \tabsrule\\
\rowcolor{clr:light_gray}S-TimeModAttn (M=12; H=0; $\varepsilon_t$=6/24) & .337$\pm$.014 & .379$\pm$.042 & .307$\pm$.016 & .698$\pm$.028 & .365$\pm$.018  \tabsrule\\
\tabbline
Parallel Deep Learning models\tabsrule\\
\rowcolor{clr:light_gray}P-TimeModAttn (M=0; H=0; $\varepsilon_t$=6/24) & .304$\pm$.015 & .328$\pm$.063 & .163$\pm$.018 & .571$\pm$.037 & .294$\pm$.015  \tabsrule\\
\rowcolor{clr:light_gray}P-TimeModAttn (M=0; H=4; $\varepsilon_t$=6/24) & \textbf{.401$\pm$.018} & \textbf{.547$\pm$.042} & \textbf{.375$\pm$.021} & \textbf{.773$\pm$.029} & \textbf{.469$\pm$.032}  \tabsrule\\
\rowcolor{clr:light_gray}P-TimeModAttn (M=12; H=0; $\varepsilon_t$=6/24) & .366$\pm$.021 & .453$\pm$.050 & .326$\pm$.024 & .724$\pm$.017 & .383$\pm$.017  \tabsrule\\
\tabbline
\end{tabular}
\end{table*}

%
\highlighttext{Using a fully invariant temporal modulation in the encoder ($M=0$) is detrimental for the performance because the irregular cadence (temporal information), from the light-curves, is not given to the encoder. Nevertheless, a degree of classification can still be achieved as the encoder has access to the observation-flux information from the complete light-curves. In fact, the attention mechanism is able to compute flexible statistics from the observation-flux distribution even if no temporal information is available. For example, the attention mechanism could learn to compute the mean and standard deviation of the observation-flux distribution, weighted sums to ignore the SN-dimming region, getting information about the SN-peak by paying attention to the maximum observation-fluxes, etc.}

\highlighttext{When bypassing the attention mechanism in the encoder ($H=0$) an even worst classification performance is obtained, as only the information of the last observation is available for the encoder. This confirms that the attention mechanism is required to capture time dependencies from the light-curves and, therefore, meaningful information from the astronomical event. Note that a degree of classification can still be achieved even in this case.}

\highlighttext{As expected, the combination of both cases ($M=0$; $H=0$) produces a model that achieved the worst classification performance. This could be explained as the information obtained from the light-curves, and handled by the encoder, is minimal and insufficient to correctly characterized the light-curves and, therefore, to solve the classification task. \customtext{As an extra note, the worst possible performance (completely random classification) can be achieved by also inducing $\x_{i,j}\gets \vtzero\odot\x_{i,j}$ in the input vector of the modulation operation, which results in a dummy encoder.}}

\subsection{TimeModRNN and CatTimeAttn Models}\label{sec:timemodrnn_cattimeattn} 
\highlighttext{In this section, two additional model settings are tested. First, we study the effect of using the proposed temporal modulation (TimeFiLM) along with RNNs. A new architecture setting is used (TimeModRNN) as follows: the $N_L$ HMSelfAttn's layers of the encoder are replaced with $N_L$ RNN's layers (e.g., GRU, LSTM).}

\highlighttext{In a second experiment, we compare the use of the temporal modulation (TimeFiLM) w.r.t. the use of a Temporal Encoding (TE) to induce the temporal information in the TimeModAttn model. With this aim, a new architecture setting is used (CatTimeAttn), where two changes are implemented: 1) The temporal modulation is completely removed (bypassed) from the encoder. 2) The encoder input vector $\x_{i,j}$, used in the TimeModAttn model, is re-defined in order to include the values of a juxtaposed TE: $\x_{i,j}\gets\cat{\x_{i,j},\t_{i,j}}$. The TE vector is defined as follows:
\begin{align}
\t_{i,j} &= f_\text{TE}\tuple{t_{i,j}}=
\begin{bmatrix}
  \sinfun{\omega_1 t_{i,j}} \\ 
  \cosfun{\omega_1 t_{i,j}} \\ 
  \vdots \\
  \sinfun{\omega_{K/2} t_{i,j}} \\
  \cosfun{\omega_{K/2} t_{i,j}}
\end{bmatrix}, \omega_k=\frac{2\pi}{T_k},
\end{align}
where the TE vector $\t_{i,j}$ has $K$ components by using $K/2$ different periods $\set{T_k}{k\seq1}{K/2}$. For the sake of a fair comparison, we initialize the TE by using the same setting's rule of harmonics used for the Fourier decomposition shown in section \refsec{sec:timefilm} (TimeFiLM). Note that the dimensionality of the TE follows the proportion $K=2M$, where $M$ is the number of harmonics used in the Fourier decomposition.}

\highlighttext{Table \ref{tab:extra_models_late} shows the late-classification performance results obtained for both, the TimeModRNN model and the CatTimeAttn model. In addition, table \ref{tab:extra_models_early} shows the early-classification performance results.}

\def\tabsrule{\rule{0pt}{10pt}\rule[0pt]{0pt}{0pt}}
\def\tabbline{\Xcline{1-6}{1.5pt}\tabsrule}
\begin{table*}[!t]
\centering
\caption{
\highlighttext{Late-classification performances for the BRF baselines, RNN baselines, and attention-based models (TimeModAttn) using \cdays multi-band light-curves. Several variants of TimeModRNN and CatTimeAttn are shown. Both, the serial (\texttt{S-model}) and parallel (\texttt{P-model}) encoders are reported along with several pre-training and data-augmentation schemes (mean$\pm$std from 5-fold cross-validation).}
}
\label{tab:extra_models_late}\vspace{.1cm}
\tiny\scriptsize\footnotesize\small\normalsize
\footnotesize
\begin{tabular}{lccccc}
\tabbline
Feature-based models & b-Precision${}_{ }^{ }$ & b-Recall${}_{ }^{ }$ & b-$F_1$score${}_{ }^{ }$ & b-AUCROC${}_{ }^{ }$ & b-AUCPR${}_{ }^{ }$ \tabsrule\\
\cmidrule{2-6}
BRF (fmode=all; training-set=[r]) & .527$\pm$.030 & .687$\pm$.052 & .525$\pm$.039 & .866$\pm$.020 & .602$\pm$.051  \tabsrule\\
BRF (fmode=all; training-set=spm-mcmc-estw[s]) & \textbf{.592$\pm$.032} & \textbf{.719$\pm$.048} & \textbf{.594$\pm$.047} & \textbf{.890$\pm$.018} & \textbf{.654$\pm$.053}  \tabsrule\\
\tabbline
Serial Deep Learning models\\
S-RNN+$\Delta t$ (cell=GRU) & .545$\pm$.034 & .706$\pm$.070 & .556$\pm$.045 & .879$\pm$.034 & .610$\pm$.066  \tabsrule\\
S-RNN+$\Delta t$ (cell=LSTM) & .550$\pm$.031 & .711$\pm$.070 & .558$\pm$.040 & .887$\pm$.033 & .621$\pm$.070  \tabsrule\\
S-TimeModRNN+$\Delta t$ (cell=GRU) & .581$\pm$.033 & .737$\pm$.061 & .596$\pm$.045 & .898$\pm$.026 & .651$\pm$.058  \tabsrule\\
S-TimeModRNN+$\Delta t$ (cell=LSTM) & .581$\pm$.024 & .749$\pm$.047 & \textbf{.597$\pm$.037} & .900$\pm$.024 & .643$\pm$.057  \tabsrule\\
S-TimeCatAttn (TE=24; H=8) & .577$\pm$.021 & .745$\pm$.046 & .581$\pm$.035 & .905$\pm$.023 & .647$\pm$.064  \tabsrule\\
\rowcolor{clr:light_gray}S-TimeModAttn (M=12; H=8; $\varepsilon_t$=6/24) & \textbf{.588$\pm$.023} & \textbf{.759$\pm$.040} & .596$\pm$.033 & \textbf{.910$\pm$.020} & \textbf{.671$\pm$.056}  \tabsrule\\
\tabbline
Parallel Deep Learning models\\
P-RNN+$\Delta t$ (cell=GRU) & .547$\pm$.030 & .697$\pm$.070 & .552$\pm$.041 & .879$\pm$.031 & .610$\pm$.055  \tabsrule\\
P-RNN+$\Delta t$ (cell=LSTM) & .541$\pm$.022 & .704$\pm$.061 & .540$\pm$.032 & .876$\pm$.029 & .606$\pm$.051  \tabsrule\\
P-TimeModRNN+$\Delta t$ (cell=GRU) & .578$\pm$.033 & .734$\pm$.063 & .589$\pm$.049 & .894$\pm$.028 & .650$\pm$.067  \tabsrule\\
P-TimeModRNN+$\Delta t$ (cell=LSTM) & \textbf{.580$\pm$.026} & .741$\pm$.058 & .592$\pm$.042 & .897$\pm$.025 & .646$\pm$.056  \tabsrule\\
P-TimeCatAttn (TE=24; H=4) & .567$\pm$.025 & .738$\pm$.054 & .575$\pm$.043 & .898$\pm$.024 & .657$\pm$.059  \tabsrule\\
\rowcolor{clr:light_gray}P-TimeModAttn (M=12; H=4; $\varepsilon_t$=6/24) & \textbf{.580$\pm$.020} & \textbf{.753$\pm$.044} & \textbf{.594$\pm$.035} & \textbf{.911$\pm$.017} & \textbf{.689$\pm$.047}  \tabsrule\\
\tabbline
\end{tabular}
\end{table*}

\def\tabsrule{\rule{0pt}{10pt}\rule[0pt]{0pt}{0pt}}
\def\tabbline{\Xcline{1-6}{1.5pt}\tabsrule}
\begin{table*}[!t]
\centering
\caption{
\highlighttext{Early-classification performances for the RNN baselines and attention-based models (TimeModAttn). The moving threshold-day Curve Average (mtdCA) is used ($\ddag$). Several variants of TimeModRNN and CatTimeAttn are shown. Both, the serial (\texttt{S-model}) and parallel (\texttt{P-model}) encoders are reported along with several pre-training and data-augmentation schemes (mean$\pm$std from 5-fold cross-validation).}
}
\label{tab:extra_models_early}\vspace{.1cm}
\tiny\scriptsize\footnotesize\small\normalsize
\footnotesize
\begin{tabular}{lccccc}
\tabbline
Serial Deep Learning models & b-Precision${}_{ }^{\ddag}$ & b-Recall${}_{ }^{\ddag}$ & b-$F_1$score${}_{ }^{\ddag}$ & b-AUCROC${}_{ }^{\ddag}$ & b-AUCPR${}_{ }^{\ddag}$ \tabsrule\\
\cmidrule{2-6}
S-RNN+$\Delta t$ (cell=GRU) & .481$\pm$.030 & .577$\pm$.044 & .454$\pm$.031 & .792$\pm$.024 & .520$\pm$.039  \tabsrule\\
S-RNN+$\Delta t$ (cell=LSTM) & .480$\pm$.023 & .590$\pm$.036 & .457$\pm$.027 & .804$\pm$.026 & .527$\pm$.043  \tabsrule\\
S-TimeModRNN+$\Delta t$ (cell=GRU) & .515$\pm$.027 & .595$\pm$.036 & .492$\pm$.031 & .813$\pm$.021 & .551$\pm$.043  \tabsrule\\
S-TimeModRNN+$\Delta t$ (cell=LSTM) & .512$\pm$.023 & .601$\pm$.025 & .491$\pm$.026 & .815$\pm$.023 & .555$\pm$.046  \tabsrule\\
S-TimeCatAttn (TE=24; H=8) & .513$\pm$.018 & .618$\pm$.025 & .484$\pm$.022 & .834$\pm$.016 & .562$\pm$.038  \tabsrule\\
\rowcolor{clr:light_gray}S-TimeModAttn (M=12; H=8; $\varepsilon_t$=6/24) & \textbf{.522$\pm$.022} & \textbf{.630$\pm$.026} & \textbf{.495$\pm$.020} & \textbf{.841$\pm$.016} & \textbf{.580$\pm$.040}  \tabsrule\\
\tabbline
Parallel Deep Learning models\\
P-RNN+$\Delta t$ (cell=GRU) & .485$\pm$.023 & .580$\pm$.041 & .462$\pm$.028 & .795$\pm$.026 & .524$\pm$.040  \tabsrule\\
P-RNN+$\Delta t$ (cell=LSTM) & .476$\pm$.018 & .586$\pm$.036 & .451$\pm$.024 & .795$\pm$.025 & .516$\pm$.034  \tabsrule\\
P-TimeModRNN+$\Delta t$ (cell=GRU) & .509$\pm$.026 & .597$\pm$.035 & .490$\pm$.029 & .810$\pm$.021 & .550$\pm$.047  \tabsrule\\
P-TimeModRNN+$\Delta t$ (cell=LSTM) & .513$\pm$.021 & .601$\pm$.033 & .490$\pm$.025 & .814$\pm$.022 & .550$\pm$.043  \tabsrule\\
P-TimeCatAttn (TE=24; H=4) & .501$\pm$.019 & .604$\pm$.034 & .482$\pm$.026 & .822$\pm$.019 & .561$\pm$.038  \tabsrule\\
\rowcolor{clr:light_gray}P-TimeModAttn (M=12; H=4; $\varepsilon_t$=6/24) & \textbf{.514$\pm$.018} & \textbf{.621$\pm$.027} & \textbf{.499$\pm$.019} & \textbf{.841$\pm$.015} & \textbf{.587$\pm$.029}  \tabsrule\\
\tabbline
\end{tabular}
\end{table*}

\highlighttext{From the reported experiments, we observe that the TimeModRNN models achieved higher performances than the RNN baselines for both, the late-classification and early-classification. This suggests that the use of the time modulation (TimeFiLM) results in a better representation of the irregular cadence for the encoder, enhancing the overall performance w.r.t. the use of the time difference information. As previously discussed in Appendix \refsec{sec:reconstruction}, this phenomenon could be related with the use of a dual representation of the time information: time modulation in the encoder and time difference in the decoder.}

\highlighttext{On the other hand, the CatTimeAttn models achieved lower performances that the TimeModAttn model for both, the late-classification and early-classification. Note that, in the late-classification scenario, the CatTimeAttn can be highly competitive w.r.t. the TimeModAttn model. In contrast, in the early-classification scenario, the performance difference is larger, suggesting that the TimeModAttn model is a well-suitable alternative for the early-classification scenario.}

\bibliography{refs}{}
\bibliographystyle{aasjournal}

\end{document}